\documentclass[12pt,a4paper,hidelinks]{article}

\usepackage[utf8]{inputenc}
\usepackage{setspace}
\usepackage[left=1in,right=1in,top=1in,bottom=1in]{geometry}
\usepackage{lscape}
\usepackage{pdflscape}
\renewcommand{\baselinestretch}{1.25}

\usepackage{algorithm}
\usepackage[noend]{algpseudocode}

\usepackage{mathtools}
\usepackage{amsmath}
\usepackage{bbm}
\usepackage{amssymb}
\usepackage{graphicx}
\usepackage{array}
\usepackage[labelfont=bf]{caption}
\usepackage{subcaption}
\usepackage{rotating}
\usepackage{graphicx}
\usepackage{placeins}

\usepackage{hyperref}
\usepackage{afterpage} 
\usepackage{longtable} 
\usepackage{comment}
\usepackage{booktabs}
\usepackage{xcolor}

\usepackage[maxbibnames=99,natbib,style=authoryear,backend=biber]{biblatex}
\DeclareLabelalphaTemplate{
  \labelelement{
    \field[final]{shorthand}
    \field{labelname}
    \field{label}
  }
  \labelelement{
    \literal{(}
  }
  \labelelement{
    \field{year}
  }
  \labelelement{
    \literal{)}
  }
}

\usepackage{charter}
\usepackage{microtype}
\usepackage{bm}

\newcommand{\bom}{\boldmath}
\newcommand{\fmbeta}{\mbox{\bom${\beta}$}}

\newcommand{\ind}{\mbox{\bom$\,{1}\!$}}
\newcommand{\iid}{\mbox{$\mathrm{iid}.\ $}}
\newcommand{\N}{\mbox{$\mathcal{N}$}}

\DeclareMathOperator{\argmax}{\arg\,\max\,}
\DeclareMathOperator{\aapprox}{\overset{a}{\sim}\,}

\DeclareMathOperator{\EX}{\mathbb{E}}
\DeclareMathOperator{\MX}{\mathbb{M}}
\DeclareMathOperator{\PX}{\mathbb{P}}

\newcommand{\subsubsubsection}[1]{\paragraph{#1}\mbox{}\\}

\begin{document}

\setlength{\abovedisplayskip}{0em}
\setlength{\abovedisplayshortskip}{0em}
\setlength{\belowdisplayskip}{1em}
\setlength{\belowdisplayshortskip}{1em}

\title{State Dependence and Unobserved Heterogeneity \\ in the Extensive Margin of Trade\thanks{A previous version of this paper was circulated under the title ``Persistent Zeros: The Extensive Margin of Trade''. 
We thank Badi Baltagi, Tanmay Belavadi, Daniel Czarnowske, Jonathan Eaton, Hartmut Egger, Miriam Frey, Mario Larch, Michael Pfaffermayr, Joel Stiebale, Yuta Watabe, Martin Weidner, and Thomas Zylkin, as well as conference and seminar participants at the Aarhus-Kiel Workshop 2018, Bayes Business School CEA Online Research Seminar 2021, Canadian Economic Association 2020, European Trade Study Group 2018 and 2019, European Economic Association 2020, F.R.E.I.T.\ Empirical Trade Online Seminar 2020, Göttingen Workshop ``International Economics'' 2020, Kiel Institute for the World Economy, NOeG 2020, Université Paris 1 Panthéon-Sorbonne, University of Bayreuth, University of Düsseldorf, and University of Potsdam for helpful comments and discussions.
\textbf{\textit{Note}}: An implementation of the estimators developed in this paper is available in the R-package \textit{alpaca} on CRAN.}}
\author{Julian Hinz\thanks{\textit{Affiliation:} Bielefeld University, Kiel Institute for the World Economy \& Kiel Centre for Globalization. \textit{Address:} Universitätsstraße 25, 33615 Bielefeld, Germany.} \quad Amrei Stammann\thanks{\textit{Affiliation:} Ruhr-University Bochum. \textit{Address:} Universitätsstraße 150, 44801 Bochum, Germany.} \quad Joschka Wanner\thanks{\textit{Affiliation:}
University of Potsdam, Kiel Institute for the World Economy. \textit{Address:}
August-Bebel-Straße 89, 14482 Potsdam, Germany.}}

\thispagestyle{empty}

\vspace{-1em}

\maketitle

\vspace{-1em}

\begin{abstract} 
\noindent We study the role and drivers of persistence in the extensive margin of bilateral trade. 
Motivated by a stylized heterogeneous firms model of international trade with market entry costs, we consider dynamic three-way fixed effects binary choice models and study the corresponding incidental parameter problem. The standard maximum likelihood estimator is consistent under asymptotics where all panel dimensions grow at a constant rate, but it has an asymptotic bias in its limiting distribution, invalidating inference even in situations where the bias appears to be small. Thus, we propose two different bias-corrected estimators. Monte Carlo simulations confirm their desirable statistical properties.
We apply these estimators in a reassessment of the most commonly studied determinants of the extensive margin of trade. Both true state dependence and unobserved heterogeneity contribute considerably to trade persistence and taking this persistence into account matters significantly in identifying the effects of trade policies on the extensive margin.

\end{abstract}

\vfill


\newpage
\renewcommand{\baselinestretch}{2}
\setcounter{page}{1}


\section{Introduction}
\label{sec:introduction}

What induces country pairs to trade? In 2019, still more than one third of potential bilateral trade relations reported zero trade flows.\footnote{According to data from IMF DOTS.} Comparing these zero trade flows with trade relations in 2018, these zeros turn out to be extremely persistent: 88.6 percent of country pairs that did not trade in 2018 did not trade in 2019 either, as can be seen in the transition matrix depicted in Table \ref{tab:transition_matrix_2018_2019}. And similarly, 92.3 percent of pairs that \textit{did} trade in 2018 continued to do so in the year after.\footnote{
Note that throughout the paper, ``country pair'' refers to a \textit{directed} pair of countries, i.e.\ Germany-France and France-Germany are two distinct country pairs.} 


\begin{table}[!ht]
    \vspace{1em}
    \centering
    \caption{Persistence in Bilateral Trade Relations (2018 -- 2019)}
    \label{tab:transition_matrix_2018_2019}
    \begin{tabular}{ccc}
        \hline
        & \multicolumn{2}{c}{Traded in 2019}   \\ 
        \cline{2-3}
        Traded in 2018 & No & Yes \\ 
        \hline
       	No & 88.6 \% & 11.4 \% \\
		Yes & ~~7.7 \% & 92.3 \% \\ 
        \hline
    \end{tabular}
\end{table}


\begin{figure}[!t]
    \centering
    \caption{Determinants of the Extensive Margin of Trade --- Gravity and Persistence.}
    \label{fig:share_non_zeros_2019}
    \includegraphics[width=\linewidth]{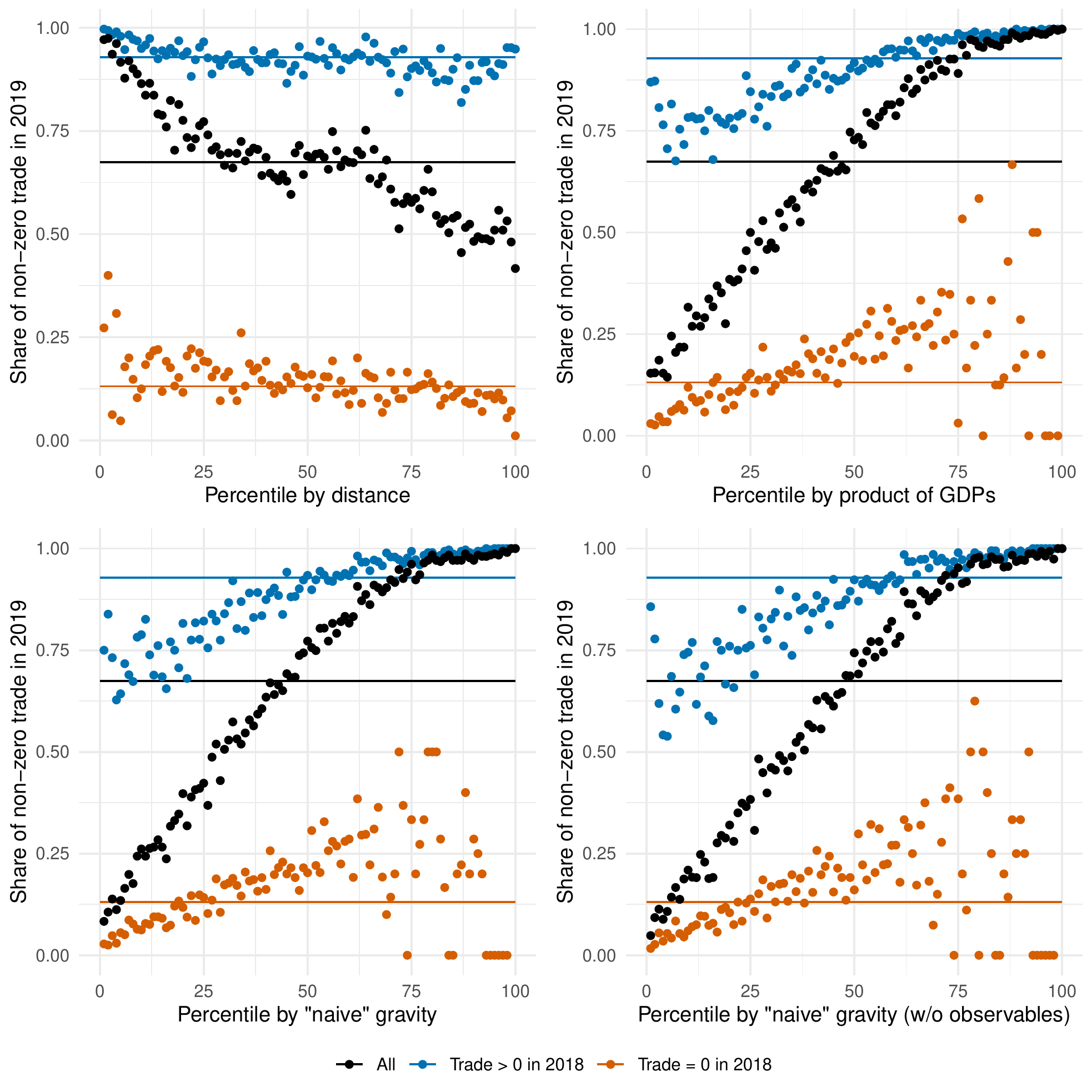}
    \begin{flushleft}
    {\footnotesize \textit{Note:} Trade data for 2018 -- 2019 come from UN Comtrade, GDP, distances and gravity variables are sourced from CEPII \citep{Conte2021}.}
    \end{flushleft}
\end{figure}


Both intuitively and based on existing theoretical and empirical insights, we would expect geographically close and economically large country pairs to have the greater bilateral trade potential and thus be more likely to engage in international trade. As distance is time-invariant and economic size does not change abruptly from one year to another, these gravity-like characteristics may explain (part of) the observed persistence. Figure \ref{fig:share_non_zeros_2019} breaks down the share of non-zero trade flows in 2019 along the percentiles of four different ad hoc indicators of trade potential: bilateral distance; product of GDPs; ``naive'' gravity, i.e.\ the product of GDPs divided by the countries' bilateral distance; and the latter when excluding country pairs in FTAs, with common currencies or common colonial history. The x-axis indicates the potential trade volume, i.e.\ the joint economic size and/or proximity of any two countries.
All four plots paint a common picture: the black dots, covering all country pairs, show a strong general relationship between trade potential and actual non-zero trade.
The blue and red dots split the country pairs according to whether the two did or did not engage in trade in the previous year. The clearly separated pattern for the two groups highlights the remarkable persistence of trade relations, even after controlling for differences in trade potential in terms of distance, size, and bilateral trade policy.
More than 50 percent of those country pairs in the lowest percentiles of trade potential trade again in 2019, provided they already did so in 2018. On the other hand, even comparably large and close pairs are likely not to trade in 2019 if they did not trade in 2018 either.\footnote{A very similar pattern emerges for other points in time (see Figure \ref{fig:share_non_zeros_2000_2001} in Appendix \ref{app:stylized_facts} where the same graph is reproduced for the years 2000--2001). If longer time intervals are considered, a similar picture remains, but the relationship becomes considerably weaker (see Figure \ref{fig:share_non_zeros_2000_2019} in Appendix \ref{app:stylized_facts} for the years 2000--2019).}\\

Two potential features of the extensive margin of trade that can generate the pattern documented by Table \ref{tab:transition_matrix_2018_2019} and Figure \ref{fig:share_non_zeros_2019} are what \citet{Heckman1981} termed ``true state dependence'' and ``spurious state dependence''. The former characterizes countries to actually be more likely to trade \textit{because} they did so in the previous period, whereas the latter describes persistence due to unobservable factors continuously driving bilateral trade potential.\\

In this paper we introduce estimators for the determinants of the extensive margin of international trade that explicitly take its persistence due to observable characteristics, true state dependence, and unobserved heterogeneity into account.
We introduce features from the firm dynamics literature into a heterogeneous firms model of international trade with bounded productivity to derive expressions for an exporting country's participation in a specific destination market in a given period. These expressions depend on partly unobserved (i) exporter-time, (ii) destination-time, and (iii) exporter-destination specific components, as well as on (iv) whether the exporter has already served the market in the previous period, and on (v) exporter-destination-time specific gravity-type trade cost determinants.
We estimate the model making use of recent computational advances in the estimation of binary choice estimators with high-dimensional fixed effects to address (i)-(iii).
The inclusion of fixed effects in a binary choice setting induces an incidental parameter problem that is potentially aggravated by the dynamics introduced by (iv).
To mitigate this bias, we propose  new analytically and jackknife bias-corrected estimators for coefficients and average partial effects in three-way fixed effects specifications. Additionally, we provide an expression for long-run partial effects.
Extensive simulation experiments demonstrate the desirable statistical properties of our proposed bias-corrected estimators.\\

The empirical application provides evidence that both unobserved bilateral factors and true state dependence due to entry dynamics contribute strongly to the high persistence. Taking this persistence into account changes the estimated effects of the most commonly studied potential determinants considerably: 
The impact of a common currency is reduced from almost 10 percentage points to less than 4 percentage points, the effect of a common regional trade agreement drops from 5.4 to 1.2 percentage points, and joint membership in the WTO/GATT increases the trading probability by 1.3 rather than 3.4 percentage points. For all three variables, only about two thirds of these effects are realized immediately, while the remaining third is added gradually over time. 
Furthermore, specifications with a lagged dependent variable and/or bilateral fixed effects yield better predictions for which country pairs will trade than specifications that fail to account for state dependence appropriately.\\

Our paper builds on recent insights from three flourishing strands of literature. First, our paper is related to the literature on the extensive margin of international trade. A number of theoretical frameworks have sought to propose mechanisms behind the decisions of firms to export, and their aggregate implications of zero or non-zero trade flows at the country pair level. Analogous to the intensive margin counterpart, these theories have established \textit{gravity}-like determinants, such as two countries' bilateral distance, a free trade agreement, a common currency and joint membership in the WTO/GATT.
\citet{Egger2011a} and \citet*{Egger2011b} append an extensive margin to an \citet{Anderson2003}-type model by assuming export participation to be determined by (homogeneous) firms weighing operating profits and bilateral fixed costs of exporting. 
\citet{Helpman2008} build a model of international trade with heterogeneous firms and bounded productivity in which a country only exports to a given destination if the most productive firm can afford to overcome the fixed costs of exporting. 
\citet{Eaton2013} move away from the arguably simplifying notion of a continuum of firms and develop a model of a finite set of heterogeneous firms. Here, no firm may export to a given market because of their individual efficiency draws.
Our model proposed in this paper directly builds on \citet{Helpman2008} and extends it by features from the literature on firm dynamics.
In this firm-level literature, \citet{Das2007} develop a dynamic discrete-choice model in which current export participation depends on previous exporting, and hence sunk costs, and observable characteristics of profits from exporting \citep[in line with previous empirical evidence by][]{Roberts1997,Bernard2004}.
\citet{Alessandria2007} 
embed the distinction between sunk costs and ``period-by-period'' fixed costs into general equilibrium.\footnote{A number of recent contributions also stress the dynamic character of firms' exporting behaviour and additionally provide alternative rationales for dynamic feedbacks beyond sunk costs of entry, such as ``demand learning'' or consumer accumulation \citep*[see e.g.][]{Bernard2017,Ruhl2017,Berman2019,Piveteau2019}.}
We aim at reconciling the estimation of the aggregate extensive margin with the insight from the firm-level literature that dynamics feature prominently in the determination of the exporting decision by deriving an econometric specification that explicitly incorporates previous export experience at the country pair level. \\


Second, our paper builds on advances in the literature on the gravity equation and the \textit{intensive} margin of international trade.
With the advent of what has now been coined \textit{structural} gravity \citep{Head2014}, the gravity framework has gained rich microfoundations. \citet{Anderson2003} and \citet{Eaton2002} each formulate an underlying structure for exporting and importing countries that in estimations can easily be captured by appropriate two-way country(-time) fixed effects, as first noted by \citet{Feenstra2004} and \citet{Redding2004}.
Since \citet{Baier2007}, it has furthermore become standard to include country pair fixed effects to tackle unobservable bilateral trade cost determinants. Additionally taking into account the multiplicative structure of the gravity equation following \citet{SantosSilva2006}, nonlinear estimation with exporter-time, importer-time, and country pair fixed effects has become the gold standard for the intensive margin.\footnote{In a linear intensive margin setting, this most general fixed effects structure was already proposed by \citet*{Baltagi2003}.}
Estimating the model introduced in this paper similarly calls for three sets of fixed effects, specific to exporters and importers in a given year, as well as to a given country pair over time. 
The binary nature of the decision whether to export to a destination market \textit{at all}, also clearly asks for a nonlinear estimator. Therefore, in this paper, we put the estimation of the extensive margin on a par with the intensive margin gold standard by introducing a respective three-way fixed effects binary choice specification. \\

Third, the paper builds on and contributes to the literature on estimating nonlinear fixed effects models.
As it is known since \citet{Neyman1948}, the inclusion of fixed effects potentially introduces an incidental parameter problem (IPP).
Although the maximum likelihood estimator (MLE) is consistent if all dimensions of the panel grow large, it has an asymptotic bias in its limiting distribution leading to invalid inference \citep[see][]{Fernandez-Val2018}.  
Recently, there have been a number of advances to deal with the IPP \citep[see][for a recent overview]{Fernandez-Val2018}.
In the context of the aggregate extensive margin, only approaches for \textit{cross-sectional} bilateral data with  importer ($j$) and exporter ($i$) fixed effects have been suggested. \citet{Cruz-Gonzalez2017} apply the  bias correction of \citet{Fernandez-Val2016a}\footnote{The bias corrections of \citet{Fernandez-Val2016a} were originally developed for classical panel data models with individual and time fixed effects and cover a wide range of nonlinear models.}, and  \citet{Charbonneau2017} proposes a conditional logit estimator. 
Many trade data sets however consist of bilateral cross-sections \textit{over time}, i.e.\ network panel data.   The theory-consistent estimation of our model includes  fixed effects for exporter-time ($it$) and importer-time ($jt$). On closer inspection, one finds that this two-way case can basically be covered by the bias corrections of \citet{Fernandez-Val2016a} for individual and time fixed effects.\footnote{Similarly, it is possible to adapt the estimator of \citet{Charbonneau2017} to the setting with exporter-time ($it$) and importer-time ($jt$) fixed effects. However, her approach has some limitations: 1. it is limited to logit models, 2. it precludes the possibility to estimate average partial effects, 3. it is computationally infeasible in cases where the number of levels per fixed effects becomes large.} However, the literature lacks a suitable method to estimate our preferred specification, which additionally includes a third, bilateral ($ij$), set of fixed effects.
Our contribution is to develop suitable  three-way fixed effects binary choice estimators for network panel data with potentially weakly exogeneous regressors under asymptotics where $I, J$ and $T$ are large. 
Therefore, our article complements the work of \citet{Weidner2020} on estimating the intensive margin of trade, who examine the IPP for the three-way fixed effects Poisson pseudo maximum likelihood (PPML) estimator under fixed $T$ asymptotics and suggest appropriate bias corrections.
\\

The remainder of the paper is structured as follows. In Section \ref{sec:theory} we build a dynamic model of the extensive margin of international trade. The model yields aggregate 
predictions that can be structurally estimated using a probit model with high-dimensional fixed effects. In Section \ref{sec:binary_response_estimator} we describe the new bias-corrected three-way fixed effects estimator. We demonstrate its performance in Monte Carlo simulations in Section \ref{sec:simulation}, before finally showing the estimator in action by estimating the model in Section \ref{sec:application}. Section \ref{sec:conclusions} concludes.


\section{An Empirical Model of the Extensive Margin of Trade}
\label{sec:theory}

We start by setting up a model of the extensive margin of trade that will later guide our econometric specification.
We consider a stylized dynamic \citet{Melitz2003}-type heterogeneous firms model of international trade. Following \citet[][henceforth HMR]{Helpman2008} we assume a bounded productivity distribution, like a truncated Pareto in HMR's case. We deviate from HMR by explicitly stating a time dimension and, unlike in the standard Melitz setting, separate fixed exporting costs into costs of entering a new market and costs of selling in a given market \citep[as in][]{Alessandria2007,Das2007}.\\

There are $N$ countries, indexed by $i$ and $j$, each of which consumes and produces a continuum of products. The representative consumer in $j$ receives utility according to a CES utility function:

\begin{align}
    u_{jt} = \left( \int_{\omega \in \Omega_{jt}}(\xi_{ijt})^{\frac{1}{\sigma}} q_{jt} (\omega)^{\frac{\sigma - 1}{\sigma}} d\omega \right)^{\frac{\sigma}{\sigma - 1}} \quad \text{with} \quad \sigma > 1.
\end{align}

where $q_{jt} (\omega)$ is $j$'s consumption of product $\omega$ in period $t$, $\Omega_{jt}$ is the set of products available in $j$, $\sigma$ is the elasticity of substitution across products, and $\xi_{ijt}$ is a log-normally distributed idiosyncratic demand shock (with $\mu_\xi=0$ and $\sigma_\xi=1$) for goods from country $i$ in country $j$ and period $t$ \citep[similar to][]{Eaton2011}. Demand in country $j$ for good $\omega$ depends on this demand shock, $j$'s overall expenditure $E_{jt}$, and the good price $p_{jt}(\omega)$ relative to the overall price level as captured by the price index $P_{jt}$:

\begin{align*}
    q_{jt} (\omega) &= \frac{p_{jt} (\omega)^{- \sigma}}{P_{jt}^{1 - \sigma}} \xi_{ijt}E_{jt}. \\ 
    \text{with} \qquad P_{jt} &= \left( \int_{\omega \in \Omega_{jt}} \xi_{ijt}p_{jt} (\omega)^{1 - \sigma} d\omega \right)^{\frac{1}{1 - \sigma}} \, . 
\end{align*}

Each country has a fixed continuum of potentially active firms that have different productivities drawn from the distribution $G_{it}(\varphi)$, where $\varphi\in(0,\varphi_{it}^{*}]$. The productivity distribution evolves over time and firms' ranks within the productivity distribution can also change from period to period, though firms that in the last period did not export to a market already served by a domestic competitor are assumed not to directly jump to being the country's most productive firm in the next period.\footnote{Note that we could in principle also allow for new firm entry into the pool of potential producers without changing our final expression for the extensive margin as long as the new entrants cannot become the country's most productive firm right away.} Each period, a firm can decide to pay a fixed cost $f_{it}^{prod}$ and start production of a differentiated variety using labour $l$ as its only input, such that $l_{t}(\omega) = f_{it}^{prod} + q_{t}(\omega) / \varphi_{t}(\omega)$. A firm's marginal cost of providing one unit of its good to market $j$ consists of iceberg trade costs $\tau_{ijt}$ and labour costs $w_{it}/\varphi_{t}(\omega)$. Firms compete with each other in monopolistic competition and charge a constant markup over marginal costs. Therefore, the price of a good $\omega$ produced in $i$ and sold in $j$ is:

\begin{align*}
    p_{ijt}(\omega)=\frac{\sigma}{\sigma-1}\frac{\tau_{ijt}w_{it}}{\varphi_{t}(\omega)}.
\end{align*}

A firm's \textit{operating} profits in market $j$ are hence given by:

\begin{align*}
    \tilde{\pi}_{ijt}(\omega)=\frac{1}{\sigma}\left(\frac{\sigma}{\sigma-1}\frac{\tau_{ijt}w_{it}}{\varphi_{t}(\omega)}\right)^{1-\sigma}P_{jt}^{\sigma-1}\xi_{ijt}E_{jt}.
\end{align*}

If a firm wants to export to a market $j$ in period $t$, it has to pay a fixed exporting cost $f_{ijt}^{exp}$. The exporting fixed cost is higher by a market entry cost factor $f^{entry}\geq 1$ if the firm has not been active in the respective market in the previous period. For tractability, the entry cost factor is assumed to be constant across countries and time. 
Capturing the export decision by a binary variable $y_{ijt}(\omega)$, i.e.\ equal to one if the firm decides to serve market $j$ in period $t$, we can formalize a firm's \textit{realized} profits in market $j$ as follows:

\begin{align*}
    \pi_{ijt}(\omega)=y_{ijt}(\omega)\left\{\tilde{\pi}_{ijt}(\omega)-f_{ijt}^{exp}(f^{entry})^{[1-y_{ij(t-1)}(\omega)]}\right\}.
\end{align*}

In the absence of entry costs, a firm would simply compare its operating profits to the fixed exporting cost and decide to serve a market if the former are greater than the latter. With market entry costs, a firm might be willing to incur a loss in the current period if expected future profits from that same market outweigh the initial loss. Firms discount future profits at a rate $\delta$ per period. To keep things tractable and allow us to derive a theory-consistent estimation expression below, we assume that firms expect their future operating profits from and fixed costs of serving a given market to be equal to today's values, i.e.\ $\EX_{t}[\tilde{\pi}_{ij(t+s)}]=\tilde{\pi}_{ijt}$ and $\EX_{t}[f^{exp}_{ij(t+s)}]=f^{exp}_{ijt}$ $\forall s\in\mathbb{N}$.\footnote{Note that our final expression for the extensive margin also holds if firms instead expect their operating profits from serving an export market to grow at a constant rate $\bar{g}<\delta$.} The current value of today's and all future operating profits from market $j$ is then given by $\sum_{s=0}^{\infty}(1-\delta)^{s}\tilde{\pi}_{ijt}=\frac{\tilde{\pi}_{ijt}}{\delta}$. A firm will decide to serve a destination market if these discounted expected profits exceed the sum of today's and discounted future fixed costs of entry and exporting, given by

\begin{align*}
    f^{exp}_{ijt} (f^{entry})^{(1-y_{ij(t-1)}(\omega))}+\sum_{s=1}^{\infty}(1-\delta)^{s}f^{exp}_{ijt}=\frac{f^{exp}_{ijt}}{\delta} \left(1 + \delta (f^{entry} - 1) \right)^{(1-y_{ij(t-1)}(\omega))} \, .
\end{align*}



Given this model setup, the question whether a country exports to another country \textit{at all} can be considered by looking at the most productive firm (with $\varphi_t^{*}$) only. Denoting that firm's product by $\omega^{*}$, we can capture the aggregate extensive margin by the binary variable $y_{ijt}$ as follows:

\begin{align}\label{eq:extensive_margin_aggregate_theory}
    y_{ijt}=y_{ijt}(\omega^{*})=\begin{cases}
    1\quad\text{if}\quad \frac{\left(\frac{1}{\sigma}\left(\frac{\sigma}{\sigma-1}\frac{\tau_{ijt}w_{it}}{\varphi_{it}^{*}}\right)^{1-\sigma}P_{jt}^{\sigma-1}\xi_{ijt}E_{jt}\right)}{f_{ijt}^{exp}\left(1+\delta(f^{entry}-1)\right)^{(1-y_{ij(t-1)})}}\geq 1,  \\
    0\quad\text{else}.
    \end{cases}
\end{align}

Country $i$ is hence more likely to export to country $j$ in period $t$ if (i) bilateral variable trade costs are lower; (ii) wages in $i$, and hence production costs, are lower; (iii) the productivity of the most productive firm is higher, again reducing production costs; (iv) competitive pressure, inversely captured by the price index, in $j$ is lower, corresponding to the idea of inward multilateral resistance coined by \citet{Anderson2003} in the intensive margin context; (v) the market in $j$ is larger; (vi) bilateral fixed costs of exporting are smaller; or (vii) $i$'s most productive firm already served market $j$ in the previous period and therefore does not have to pay the market entry cost. Note that (i) to (iv) all act via higher operating profits and depend on the elasticity of substitution between goods. The higher this elasticity, the stronger the reaction of profits to changes in any of these factors. At the same time, a higher elasticity reduces the mark-up firms can charge and hence makes it generally harder to earn enough profits to mitigate the fixed costs of exporting. Further note that the importance of the entry costs depends on the discount factor. Intuitively, if agents are more patient, the one-time entry costs matter less compared to the repeatedly earned profits. Empirically, (vii) induces true state dependence. As previous exporters do not have to incur entry costs, they are more likely to stay active in the destination market and the extensive margin becomes more persistent than would be implied merely by the persistence of productivity, market potential, and trade costs.\\

In order to turn equation \eqref{eq:extensive_margin_aggregate_theory} into the empirical expression that we will bring to the data, we take the natural logarithm and group all exporter-time and importer-time specific components and capture them with corresponding sets of fixed effects. Further, we need to specify the fixed and variable trade costs. In keeping with the existing literature, we model them as a linear combination of different observable bilateral variables, such as geographic distance, whether $i$ and $j$ are both WTO/GATT members, and whether $i$ and $j$ share a common currency. In our most general specification, we additionally include country pair fixed effects. Following \citet{Baier2007}, this is common practice in the estimation of the determinants of the intensive margin of trade in order to avoid endogeneity due to unobserved heterogeneity. Further, these bilateral fixed effects may capture (part of) the strong persistence documented above.\footnote{If the trade costs further include any exporter(-time) or importer(-time) specific components, these are captured by the aforementioned corresponding sets of fixed effects.} Note, however, that the nature of the persistence captured by these fixed effects is different from the one that is due to the entry dynamics. This additional state dependence is ``spurious'' in the sense that countries are not actually more likely to export to a destination because of the prior experience, but because they keep incorporating the same unobserved factors over time. With the three sets of fixed effects and our parametrization for time-varying trade cost determinants, we arrive at the following econometric model:

\begin{align}\label{eq:empirical_three-way}
    y_{ijt}=\begin{cases}
    1\quad\text{if}\quad\kappa+\lambda_{it}+\psi_{jt}+\beta_{y} y_{ij(t-1)}+\mathbf{z}_{ijt}'\fmbeta_{z}+\mu_{ij}\geq\zeta_{ijt}, \\
    0\quad\text{else},
    \end{cases}
\end{align}

where $\kappa=-\sigma\log(\sigma)-(1-\sigma)\log(\sigma-1)-\log(1+\delta (f^{entry}-1))$, $\lambda_{it}=(1-\sigma)(\log(w_{it})-\log(\varphi_{it}^{*}))$, $\psi_{jt}=(\sigma-1)\log(P_{jt})+\log(E_{jt})$, $\beta_{y}=\log(1+\delta (f^{entry}-1))$, $\mathbf{z}_{ijt}'\fmbeta_{z}+\mu_{ij}=(1-\sigma)\log(\tau_{ijt})-\log(f_{ijt}^{exp})$, and $\zeta_{ijt}=-\log(\xi_{ijt})\sim \mathcal{N}(0,1)$. The error term distribution implies that a probit estimator is the appropriate choice to estimate our model. Alternatively, we could deviate from \citet{Eaton2011} and assume a log-logistic distribution for the idiosyncratic demand shocks, which would lead to a logit specification. 
As mentioned above, we capture the three sets of unobserved components by introducing according sets of fixed effects. A supposed alternative using random effects is actually not possible, at least for the $it$ and $jt$ effects, as they are implied by the theoretical model. We therefore cannot make the distributional assumptions required in a random effects setting. The theory is silent about the exact form of the bilateral heterogeneity. We decide for a third set of fixed effects as the most general option in order to avoid assumptions on its distribution or its correlation to observed factors. \\

Our theoretical framework implies a flexible empirical specification that can reconcile the extensive margin estimation with the stylized fact presented in Section \ref{sec:introduction}. 
Note that we chose to make a number of simplifying assumptions in order to achieve the clear theory-consistent interpretation of specification \eqref{eq:empirical_three-way}.
An alternative interpretation of equation \eqref{eq:empirical_three-way} as a reduced-from representation of a more elaborate and realistic model \citep[similar e.g. to how][motivate their empirical consideration]{Roberts1997} is equally justifiable. At the same time, while our model is written along the lines of \citet{Helpman2008}, which remains the benchmark for the empirical assessment of the (aggregate) extensive margin of trade, it is not decisive for our empirical specification that zero trade flows result from a truncated productivity distribution instead of a discrete number of firms \citep[as in][]{Eaton2013} or from fixed exporting costs in a \citet{Krugman1980}-type homogeneous firms setting \citep[as in][]{Egger2011a,Egger2011b}.

\section{Binary Choice Estimators with Three-Way Fixed \mbox{Effects}}
\label{sec:binary_response_estimator}

Having set up the empirical framework, we now turn to the estimation. As equation \eqref{eq:empirical_three-way} demands three-way fixed effects to capture unobservable characteristics, we describe how to implement suitable binary choice estimators. In a first step, we review the standard maximum likelihood estimator (MLE) for probit and logit models with  fixed effects. In a second step, we explain the consequences of the incidental parameter problem  on the MLE and characterize new bias corrections to address the induced incidental parameter problem. 


\subsection{Model and Maximum Likelihood Estimation}

Our empirical model \eqref{eq:empirical_three-way} can be be written in a general way by the following three-way fixed effects binary choice model:

\begin{align}\label{eq:general_three-way}
    y_{ijt}=
    1(\mathbf{x}_{ijt}'\fmbeta+\psi_{jt}+\lambda_{it}+ \mu_{ij}\geq\zeta_{ijt}), \qquad \zeta_{ijt} \mid \mathbf{x}_{ij}^t,  \boldsymbol{\psi}, \boldsymbol{\lambda},  \boldsymbol{\mu} \sim F_{\zeta}
\end{align}

where $i = 1, \dots, I$, $j = 1, \dots, J$, and $t = 1, \dots, T$ are the indexes denoting exporters, importers, and time periods, respectively, $\mathbf{x}_{ijt}$ is a $p$-dimensional vector of regressors, $\fmbeta$ are the  corresponding structural parameters, $\boldsymbol{\psi} = (\psi_{11}, \dots, \psi_{JT})$, $\boldsymbol{\lambda} = (\lambda_{11}, \dots, \lambda_{IT})$, and $\boldsymbol{\mu} = (\mu_{11}, \dots, \mu_{IJ})$ are the incidental parameters, $\mathbf{x}_{ij}^t = (\mathbf{x}_{ijt}, \mathbf{x}_{ij(t-1)}, \dots, \mathbf{x}_{ij1})$, and
$F_{\zeta}$ is either the logistic or standard normal cumulative distribution function.\footnote{To keep notation simple, we abstract from the cases of no-self flows, $i \neq j$, and unbalanced panels. The no-self flows are unproblematic for our estimators. For unbalanced panels, we have to additionally assume, that the attrition process is random conditional on $\mathbf{x}_{ij}^t$ and the fixed effects.} Note that we allow $\mathbf{x}_{ijt}$ to contain weakly exogenous regressors. This is important, because our empirical model \eqref{eq:empirical_three-way} contains a lagged dependent variable, which violates the strict exogeneity condition, i.e. $\zeta_{ijt} \mid \mathbf{x}_{ij}^T,  \boldsymbol{\psi}, \boldsymbol{\lambda},  \boldsymbol{\mu} \sim F_{\zeta}$ with $\mathbf{x}_{ij}^T = (\mathbf{x}_{ijT}, \mathbf{x}_{ij(T-1)}, \dots, \mathbf{x}_{ij1})$. 
Even in the absence of a lagged dependent variable, the strict exogeneity condition is often too restrictive, since it rules out any dynamic feedback between regressors and the dependent variable.\footnote{For instance, one can imagine that country pairs that do not trade with each other in period $t$, might sign a trade agreement in the future, because they did not trade in $t$.}\\



A standard estimator for the parameters of interest $\fmbeta$ and the incidental parameters $\boldsymbol{\alpha} = (\boldsymbol{\lambda}, \boldsymbol{\mu}, \boldsymbol{\psi})$  is the following maximum likelihood estimator (MLE)

\begin{equation}
    \hat{\boldsymbol{\theta}} = 	(\hat{\boldsymbol{\beta}}, \hat{\boldsymbol{\alpha}}) = \underset{ \boldsymbol{\beta} \in \mathbb{R}^{p}, \boldsymbol{\alpha} \in \mathbb{R}^{IT+JT+IJ}}{\argmax} \sum_{i=1}^{I} \sum_{j=1}^{J} \sum_{t=1}^{T} \ell_{ijt}(\boldsymbol{\beta}, \psi_{jt}, \lambda_{it}, \mu_{ij})
    \label{eq:MLE}
\end{equation}

where $\ell_{ijt}(\boldsymbol{\beta}, \psi_{jt},  \lambda_{it}, \mu_{ij} ) = y_{ijt} \log(F_{ijt}) + (1-y_{ijt}) \log(1-F_{ijt})$ is the log-likelihood contribution of the $ijt$-th observation, and $F_{ijt}$ denotes the cumulative distribution function chosen for $\zeta_{ijt}$ evaluated at the linear index $\eta_{ijt} = \mathbf{x}_{ijt}'\fmbeta+ \psi_{jt}+\lambda_{it}+\mu_{ij}$.\footnote{Note that the incidental parameters $\mu_{ij}, \lambda_{it}, \psi_{jt}$ are not uniquely identified due to collinearity issues and we need to impose some restrictions. This can be achieved by adding a penalty on the log-likelihood which imposes a normalization on the incidental parameters.}\\

The brute-force estimation of equation \eqref{eq:MLE} quickly becomes computationally demanding, if not impossible due to the large number of parameters that need to be estimated. For example, in a balanced data set ($I=J=N$) without self flows, the number of parameters to be estimated is $\approx N (N-1) + 2 N T$. In a trade panel data set with 200 countries and 50 years, the number of fixed effects in this case amounts to 59,800 parameters. However, recent developments in computational econometrics reduce this computational burden by employing  a straightforward strategy called pseudo-demeaning, which mimics the well-known within transformation for linear regression models \citep[see][]{Stammann2018}. We give a brief summary of the pseudo-demeaning algorithm in Appendix \ref{sec:compdetails}. \\

The parameters of the variables of interest can only be directly interpreted in terms of their signs and relative magnitudes, but are not informative in themselves about the absolute strength of the effects they represent. Another quantity of interest are therefore the average partial effects:

\begin{align}
    \delta_k &=\frac{1}{IJT} \sum_{i=1}^{I} \sum_{j=1}^{J} \sum_{t=1}^{T} \Delta_{ijt}^k \; ,
    \label{eq:APE_direct}
\end{align}

where the partial effect of the $k$-th regressor $\Delta_{ijt}^k$ is either $\Delta_{ijt}^k = \partial F_{ijt} / \partial x_{ijtk}$ in the case of a non-binary regressor or $\Delta_{ijt}^k =  F_{ijt}|_{x_{ijtk = 1}} - F_{ijt}|_{x_{ijtk = 0}}$ in the case of binary regressors. Here $F_{ijt}|_{x_{ijtk = z}}$ indicates that all values of the $k$-th regressor are replaced by $z$. In dynamic models, the simple average partial effect $\delta_k$ does not provide the full picture of how the export probability is affected by a change in a regressor. Rather, there are additional feedback effects.\footnote{In our context, the introduction of a permanent trade policy that increases the probability to export to a destination implies that in the next period, entry costs are more likely to have already been paid, and hence the impact becomes higher with increasing duration of the policy.} To derive expressions for long-run effects, we make use of the long-run probability of $y_{ijt}=1$ for a given set of regressors and fixed effects, also mentioned in \citet{Carro2003} and \citet{Browning2010}:

\begin{equation}
    \tilde{F}_{ijt} = \frac{F_{ijt}|_{y_{ij(t-1) = 0}}}{1-\Delta_{ijt}^y},
\end{equation}

where $\Delta_{ijt}^y =  F_{ijt}|_{y_{ij(t-1) = 1}} - F_{ijt}|_{y_{ij(t-1) = 0}}$.
Long-run average partial effects are then given by

\begin{equation}
    \delta_k^{LR}=\frac{1}{IJT}\sum_{i=1}^{I}\sum_{j=1}^{J}\sum_{t=1}^{T} \Delta_{ijt}^{k,LR},\text{ with } \Delta_{ijt}^{k,LR} = \frac{\partial \tilde{F}_{ijt}}{\partial x_{ijtk}} \text{ or } \Delta_{ijt}^{k, LR} =  \tilde{F}_{ijt}|_{x_{ijtk = 1}} - \tilde{F}_{ijt}|_{x_{ijtk = 0}},
    \label{eq:APE_long_run}
\end{equation}

in the case of non-binary or binary regressors, respectively. Estimators of (\ref{eq:APE_direct}) and (\ref{eq:APE_long_run}) can be formed by plugging in the MLE defined in (\ref{eq:MLE}).


\subsection{Incidental Parameter Bias Correction}
\label{sec:bias_correction}


The MLE of many nonlinear fixed effects models --- including binary choice models ---  suffers from the well-known incidental parameter problem (IPP) first identified by \citet{Neyman1948}. The problem stems from the necessity to estimate many nuisance parameters, which contaminate the estimator of the structural parameters and average partial effects. It can be further amplified by the inclusion of predetermined regressors like a lagged dependent variable. This amplification  can be interpreted as a Nickell-type bias  \citep{Nickell1981}. Although the MLE is consistent (under appropriate asymptotics), it has an asymptotic bias in its limiting distribution leading to invalid inference \citep[see][]{Fernandez-Val2018}.  The literature suggests different types of bias corrections to reduce these incidental parameter and Nickell-type biases \citep[][for an overview]{Fernandez-Val2018}.
Jackknife corrections, like the leave-one-out jackknife proposed by \citet{Hahn2004}, or the split-panel jackknife (SPJ) introduced by \citet{Dhaene2013}, are the simplest approaches to obtain a bias correction, at the expense of being computationally costly. In contrast to analytical corrections, their application only requires knowledge of the order of the bias components to form appropriate subpanels that are used to reestimate the model and to form an estimator of the bias terms.
For analytical bias correction (ABC), it is necessary to derive the asymptotic distribution of the MLE, in order to obtain an explicit expression of the asymptotic bias. This is then used to form a suitable estimator for the bias terms. \\


The IPP affects all quantities of interest mentioned in the previous subsection, i.e.   $\hat{\boldsymbol{\beta}}$, $\hat{\delta}_k$, and $\hat{\delta}_k^{LR}$.\footnote{Note that we do not make a notational distinction between direct and long-run APEs in the following. In all expressions for APEs, one can move between the direct and long-run case by substituting $\Delta_{ijt}^{LR}$ for $\Delta_{ijt}$.}  Its consequences become clear when looking at the asymptotic distribution of the estimators.
Under asymptotic sequences, where all panel dimensions grow at a constant rate as $I, J, T \rightarrow \infty$, the MLE is consistent but asymptotically biased because its asymptotic distribution is not centered around the true parameter value $\boldsymbol{\beta}^0$:

\begin{equation*}
    \hat{\boldsymbol{\beta}} \aapprox \N (\boldsymbol{\beta}^0 +  \overline{\mathbf{b}}_{ \infty}^{\beta}, \; \overline{\mathbf{V}}_{\infty}^{\beta}) \, ,
\end{equation*}

where $\overline{\mathbf{b}}_{ \infty}^{\beta}$ denotes the asymptotic bias and $\overline{\mathbf{V}}_{\infty}^{\beta}$ is the covariance matrix \citep[see][]{Fernandez-Val2018}.
\citet{Fernandez-Val2018}  derive a simple heuristic to determine the order of the bias induced by the incidental parameters: $bias \sim p / n$, where $p$ is the number of the incidental parameters and $n$ is the sample size.  Based on this heuristic \citet{Fernandez-Val2018} conjecture that the bias of the three-way fixed effects estimator we consider in this paper is of order $(IT+JT+IJ)/(IJT)$ and of the form $B_1 / I + B_2 / J + B_3 / T$. In line with the bias structure in two-way error component models, the inclusion of importer-time and exporter-time fixed effects entails two bias terms of order $1 / I$ and $1 / J$, respectively.
Intuitively, the inclusion of dyadic fixed effects induces another bias of order $1/T$ because there are only $T$ informative observations per additionally included parameter.
Although the bias decreases with increasing $I,J,T$, its order is larger than the order of the standard deviation of the MLE, $1/\sqrt{IJT}$.\footnote{As mentioned by \citet{Fernandez-Val2018}, even if $I = J = T$, the order of the bias is $1/I$ and the order of the standard deviation is $\sqrt{1/I^3}$.} As a consequence, confidence intervals do not have the desired nominal confidence level, i.e. they undercover, leading to invalid inference \citep[see][]{Fernandez-Val2018}.
The asymptotic distribution of the APEs is affected similarly by the IPP:

\begin{equation*}
    \hat{\delta}_k \aapprox \N (\delta_k^0 +  \overline{b}_{k, \infty}^{\delta}, \; \overline{V}_{k,\infty}^{\delta}) \, ,
\end{equation*}

where $\overline{b}_{k, \infty}^{\delta}$ denotes the asymptotic bias and $\overline{V}_{k, \infty}^{\delta}$ is the variance. 
The conjecture of \citet{Fernandez-Val2018} does not incorporate explicit expressions for bias-corrected estimators.
Thus, based on their conjecture, we propose novel analytical and jackknife bias corrections for three-way fixed effects models, which deal with the IPP (including the Nickell-type bias) and the associated inference problem.\footnote{In  Appendix \ref{sec:asymptotic}, we formulate the asymptotic  distributions of $\hat{\boldsymbol{\beta}}$ and $\hat{\delta}_k$. In Appendix \ref{sec:neyman}, we illustrate the statistical problem and the working of bias corrections with a version of the prominent \citet{Neyman1948} variance example.}
\\

In the following we adapt and extend the  analytical and split-panel jackknife bias corrections  proposed  by \citet{Fernandez-Val2016a}  in the context of nonlinear models with individual and time fixed effects to our three-way error component structure.\footnote{In Appendix \ref{sec:asymptotic}, we also derive the bias corrections for a two-way fixed effects model in our $ijt$ network panel structure. Previous two-way bias corrections considered either classical $it$ panel structures or $ij$ pseudo-panels.}$^{,}$\footnote{We do not elaborate on  the leave-one-out jackknife bias correction because it requires all variables to be independent over time and thus rules out predetermined and serially-correlated regressors \citep{Fernandez-Val2018}.}

For the split-panel jackknife bias correction, the aforementioned three-part bias structure implies that we need to split our panel across three dimensions, leading to the following estimator for the structural parameters:

\begin{align}
    \widehat{\fmbeta}^{sp}&=4\widehat{\fmbeta}-\widehat{\fmbeta}_{I/2,J,T}-\widehat{\fmbeta}_{I,J/2,T}-\widehat{\fmbeta}_{I,J,T/2}, \quad\text{with} \label{eq:spj_beta_3way} \\
    \widehat{\fmbeta}_{I/2,J,T} &= \frac{1}{2}\Big[\widehat{\fmbeta}_{\{i:i\leq \lfloor I/2 \rfloor,J,T\}}+\widehat{\fmbeta}_{\{i:i \geq \lceil I/2 + 1 \rceil,J,T\}}\Big], \notag \\
    \widehat{\fmbeta}_{I,J/2,T} &= \frac{1}{2}\Big[\widehat{\fmbeta}_{\{I,j:j\leq \lfloor J/2 \rfloor,T\}}+\widehat{\fmbeta}_{\{I,j:j \geq \lceil J/2 + 1\rceil,T\}}\Big], \notag\\
    \widehat{\fmbeta}_{I,J,T/2} &= \frac{1}{2}\Big[\widehat{\fmbeta}_{\{I,J,t:t\leq \lfloor T/2 \rfloor \}}+\widehat{\fmbeta}_{\{I,J,t:t \geq \lceil T/2 + 1\rceil \}}\Big]. \notag
\end{align}

where $\lfloor \cdot \rfloor$ and $\lceil \cdot \rceil$ denote the floor and ceiling functions.
To clarify the notation, the subscript ${\{i:i\leq \lceil I/2 \rceil \},J,T}$ denotes that the estimator is based on a subsample, which contains all importers and time periods, but only the first half of all exporters.  Note that similar to \citet{Fernandez-Val2016a}, the split-panel jackknife bias correction requires a homogeneity assumption of the distribution of $y_{ijt}$ and $\mathbf{x}_{ijt}$ across the dimensions $I$, $J$, and $T$. For instance, this assumption is violated in the case of time trends or structural breaks, where the subsample estimates of splitting dimension $T$, i.e. $\widehat{\fmbeta}_{\{I,J,t:t\leq \lfloor T/2 \rfloor \}}$ and $\widehat{\fmbeta}_{\{I,J,t:t \geq \lceil T/2 + 1\rceil \}}$, are systematically different \citep[see][]{Fernandez-Val2016a}.\footnote{The homogeneity assumption can be tested with a Wald test \citep[see][]{Dhaene2013, Fernandez-Val2016a}.} This additional homogeneity assumption is not required for the analytical bias correction. \\

\begin{table}[!ht]
	\begin{center}
	\caption{Expressions and Derivatives for Logit and Probit Models}
	\label{tab:expression}
	\renewcommand{\arraystretch}{1.25}
	\begin{tabular}{lll}
		\toprule
		& Logit & Probit  \\ 
		\midrule
		$F_{ijt}$ & $(1+\exp(-\eta_{ijt}))^{-1}$  & $\Phi(\eta_{ijt})$\\
		$\partial_{\eta} F_{ijt}$ & $F_{ijt}(1-F_{ijt})$ & $\phi(\eta_{ijt})$\\
		$\partial_{\eta^2} F_{ijt}$ & $\partial_{\eta} F_{ijt} (1-2F_{ijt})$ & $-\eta_{ijt}\phi(\eta_{ijt})$\\
		$\nu_{ijt}$ & $(y_{ijt}-F_{ijt})/\partial_{\eta} F_{ijt}$&  $(y_{ijt}-F_{ijt})/\partial_{\eta} F_{ijt}$\\
		$H_{ijt}$ & $1$ & $\partial_{\eta} F_{ijt} / (F_{ijt}(1-F_{ijt}))$ \\
		$\omega_{ijt}$ & $\partial_{\eta} F_{ijt}$ & $H_{ijt} \partial_{\eta} F_{ijt}$\\
		$\partial_{\eta} \ell_{ijt}$ & $ y_{ijt} - F_{ijt}$ & $H_{ijt} (y_{ijt} - F_{ijt})$\\
		\bottomrule
		\multicolumn{3}{p{0.55\textwidth}}{\footnotesize{\textit{Note}: 
		$\eta_{ijt} = \mathbf{x}_{ijt}^{\prime} \boldsymbol{\beta}+\lambda_{it} + \psi_{jt} + \mu_{ij}$  is the linear predictor.}}
	\end{tabular}
\end{center}
\end{table}

To characterize the analytical bias correction, we need to introduce some additional notation. Let $\partial_{\iota^r}g(\cdot)$ denote the $r$-th order partial derivative of an arbitrary function $g(\cdot)$ with respect to some parameter $\iota$. A collection of further required expressions are reported in Table \ref{tab:expression}.
Let $\mathbf{D}$ denote the dummy matrix corresponding to the fixed effects and $\mathbf{X}$ denote the matrix with the regressors of interest. Define the residual projection
$\widehat{\MX} = \mathbf{I}_{IJT} - \widehat{\PX} = \mathbf{I}_{IJT} - \mathbf{D}(\mathbf{D}^{\prime} \widehat{\boldsymbol{\Omega}} \mathbf{D})^{-1} \mathbf{D}^{\prime} \widehat{\boldsymbol{\Omega}}$, where $\mathbf{I}_{IJT}$ is an ${IJT} \times {IJT}$ identity matrix, and $\widehat{\boldsymbol{\Omega}}$ is a diagonal weighting matrix with elements $\hat{\omega}_{ijt}=\partial_{\eta}\hat{F}_{ijt}$.\\

Combining insights from the classical panel structure in \citet{Fernandez-Val2016a}, the pseudo-panel setting in \citet{Cruz-Gonzalez2017}, and the three-way fixed effects conjecture by \citet{Fernandez-Val2018}, we
propose the following analytical bias correction:

\begin{align}
    \tilde{\boldsymbol{\beta}}^a &= \hat{\boldsymbol{\beta}} - \frac{\widehat{\mathbf{B}}_1^{\beta}}{I} - \frac{\widehat{\mathbf{B}}_2^{\beta}}{J} -  \frac{\widehat{\mathbf{B}}_3^{\beta}}{T}, \quad \text{with} \quad \widehat{\mathbf{B}}_1^{\beta} = \widehat{\mathbf{W}}^{-1}\widehat{\mathbf{B}}_1, \widehat{\mathbf{B}}_2^{\beta} =\widehat{\mathbf{W}}^{-1}  \widehat{\mathbf{B}}_2,\widehat{\mathbf{B}}_3^{\beta} =\widehat{\mathbf{W}}^{-1} \widehat{\mathbf{B}}_3, \label{eq:abc_three-way}\\
    \widehat{\mathbf{B}}_1 &= - \frac{1}{2JT} \sum_{j = 1}^{J} \sum_{t = 1}^{T} \frac{\sum_{i = 1}^{I} \widehat{H}_{ijt} \partial_{\eta^{2}} \widehat{F}_{ijt} \left(\widehat{\MX}\mathbf{X}\right)_{ijt}}{\sum_{i = 1}^{I} \hat{\omega}_{ijt}} \, , \nonumber \\
    \widehat{\mathbf{B}}_2 &= - \frac{1}{2IT} \sum_{i = 1}^{I} \sum_{t = 1}^{T} \frac{\sum_{j = 1}^{J} \widehat{H}_{ijt} \partial_{\eta^{2}} \widehat{F}_{ijt} \left(\widehat{\MX}\mathbf{X}\right)_{ijt}}{\sum_{j = 1}^{J} \hat{\omega}_{ijt}} \, , \nonumber \\
    \widehat{\mathbf{B}}_3 &= - \frac{1}{2IJ} \sum_{i = 1}^{I} \sum_{j = 1}^{J} \left( \sum_{t = 1}^{T} \hat{\omega}_{ijt} \right)^{-1} \left( \sum_{t = 1}^{T} \widehat{H}_{ijt} \partial_{\eta^{2}} \widehat{F}_{ijt} \left(\widehat{\MX}\mathbf{X}\right)_{ijt} \right. \nonumber \\
    & \qquad \qquad \left. + 2 \sum_{l=1}^{L}(T/(T-L)) \sum_{t=l+1}^{T} \partial_{\eta} \hat{\ell}_{ijt-l} \hat{\omega}_{ijt} \left(\widehat{\MX}\mathbf{X}\right)_{ijt} \right) \, ,\nonumber \\
    \widehat{\mathbf{W}} &= \frac{1}{IJT} \sum_{i = 1}^{I} \sum_{j = 1}^{J} \sum_{t = 1}^{T} \hat{\omega}_{ijt} \left(\widehat{\MX}\mathbf{X}\right)_{ijt} \left(\widehat{\MX}\mathbf{X}\right)_{ijt}^{\prime} \, . \nonumber
\end{align}

The first two correction terms in equation \eqref{eq:abc_three-way} are generalizations of the corresponding components in the $ij$ pseudo-panel structure of \citet{Cruz-Gonzalez2017} to our $ijt$ structure. The inclusion of a third set of ($ij$) fixed effects additionally leads to the third correction term that mimics the correction for individual fixed effects in an $it$-panel setting. In contrast to $\widehat{\mathbf{B}}_1$ and $\widehat{\mathbf{B}}_2$, the expression $\widehat{\mathbf{B}}_3$ includes a part to correct the Nickell-type bias that arises by imposing the weak exogeneity condition instead of the strict exogeneity condition.
The parameter $L$ is a bandwidth  used for the estimation of truncated spectral densities \citep{Hahn2007}. In a model in which all regressors are strictly exogenous, $L$ is set to zero, such that the second part in the numerator of $\widehat{\mathbf{B}}_3$ vanishes. Note that the strict exogeneity condition is often too strong in the context of  panel data, because it rules out any dynamic feedback between regressors and the dependent variable. 
In case of weakly exogeneous regressors, for instance if one of the regressors is the lagged dependent variable, \citet{Fernandez-Val2016a} suggest conducting a sensitivity analysis with $L \in \{1,2,3,4\}$.\\

Moving to the APEs, the split-panel jackknife estimator is formed by replacing the estimators for the structural parameters with estimators for the APEs in formula \eqref{eq:spj_beta_3way}. The analytically bias-corrected estimator is given by

\begin{align}\label{eq:ana_ape_3way}
    \tilde{\boldsymbol{\delta}}^a & = \hat{\boldsymbol{\delta}} - \frac{\widehat{\mathbf{B}}_1^{\delta}}{I} - \frac{\widehat{\mathbf{B}}_2^{\delta}}{J} - \frac{\widehat{\mathbf{B}}_3^{\delta}}{T}, \quad \text{with} \\
    \widehat{\mathbf{B}}_1^{\delta} &= \frac{1}{2JT} \sum_{j = 1}^{J} \sum_{t = 1}^{T}  \frac{\sum_{i = 1}^{I} - \widehat{H}_{ijt} \partial_{\eta^{2}} \widehat{F}_{ijt} \left(\widehat{\PX}\widehat{\boldsymbol{\Psi}}\right)_{ijt} + \partial_{\eta^{2}} \widehat{\boldsymbol{\Delta}}_{ijt}}{\sum_{i = 1}^{I}  \hat{\omega}_{ijt}} \, , \nonumber \\
    \widehat{\mathbf{B}}_2^{\delta} &= \frac{1}{2IT} \sum_{i = 1}^{I} \sum_{t = 1}^{T}  \frac{\sum_{j = 1}^{J} - \widehat{H}_{ijt} \partial_{\eta^{2}} \widehat{F}_{ijt} \left(\widehat{\PX}\widehat{\boldsymbol{\Psi}}\right)_{ijt} + \partial_{\eta^{2}} \widehat{\boldsymbol{\Delta}}_{ijt}}{\sum_{j = 1}^{J}  \hat{\omega}_{ijt}} \, , \nonumber \\
    \widehat{\mathbf{B}}_3^{\delta} & =  \frac{1}{2IJ} \sum_{i = 1}^{I} \sum_{j = 1}^{J} \left( \sum_{t = 1}^{T} \hat{\omega}_{ijt} \right)^{-1} \left(
    \sum_{t = 1}^{T} - \widehat{H}_{ijt} \partial_{\eta^{2}} \widehat{F}_{ijt} \left(\widehat{\PX}\widehat{\boldsymbol{\Psi}}\right)_{ijt} + \partial_{\eta^{2}} \widehat{\boldsymbol{\Delta}}_{ijt} \right. \nonumber \\
    & \qquad \qquad \left. + 2 \sum_{l = 1}^{L} \left(T / \left(T - l\right)\right)  \sum_{t = l + 1}^{T} \partial_{\eta} \hat{\ell}_{ijt-l} \hat{\omega}_{ijt} \left(\widehat{\MX}\widehat{\boldsymbol{\Psi}}\right)_{ijt} \right) \, , \nonumber 
\end{align}

where $\widehat{\boldsymbol{\Psi}}_{ijt} = \partial_{\eta} \widehat{\boldsymbol{\Delta}}_{ijt}/\hat{\omega}_{ijt}$.
The last part in the numerator of $\widehat{\mathbf{B}}_3^{\delta}$ is again dropped if all regressors are assumed to be strictly exogenous. Note that all quantities are  evaluated at bias-corrected structural parameters and the corresponding estimates of the fixed effects.\footnote{For this purpose, we use a computationally efficient offset algorithm as in \citet{Czarnowske2019}.} 
An appropriate covariance estimator for the APEs of the three-way fixed effects model is

\begin{align}
    \widehat{\mathbf{V}}^{\delta} &= \frac{1}{I^{2}J^{2}T^{2}} \left( \underbrace{\left(\sum_{i = 1}^{I} \sum_{j =  1}^{J} \sum_{t = 1}^{T}  \widehat{\bar{\boldsymbol{\Delta}}}_{ijt}\right)\left( \sum_{i = 1}^{I} \sum_{j =  1}^{J} \sum_{t = 1}^{T}  \widehat{\bar{\boldsymbol{\Delta}}}_{ijt} \right)^{\prime}}_{v_1}  + \underbrace{\sum_{i = 1}^{I} \sum_{j = 1}^{J} \sum_{t = 1}^{T}\widehat{\boldsymbol{\Gamma}}_{ijt} \widehat{\boldsymbol{\Gamma}}_{ijt}^{\prime}}_{v_2} \right. \notag\\
        & \qquad\qquad \left.  
        + \underbrace{2  \sum_{i = 1}^{I} \sum_{j = 1}^{J} \sum_{s > t}^{T} \widehat{\bar{\boldsymbol{\Delta}}}_{ijt} \widehat{\boldsymbol{\Gamma}}_{ijs}^{\prime}}_{v_3} \right)  \, , \label{eq:estcov3way}
\end{align}

where $\widehat{\bar{\boldsymbol{\Delta}}}_{ijt} = \widehat{\boldsymbol{\Delta}}_{ijt} - \hat{\boldsymbol{\delta}}$, $\widehat{\boldsymbol{\Delta}}_{ijt} = [\widehat{\Delta}_{ijt}^1, \dots, \widehat{\Delta}_{ijt}^m]^{\prime}$, $\hat{\boldsymbol{\delta}} = [\hat{\delta}_1, \dots, \hat{\delta}_m]^{\prime}$, and


\begin{align*}
    \widehat{\boldsymbol{\Gamma}}_{ijt} &= \left(\sum_{i = 1}^{I} \sum_{j = 1}^{J} \sum_{t = 1}^{T} \partial_{\beta} \widehat{\boldsymbol{\Delta}}_{ijt} - \left(\widehat{\PX} \mathbf{X}\right)_{ijt} \partial_{\eta} \widehat{\boldsymbol{\Delta}}_{ijt} \right)^{\prime} \widehat{\mathbf{W}}^{- 1} \left(\widehat{\MX} \mathbf{X}\right)_{ijt} \hat{\omega}_{ijt} \hat{\boldsymbol{\nu}}_{ijt} + \left(\widehat{\PX} \widehat{\boldsymbol{\Psi}}\right)_{ijt} \partial_{\eta} \hat{\ell}_{ijt} \, .
\end{align*}

Note that the term $v_2$ refers to the variation induced by $\hat{\boldsymbol{\theta}}$. The terms $v_1$ and $v_3$ are in the spirit of \citet{Fernandez-Val2016a}  to improve the finite sample properties. These are, on the one hand, the variation induced by replacing population by sample means ($v_1$). On the other hand, if we are concerned about the strict exogeneity assumption  (as we are in the case of  dynamic three-way error structure models), the covariance between $v_1$ and $v_2$   should be incorporated ($v_3$).\\

Just as the calculation of the MLE, the  application of the bias corrections is computationally demanding due to the high number of parameters. Therefore, we suggest to use the algorithms developed by \citet{Stammann2018} and \citet{Czarnowske2019}.


\section{Monte Carlo Simulations}
\label{sec:simulation}

In this section, we conduct extensive simulation experiments to investigate the properties of different estimators for both the structural parameters and the APEs. The estimators we study are MLE, ABC and SPJ.\footnote{We do not include OLS as an alternative estimator for APEs in our discussion since --- apart from a Nickell-bias in a dynamic three-way fixed effects model --- it suffers from a kind of misspecification bias. In a standard gaussian data generating process, where the dependent variable is continuous instead of binary, the bias corrected ordinary least squares estimator proposed by \citet{Hahn2002} works as desired, however it fails in our probit data generating process. The reason is that a large share of the predicted probabilities obtained by OLS exceed the boundaries of the unit interval, such that estimates of APEs become heavily biased. This kind of misspecification bias already becomes evident, if we consider a dynamic two-way fixed effects data generating process, which does not suffer from a Nickell bias. We investigated this issue in additional extensive simulation experiments and provide these results on request.}
Our main focus are the biases and inference accuracies. To this end, we compute the relative bias and standard deviation (SD) in percent, the ratio between bias and standard error, the ratio between standard error and standard deviation (SE/SD), and the coverage probabilities (CPs) at a nominal level of 95 percent.\\

For the simulation experiments we adapt the design for a dynamic probit model of \citet{Fernandez-Val2016a} to our $ijt$-panel structure with three-way fixed effects.\footnote{Further simulation experiments including dynamic panel models with two-way fixed effects and static panel models with three-way fixed effects are presented in Appendices \ref{app:monte_carlo_dynamic_twoway} and \ref{sec:montecarlo_extended}. In an earlier version of this article we additionally report simulations results for static two-way fixed effects models.} 
In line with our theoretical model, the simulations 
include unobserved components captured by fixed effects in the $it$, $jt$, and $ij$ dimensions, as well as the lagged dependent variable. 
Specifically, we generate data according to 

\begin{align*}
    y_{ijt} =& \ind \left[\beta_{y} y_{ijt-1} + \beta_{z} z_{ijt} +  \psi_{jt} + \lambda_{it} +\mu_{ij} \geq \epsilon_{ijt} \right] \, , \\
    y_{ij0} =& \ind \left[\beta_{z} z_{ij0}  +   \psi_{j0} + \lambda_{i0} + \mu_{ij} \geq \epsilon_{ij0} \right] \, ,
\end{align*}

where $i=1,\ldots, N$, $j = 1, \dots, N$,  $t = 1, \ldots, T$, $\beta_{y} = 0.5$, $\beta_{z} = 1$, $\psi_{jt} \sim \iid \N(0, 1 / 24)$, $\lambda_{it} \sim \iid \N(0, 1 / 24)$,  $\mu_{ij} \sim \iid \N(0, 1 / 24)$, and $\epsilon_{ijt} \sim \iid \N(0, 1)$.\footnote{We again follow \citet{Fernandez-Val2016a} and incorporate the information that $\{\lambda_{it}\}_{IT}$, $\{\psi_{jt}\}_{JT}$, and $\{\mu_{ij}\}_{IJ}$ are  independent sequences, and $\lambda_{it}$, $\psi_{jt}$, and $\mu_{ij}$  are independent for all $it$, $jt$, $ij$ in the covariance estimator for the APEs. The explicit expression is provided in Appendix \ref{sec:asymptotic}.} The exogenous regressor is modeled as an AR-1 process, $z_{ijt} = 0.5z_{ijt-1} +  \psi_{jt} + \lambda_{it} +  \mu_{ij} +  \nu_{ijt}$, where $\nu_{ijt} \sim \iid \N(0, 0.5)$ and $z_{ij0} \sim \iid \N(0, 1)$. We consider different sample sizes, specifically $N \in \{50, 100, 150\}$ and $T \in \{10, 20, 30, 40 , 50\}$ and generate 1,000 data sets for each.\\

\begin{figure}[p]
	\centering
	\caption{Dynamic: Three-way Fixed Effects ---  Predetermined Regressor}
	\label{fig:dyn3way_y}
	\includegraphics[height = 0.45\textheight]{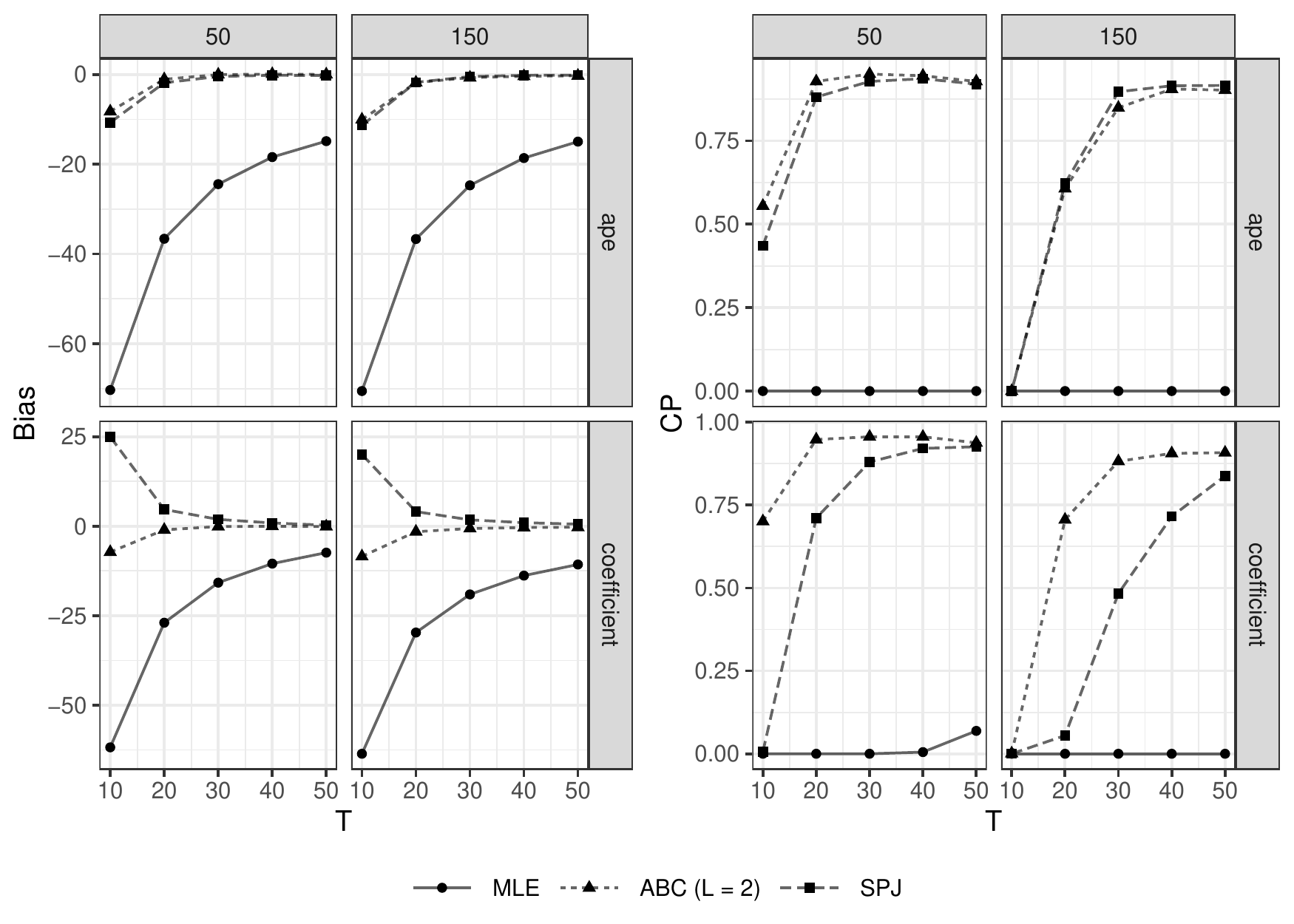}
	\vspace{1em}
    \caption{Dynamic: Three-way Fixed Effects --- Exogenous Regressor}
	\label{fig:dyn3way_x}
	\includegraphics[height = 0.45\textheight]{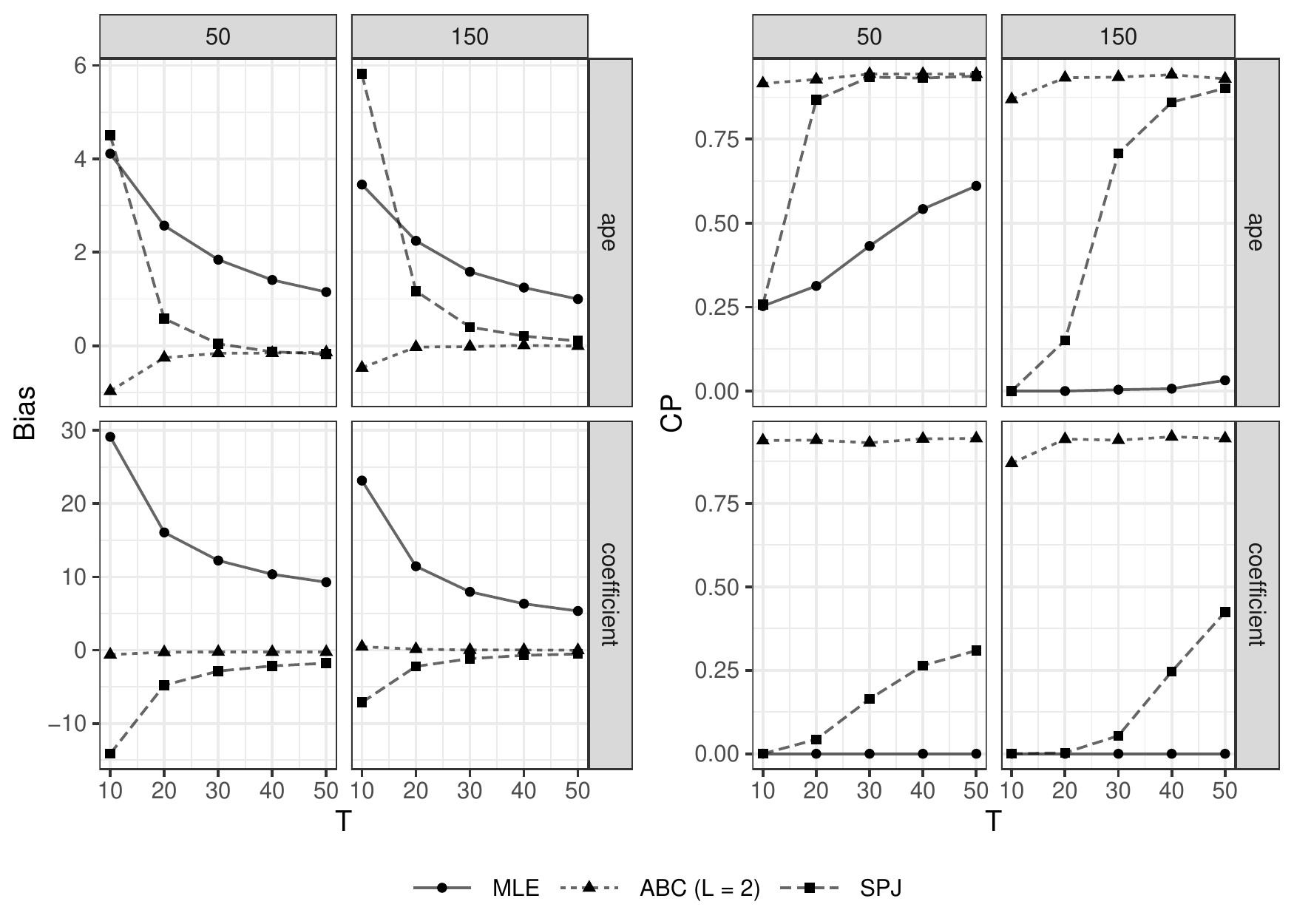}
\end{figure}

\begin{figure}[ht]
	\centering
	\caption{Dynamic: Three-way Fixed Effects ---  Exogenous Regressor (Long-Run)}
	\label{fig:dyn3way_xlong}
	\includegraphics[height = 0.45\textheight]{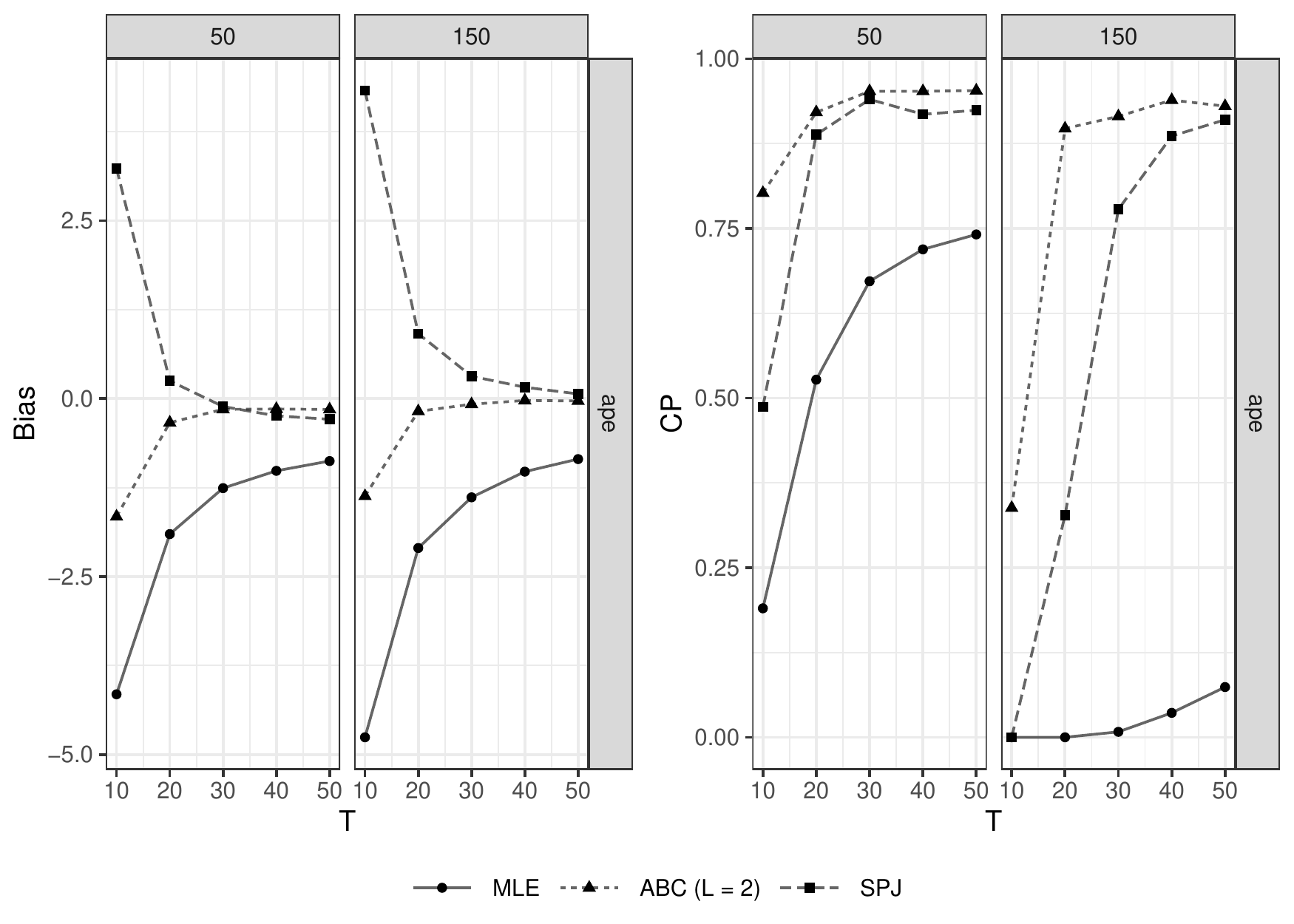}
\end{figure}

Tables \ref{tab:dyn3way_xn50} -- \ref{tab:dyn3waylong_xn150} in Appendix \ref{app:monte_carlo_dynamic_threeway} summarize the extensive simulation results. 
For ABC  we report two different choices of the bandwidth parameter, $L=1$ and $L=2$, which are indicated by values in parentheses.
In the following, we discuss the biases and coverage probabilities for $N \in \{50, 150\}$ which are shown in Figures \ref{fig:dyn3way_y} -- \ref{fig:dyn3way_xlong} for the sake of clarity. We focus on these two statistics in particular because they give us a good idea about the quality of the inference of an estimator. \\

We start by considering the different estimators for the structural parameters summarized in Figures \ref{fig:dyn3way_y} and \ref{fig:dyn3way_x}. For both kinds of regressors, MLE exhibits a severe bias that decreases with increasing $T$. However, even with $N=150$ and $T=50$, the estimator shows a distortion of 11 percent in the case of the predetermined regressor and 5 percent in the case of the exogenous regressor. We also find that the inference is not valid, since the CPs are zero or close to zero.
The bias corrections bring a substantial improvement. First, they reduce the bias considerably. For example, the MLE estimator of the predetermined regressor shows a distortion of 64 percent for $T=10$ and $N=150$. ABC  reduces the bias to 8 percent and SPJ to 20 percent. In the case of the exogenous regressor, MLE exhibits a bias of 23 percent, whereas ABC has a bias of 1 percent and SPJ of 7 percent. Irrespective of the type of the regressor, both bias-corrected estimators also converge quickly to the true parameter value  with growing $T$. Second, the bias corrections improve the CPs. 
For the exogenous regressor the CPs of ABC are close to the desired level of 95 percent for all $T$, whereas SPJ remains far away from 95 percent even at $T = 50$.
In the case of the predetermined regressor, the CPs of both corrections approach  the nominal level when $T$ rises. This happens faster for ABC.\\

We proceed with the APEs, where we distinguish between direct and long-run APEs. We first look at the direct APEs in Figures \ref{fig:dyn3way_y} and \ref{fig:dyn3way_x}. Overall, we obtain similar findings as for the structural parameters.
MLE is distorted over all settings, but the bias decreases as  $T$ increases. The distortion is especially severe in the case of the  the predetermined regressor, where it ranges from roughly $70$ percent for the settings with $T=10$ to $15$ percent for $T=50$. In contrast, the bias of the MLE for the exogenous regressor ranges from roughly $4$ percent ($T=10$) to $1$ percent ($T=50$). The bias corrections bring a substantial reduction in both cases. Whereas ABC shows only a small distortion of less than $1$ percent in the case of the exogenous regressor at $T=10$, SPJ is even more heavily distorted than MLE. However, with increasing $T$, both SPJ and ABC quickly converge to the true APE. Looking at inference,   unlike ABC, SPJ needs a sufficiently large number of time periods to get its CPs close to 95 percent. For the predetermined regressor, these convergence processes last longer than for the exogenous regressor. In contrast, the coverage probabilities of the MLE are always far below the desired nominal level. This is especially noteworthy for the exogenous regressor with $T \geq 40$, where the bias of MLE appears to be negligibly small for practical work. However,  the bias of the MLE is still large compared to the standard error. This becomes evident, when looking at the Bias/SE ratio in Tables \ref{tab:dyn3way_xn50} -- \ref{tab:dyn3way_xn150}, which is much higher for MLE than for the bias corrected estimators. Turning to the results of the long-run APEs reported in Figure \ref{fig:dyn3way_xlong}, we note that the results are qualitatively similar to the results of the direct APE of the exogenous regressor. As $T$ increases, both bias corrections quickly bring the bias towards zero and the coverage probabilities quickly reach the nominal level. Again, the convergence of ABC is faster and inference of MLE remains invalid despite small biases in settings with larger $T$.
\\

Overall, our three-way fixed effects simulation results confirm the conjecture of \citet{Fernandez-Val2018} about the general form of the bias and lend support to our bias corrections.
First, we find that the bias corrections indeed substantially mitigate the bias. Second, as already found in other studies, analytical bias corrections outperform split-panel jackknife bias corrections (see among others \cite{Fernandez-Val2016a}, and \cite{Czarnowske2019}). For samples with shorter time horizons, ABC is often less distorted and its dispersion is generally lower. This is also reflected by better CPs. Generally, in the three-way fixed effects setting, a sufficiently large number of time periods appears to be crucial to obtain reliable results for the bias-corrected estimators.
As a main takeaway, our simulation results suggest that estimates based on MLE should be treated with great caution, because even in situations where we expect small biases, the inference may be invalid.

\section{Determinants of the Extensive Margin of Trade}
\label{sec:application}


Having described the estimation and bias correction procedures, we now turn to the estimation of the determinants of the extensive margin of international trade outlined in Section \ref{sec:theory}.\\

Recall equation \eqref{eq:empirical_three-way} that relates the incidence of non-zero aggregate trade flows to exporter-time and importer-time specific characteristics, trade in the previous period, time-invariant unobservable trade barriers and bilateral trade policy variables:

\begin{align*}
    y_{ijt} = \begin{cases}
    1 \quad \text{if} \quad \kappa + \lambda_{it} + \psi_{jt} + \beta_{y} y_{ij(t-1)} + \mathbf{z}_{ijt}'\fmbeta_{z} +\mu_{ij} \geq \zeta_{ijt}, \\
    0\quad\text{else} \, .
    \end{cases}
\end{align*}

This yields the following dynamic three-way fixed effects probit model:

\begin{align}
    \Pr ( y_{ijt} = 1 | y_{ij(t-1)}, \mathbf{z}_{ijt}, \lambda_{it}, \psi_{jt} , \mu_{ij} ) &= F \left(\beta_{y} y_{ij(t-1)} + \mathbf{z}_{ijt}'\fmbeta_{z} + \lambda_{it} + \psi_{jt} + \mu_{ij} \right) \, ,
    \label{eq:estimation_threeway}
\end{align}

$y_{ij(t-1)}$ is the lagged dependent variable, $\mathbf{z}$ is a vector of observable bilateral variables, and $\beta_{y}$ and $\beta_{z}$ are the corresponding parameters.
We largely follow \citet{Helpman2008} and the wider literature on the determinants of the \textit{intensive} margin of trade \citep[compare][]{Head2014} in the choice of these variables: distance, a common land border, the same origin of the legal system, common language, previous colonial ties, a joint currency, an existing free trade agreement, or joint membership in the WTO/GATT. The effect of all time-invariant variables will only be identified in specifications in which we omit the bilateral fixed effects.\\

For data, we turn to the updated Gravity dataset provided by CEPII \citep{Conte2021}, which encompasses annual information on bilateral trade flows and these variables of interest for 198 countries from 1948 -- 2019. As only non-zero flows are reported in the data, we construct zeros analogous to \citet{Head2010}: Whenever an exporter reports at least one non-zero flow \textit{and} an importer reports at least one non-zero flow, their bilateral flow --- if not non-zero already --- is coded as zero. This procedure results in an unbalanced data set with 1,652,296 observations.\footnote{The updated Gravity dataset by \citet{Conte2021} supersedes the gravity dataset provided alongside \citet{Head2010}, which provided similar data from 1948 -- 2006. In the latter, some non-zero flows are coded as zero, in order to correct for implausible values reported in the original IMF DOTS data \citep[][p. 4,12]{Head2010}.  
} \\

Out of the 38,912 country pairs included in our analysis, 9,984 never switch their trading status, 4,609 switch a single time, and the average (median) number of switches is 3.9 (3), ranging up to 29 for trade from Benin to Sweden and 32 from Barbados to Bolivia.\footnote{See Figure \ref{fig:frequency_switches} in Appendix \ref{app:application} for the full distribution.}


\subsection{Main Results}
\label{sec:results_main}

Before turning to the regression results, we repeat the descriptive analysis about the persistence of the bilateral trade flows from Section 1, now considering the transition probabilities into export from period $t-1$ to $t$ for the time horizon from 1948 -- 2019. Table \ref{tab:transition_full_sample} confirms the high level of persistence:
89.7 percent of the country pairs which did not trade in the previous year did not trade in the following year and 91.5 percent of the pairs which \textit{did} trade in the previous year continued to trade in the following year. 
Thus, the  probability to export in period $t$ is 81.2 percentage points higher for those country pairs that already engaged in trade in period $t-1$ compared to those that did not.\footnote{The number is computed as the difference between the probability of exporting in period $t$ conditional on exporting and not exporting in period $t-1$.} However, Table \ref{tab:transition_full_sample} does not reveal any information about the kind of persistence. In the following analysis we will investigate the importance of using dynamic model specifications which allow us to disentangle the observed persistence into two sources: (i) true state dependence and (ii) observed and unobserved heterogeneity.\\

\begin{table}[ht]
	\vspace{1em}
	\centering
	\caption{Transition Probabilities (1948 -- 2019)}
	\label{tab:transition_full_sample}
	\begin{tabular}{rrr}
		\hline
		& $y_{ijt} = 0$ & $y_{ijt} = 1$ \\ 
		  \hline
		$y_{ij(t-1)} = 0$ & 89.7 \% & 10.3 \% \\ 
		$y_{ij(t-1)} = 1$ & 8.6 \% & 91.5 \% \\ 
		\hline
	\end{tabular}
\end{table}


In the following analysis we only consider analytically bias-corrected estimates, mainly for two reasons: (1) the simulation results demonstrate that ABC performs much better than SPJ; (2) SPJ requires the additional homogeneity assumption, which is likely to be violated in our application. Progressing economic integration over time --- be it through an increasing number of WTO/GATT memberships or through bilateral and multilateral integration in the form of currency unions and free trade agreements --- is at odds with the required stationary pattern.\footnote{See the corresponding homogeneity tests for model specifications (2) -- (5) of table \ref{tab:application_probit_ape} in table \ref{tab:homogeneity} in Appendix \ref{app:application}. We find evidence that the homogeneity assumption is violated across all splitting dimensions.}\\

Table \ref{tab:application_probit_ape} reports average partial effects of several static and dynamic fixed effects probit specifications.\footnote{Coefficient estimates are reported in Table \ref{tab:application_probit_coefficient} in the Appendix.} Bias-corrected estimates and their corresponding standard errors are printed in bold. For comparison, the uncorrected estimates are also shown. In column (1) we first mimic the static specification estimated by \citet{Helpman2008}.\footnote{\citet{Helpman2008} use a dataset that ranges from 1970 to 1997. They also include dummy variables for whether both countries are landlocked or islands, or follow the same religion. Hence our estimates deviate somewhat from theirs, while remaining qualitatively similar.} Their specification includes exporter, importer, and time fixed effects.\footnote{Note that following \citet{Fernandez-Val2018} the incidental bias problem is small enough to ignore in this setting with $i$, $j$ and $t$ fixed effects, since the order of the bias is $1/IT + 1/JT + 1/IJ$, which in our case becomes negligible small since $I$, $J$ and $T$ are large.}
All average partial effects have the expected sign, indicating a negative impact of distance on the probability to trade, while having a common border, the same origin of the legal system, a shared language, or a joint colonial history are all estimated to have a positive impact. Also note the strong and highly significant impact of a common currency, free trade agreement or joint membership of the WTO/GATT. Ceteris paribus, each is estimated to increase the probability of non-zero flows by between 5.2 and 7.1 percentage points. \\

\begin{table}[p]
	\centering
	\small
	\caption{Probit Estimation: Average partial effects}
	\label{tab:application_probit_ape}
	\resizebox{\linewidth}{!}{
		\begin{tabular}{l*{8}{c}}
			\toprule
			& \multicolumn{7}{c}{Dependent variable: $y_{ijt}$} \\
			\cmidrule{2-8}
			& (1) & (2) & \multicolumn{2}{c}{(3)} & (4) & \multicolumn{2}{c}{(5)} \\
			&  &  & \textit{direct} & \textit{long-run} &  & \textit{direct} & \textit{long-run} \\
			\midrule
$y_{ij(t-1)}$ & - & - & \textbf{0.394}*** & - & - & \textbf{0.187}*** & - \\
& (-)  & (-)  & (\textbf{0.001}) & (-) & (-)  & (\textbf{0.001}) & (-) \\
 & - & - & 0.393*** & - & - & 0.148*** & -\\
& (-)  & (-)  & (0.001) & (-) & (-)  & (0.001) & (-)\\[0.25em] 
log(Distance) & - & \textbf{-0.141}*** & \textbf{-0.066}*** & \textbf{-0.141}*** & - & -  & - \\
& (-)  & (\textbf{0.000}) & (\textbf{0.000}) & (\textbf{0.001}) & (-) & (-)  & (-) \\
 & -0.141*** & -0.141*** & -0.066*** & -0.141*** & - & - & - \\
& (0.001) & (0.000) & (0.000) & (0.001) & (-)  & (-)  & (-)\\[0.25em] 
Land border & - & \textbf{0.024}*** & \textbf{0.010}*** & \textbf{0.021}*** & - & - & - \\
& (-) & (\textbf{0.004}) & (\textbf{0.002}) & (\textbf{0.004}) & (-)  & (-)  & (-) \\
 & 0.034*** & 0.025*** & 0.011*** & 0.023*** & - & - &  -\\
& (0.004) & (0.004) & (0.002) & (0.004) & (-)  & (-)  & (-) \\[0.25em] 
Legal & - & \textbf{0.006}*** & \textbf{0.003}*** & \textbf{0.006}*** & - & - & - \\
& (-)  & (\textbf{0.001}) & (\textbf{0.001}) & (\textbf{0.001}) & (-) & (-)  & (-) \\
 & 0.005*** & 0.006*** & 0.003*** & 0.006*** & - & -  & - \\
& (0.001) & (0.001) & (0.001) & (0.001) & (-) & (-)  & (-) \\[0.25em] 
Language & - & \textbf{0.071}*** & \textbf{0.034}*** & \textbf{0.073}*** & -  & -  & -\\
& (-)  & (\textbf{0.001}) & (\textbf{0.001}) & (\textbf{0.002}) & (-)  &  (-) & (-) \\
 & 0.074*** & 0.072*** & 0.034*** & 0.073*** & - & - & -\\
& (0.001) & (0.001) & (0.001) & (0.002) & (-)  & (-)   & (-) \\[0.25em] 
Colonial ties & -  & \textbf{0.051}*** & \textbf{0.023}*** & \textbf{0.049}*** & -  & - &  -\\
& (-) & (\textbf{0.001}) & (\textbf{0.001}) & (\textbf{0.002}) & (-)  & (-)  & (-) \\
& 0.053*** & 0.051*** & 0.023*** & 0.048*** & -  & - & - \\
& (0.001) & (0.001) & (0.001) & (0.002) & (-)  & (-)  & (-) \\[0.25em] 
Currency union & - & \textbf{0.097}*** & \textbf{0.045}*** & \textbf{0.095}*** & \textbf{0.034}*** & \textbf{0.022}*** & \textbf{0.034}***\\
& (-)  & (\textbf{0.003}) & (\textbf{0.002}) & (\textbf{0.004}) & (\textbf{0.004}) & (\textbf{0.004}) & (\textbf{0.006})\\
 & 0.071*** & 0.097*** & 0.046*** & 0.096*** & 0.034*** & 0.023*** & 0.034***\\
& (0.003) & (0.003) & (0.002) & (0.004) & (0.004) & (0.004) & (0.005)\\[0.25em] 
FTA & - & \textbf{0.054}*** & \textbf{0.025}*** & \textbf{0.054}*** & \textbf{0.012}*** & \textbf{0.008}*** & \textbf{0.012}***\\
& (-) & (\textbf{0.001}) & (\textbf{0.001}) & (\textbf{0.002}) & (\textbf{0.002}) & (\textbf{0.002}) & (\textbf{0.002})\\
 & 0.053*** & 0.054*** & 0.025*** & 0.052*** & 0.011*** & 0.007*** & 0.011***\\
& (0.001) & (0.001) & (0.001) & (0.002) & (0.002) & (0.002) & (0.002)\\[0.25em] 
WTO / GATT & - & \textbf{0.034}*** & \textbf{0.016}*** & \textbf{0.034}*** & \textbf{0.013}*** & \textbf{0.008}*** & \textbf{0.013}***\\
& (-) & (\textbf{0.001}) & (\textbf{0.001}) & (\textbf{0.002}) & (\textbf{0.002}) & (\textbf{0.001}) & (\textbf{0.002})\\
 & 0.052*** & 0.034*** & 0.016*** & 0.034*** & 0.012*** & 0.009*** & 0.013***\\
& (0.001) & (0.001) & (0.001) & (0.002) & (0.002) & (0.001) & (0.002)\\
\midrule
Fixed effects  & $i,j,t$ & $it,jt$ & \multicolumn{2}{c}{$it,jt$} & $it,jt,ij$ & \multicolumn{2}{c}{$it,jt,ij$} \\
Sample size & 1652296 & 1652296 & \multicolumn{2}{c}{1613354}  & 1652296    & \multicolumn{2}{c}{1613354 }    \\
Deviance    & 11.65$\times$10\textsuperscript{4}  & 9.90$\times$10\textsuperscript{4}  & \multicolumn{2}{c}{7.17$\times$10\textsuperscript{4}}     & 6.50$\times$10\textsuperscript{4}     & \multicolumn{2}{c}{5.76$\times$10\textsuperscript{4}}     \\
\midrule
\multicolumn{8}{p{1.07\textwidth}}{\footnotesize \textit{Notes:}  Column (1) uncorrected average partial effects, columns (2) - (5) bias-corrected average partial effects (bold font) and  uncorrected average partial effects (normal font). Columns (4) and (5) bias-corrected with $L = 2$. Standard errors in parenthesis. $^{***}p<0.01$, $^{**}p<0.05$, $^*p<0.1$} \\
\bottomrule
\end{tabular}
}
\end{table}

Column (2) introduces a stricter set of fixed effects, namely at the exporter-time and importer-time level. This specification can be considered a theory-consistent estimation of the model by HMR, and of our model if entry costs are zero and other bilateral trade cost determinants are fully observable. The average partial effects are qualitatively the same and quantitatively similar for most variables to those in column (1). However, e.g.\ the estimated effects of contiguity and joint WTO/GATT membership are reduced by more than a third, while the estimated effect of sharing a common currency increases by 2.6 percentage points or 37 percent. \\

Specification (3) keeps the same fixed effects, but adds a lagged dependent variable and thus controls for one type of persistence. Assuming no unobservable bilateral heterogeneity, this specification correctly estimates the model set up in Section \ref{sec:theory}. As the partial effects of static and dynamic models are not directly comparable due to the feedbacks involved in the latter, we show two types of average partial effects for our dynamic specifications: the usual \textit{direct} effects and the \textit{long-run} effects described in Section \ref{sec:binary_response_estimator}. The first result to note for the third specification is the highly significant average partial effect for the lagged dependent variable, which reflects the strong impact of previous non-zero trade flows on current ones. Ceteris paribus, the average partial effect shows a 39.4 percentage points higher probability of non-zero trade, given the two countries were also engaged in trade in the previous year. This implies that 48.5 percent of the observed persistence are attributed to true state dependence in this first dynamic specification.\footnote{This value is calculated as the ratio of the estimated average partial effect and the unconditional effect of trade in the previous period on the exporting probability ($39.4  / 81.2$).}  In terms of our model, this suggests a vast effect of market entry costs on the aggregate extensive margin. The second observation is that direct APEs are about 50 percent smaller than in column (2) across the board. However, once dynamic adjustments are taken into account, the average partial effects resulting from specifications (2) and (3) become very similar, suggesting that accounting for the market entry dynamics mainly matters for getting the \textit{timing} of trade policy effects right, rather than for the overall magnitude of the effects. \\

Specification (4) takes one step back and one forward. While not including the lagged dependent variable in the estimation, it introduces a bilateral fixed effect that controls for a second type of persistence --- bilateral unobserved heterogeneity.\footnote{Note that we do not want to make the assumption of strict exogeneity in this specification, but instead we allow the regressors to be weakly exogenous. For instance, strict exogeneity is too restrictive if we want to allow that country pairs sign a trade agreement in the future, \textit{because} they did not trade with each other in period $t$.} This also follows the important insight by \citet{Baier2007}, who show that controlling for unobserved bilateral heterogeneity produces a considerably different estimated impact of free trade agreements, among other variables, on the intensive margin of trade. While now an identification of many of the variables of interest is no longer possible because of their time invariance, this specification reveals a much reduced estimated impact of the time-varying variables. The impact of a common currency on the probability of exporting is reduced to 3.4 percentage points, while those of a common free trade agreement and WTO/GATT are decreased to 1.2 and 1.3 percentage points, respectively. These results highlight the importance of controlling for unobserved country pair heterogeneity to avoid endogeneity problems associated with trade policy variables. \\ 

Finally, in the last two columns we present the results from our preferred specification (5), which implements equation \eqref{eq:estimation_threeway}. The estimation again includes the ``full set'' of fixed effects, i.e.\ exporter-time, importer-time and bilateral fixed effects, now combined with the lagged dependent variable, and therefore controls for both kinds of persistence simultaneously. 
Again, the average partial effect on the lagged dependent variable is highly significant. It now entails a partial effect of about 18.7 percentage points, i.e.\ roughly 23 percent of the observed persistence can be attributed to true state dependence and 77 percent to other observed factors and unobserved heterogeneity. Failure to account for unobserved heterogeneity in specification (3) hence overestimated the importance of the lagged dependent variable (corresponding to entry costs in our model) roughly by a factor of two and therefore mislabelled a substantial part of spurious as true state dependence. 
Considering the effects of the time-varying trade policy variables, small but statistically significant direct average partial effects in column (5) are estimated for a common currency at 2.2 percentage points and for both joint FTA and WTO memberships at 0.8 percentage points. Just as in the comparison between specifications (2) and (3), the average partial effects again become very similar when the static effects are compared to the long-run effects in the dynamic specification. \\

When comparing bias-corrected and uncorrected average partial effects throughout Table \ref{tab:application_probit_ape}, it is noticeable that both differ only slightly for the exogenous regressors within the different specifications. The most significant impact is observed on the average partial effect for the predetermined variable, which in specification (5) differs by more than 26 percent.  These results are in line with the theoretical properties of the estimators and the findings of our simulation study.\footnote{For the two-way models in columns (2) - (3) the order of the bias is $1/I + 1/J$, i.e.\ the bias only depends on the number of exporters/importers. For the three-way models in columns (4) - (5) the order of the bias is $1/I + 1/J + 1/T$, i.e.\ the bias additionally depends on the number of time periods. In our application the number of exporters/importers is relatively large but the number of time periods is substantially lower.}  Despite the small biases for the exogenous variables, a bias correction is still necessary for our three-way fixed effects specifications because the biases are not negligible relative to the standard errors and thus inference of the uncorrected estimator is invalid. Note that for applications with a shorter time horizon, the biases will be more evident.



\subsection{Predictive Analysis}

After we have seen that the presented innovations matter significantly for the estimation of extensive margin determinants, we now consider the predictive performance of the different specifications discussed above. As an additional benchmark, we also look at a naive --- purely descriptive --- approach that predicts export decisions in period $t$ solely based on the export decision in period $t-1$. For the resulting altogether six options, we evaluate the predictive power using the following measures: the total share of correctly predicted export decisions (accuracy),  the share of correctly predicted decisions conditional on not exporting (true negative rate), and  the share of correctly predicted decisions conditional on exporting (true positive rate). As becomes clear in Table \ref{tab:application_prediction}, the naive approach already works very well for predictive purposes. Close to 90 percent of the exporting decisions in period $t$ are correctly predicted by simply reproducing exporting decisions from $t-1$, irrespective of the measure we are considering. As it turns out, both the static model with $i$, $j$, and $t$ fixed effects (specification (1)) and the static model with $it$ and $jt$ fixed effects (specification (2)) produce poorer predictions than the naive approach --- particularly clearly so by about six percentage points in the former case. However, this changes once persistence is explicitly taken into account in specifications (3) to (5). All of them at least slightly improve the predictions of the naive approach. Whether persistence is incorporated \textit{only} via the lagged dependent variable \textit{or} via pair fixed effects turns out to yield very similar predictive quality. Combining both in our preferred specification --- the dynamic three-way fixed effects models --- yields the highest predictive power.  We are able to correctly predict 92.6 percent of the export decisions, 91.4 percent of the decisions conditional on not exporting, and 93.5 percent of the decisions conditional on exporting, implying an improvement compared to the naive approach by around 2 percentage points.
All in all, the predictive analysis underlines the importance of controlling for true state dependence and unobserved bilateral factors simultaneously.

\begin{table}[!ht]
	\centering
	\small
	\caption{Predictive Analysis}
	\label{tab:application_prediction}
    \begin{tabular}{l*{6}{c}}
      \toprule
           & Naive  & (1) & (2) & (3) & (4) & (5) \\
      \midrule
	Accuracy & 90.6 \% & 84.5 \% & 86.9 \% & 91.4 \% & 91.5 \% & 92.6 \% \\ 
	True negative rate & 91.5 \% & 85.6 \% & 87.7 \% & 92.3 \% & 92.5 \% & 93.5 \% \\ 
	True positive rate & 89.7 \% & 83.3 \% & 86.0 \% & 90.4 \% & 90.6  \% & 91.7 \%\\ 
    \midrule
    \multicolumn{7}{p{0.8\textwidth}}{\footnotesize \textit{Notes:} Column (1) uncorrected probit model, columns (2) -- (5) bias-corrected probit model.} \\
    \bottomrule
    \end{tabular}
\end{table}


\subsection{Robustness Checks}

\begin{table}[p]
	\centering
	\small
	\caption{Logit Estimation: Average partial effects}
	\label{tab:application_logit_ape}
	\resizebox{\linewidth}{!}{
		\begin{tabular}{l*{8}{c}}
			\toprule
			& \multicolumn{7}{c}{Dependent variable: $y_{ijt}$} \\
			\cmidrule{2-8}
			& (1) & (2) & \multicolumn{2}{c}{(3)} & (4) & \multicolumn{2}{c}{(5)} \\
			&  &  & \textit{direct} & \textit{long-run} &  & \textit{direct} & \textit{long-run} \\
			\midrule
$y_{ij(t-1)}$ & - & - & \textbf{0.373}*** & - & -  & \textbf{0.169}*** & -\\
& (-)  & (-)  & (\textbf{0.001}) & (-) & (-) & (\textbf{0.001}) & (-)\\	
& - & -  & 0.375*** & - & - & 0.136*** & -\\
& (-) & (-) & (0.001) & (-) & (-)  & (0.001) & (-)\\[0.25em] 			
log(Distance) & - & \textbf{-0.144}*** & \textbf{-0.067}*** & \textbf{-0.142}*** & - & - & - \\
& (-) & (\textbf{0.000}) & (\textbf{0.000}) & (\textbf{0.001}) & (-) & (-)  & (-) \\
& -0.144*** & -0.144*** & -0.067*** & -0.142*** & - & - & -\\
& (0.000) & (0.000) & (0.000) & (0.001) & (-) & (-)  & (-) \\[0.25em] 
Land border & - & \textbf{0.053}*** & \textbf{0.011}*** & \textbf{0.022}*** & - & - & - \\
& (-) & (\textbf{0.003}) & (\textbf{0.002}) & (\textbf{0.004}) & (-) & (-)  & (-) \\
& 0.057*** & 0.054*** & 0.012*** & 0.025*** & -  & - & - \\
& (0.003) & (0.003) & (0.002) & (0.004) & (-) & (-)  & (-) \\[0.25em] 
Legal & - & \textbf{0.007}*** & \textbf{0.003}*** & \textbf{0.006}*** & - & - & -\\
& (-) & (\textbf{0.001}) & (\textbf{0.001}) & (\textbf{0.001}) & (-)  & (-)  & (-) \\
& 0.007*** & 0.007*** & 0.003*** & 0.006*** & - & - & - \\
& (0.001) & (0.001) & (0.001) & (0.001) & (-) & (-)  & (-) \\[0.25em] 	
Language & - & \textbf{0.069}*** & \textbf{0.035}*** & \textbf{0.073}*** & - & - & - \\
& (-) & (\textbf{0.001}) & (\textbf{0.001}) & (\textbf{0.002}) & (-)  &  (-) & (-)\\
& 0.072*** & 0.070*** & 0.035*** & 0.073*** & - & - & - \\
& (0.001) & (0.001) & (0.001) & (0.002) & (-) & (-) & (-) \\[0.25em] 
Colonial ties & - & \textbf{0.052}*** & \textbf{0.024}*** & \textbf{0.049}*** & - & - & - \\
& (-) & (\textbf{0.001}) & (\textbf{0.001}) & (\textbf{0.002}) & (-)  & (-)  & (-) \\	
& 0.055*** & 0.052*** & 0.023*** & 0.049*** & -  & - & - \\
& (0.001) & (0.001) & (0.001) & (0.002) & (-) & (-)  & (-) \\[0.25em] 
Currency union &  & \textbf{0.097}*** & \textbf{0.046}*** & \textbf{0.096}*** & \textbf{0.034}*** & \textbf{0.022}*** & \textbf{0.035}***\\
& (-) & (\textbf{0.003}) & (\textbf{0.002}) & (\textbf{0.004}) & (\textbf{0.004}) & (\textbf{0.004}) & (\textbf{0.006})\\	
& 0.071*** & 0.098*** & 0.047*** & 0.097*** & 0.034*** & 0.024*** & 0.035***\\
& (0.002) & (0.003) & (0.002) & (0.004) & (0.004) & (0.004) & (0.006)\\[0.25em] 
FTA & - & \textbf{0.056}*** & \textbf{0.026}*** & \textbf{0.054}*** & \textbf{0.015}*** & \textbf{0.008}*** & \textbf{0.013}***\\
& (-) & (\textbf{0.001}) & (\textbf{0.001}) & (\textbf{0.002}) & (\textbf{0.002}) & (\textbf{0.002}) & (\textbf{0.002})\\		
& 0.054*** & 0.056*** & 0.025*** & 0.053*** & 0.014*** & 0.008*** & 0.012***\\	
& (0.001) & (0.001) & (0.001) & (0.002) & (0.002) & (0.002) &  (0.002)\\[0.25em] 	
WTO/GATT & - & \textbf{0.028}*** & \textbf{0.016}*** & \textbf{0.034}*** & \textbf{0.012}*** & \textbf{0.009}*** & \textbf{0.014}***\\
& (-) & (\textbf{0.001}) & (\textbf{0.001}) & (\textbf{0.002}) & (\textbf{0.002}) & (\textbf{0.001}) & (\textbf{0.002})\\
& 0.049*** & 0.028*** & 0.016*** & 0.034*** & 0.012*** & 0.009*** & 0.013***\\
& (0.001) & (0.001) & (0.001) & (0.002) & (0.002) & (0.002) & (0.002)\\
			\midrule
			Fixed effects  & $i,j,t$ & $it,jt$ & \multicolumn{2}{c}{$it,jt$} & $it,jt,ij$ & \multicolumn{2}{c}{$it,jt,ij$} \\
		Sample size & 1652296 & 1652296 & \multicolumn{2}{c}{1613354}  & 1652296    & \multicolumn{2}{c}{1613354 }    \\
			Deviance    & 11.57$\times$10\textsuperscript{5}  & 9.81$\times$10\textsuperscript{5}  & \multicolumn{2}{c}{7.20$\times$10\textsuperscript{5}}     & 6.44$\times$10\textsuperscript{5}     & \multicolumn{2}{c}{5.75$\times$10\textsuperscript{5}}     \\
			\midrule
			\multicolumn{8}{p{1.07\textwidth}}{\footnotesize \textit{Notes:}  Column (1) uncorrected average partial effects, columns (2) - (5) bias-corrected average partial effects (bold font) and  uncorrected average partial effects (normal font). Columns (4) and (5) bias-corrected with $L = 2$. Standard errors in parenthesis. $^{***}p<0.01$, $^{**}p<0.05$, $^*p<0.1$} \\
			\bottomrule
		\end{tabular}
	}
\end{table}

We next consider two robustness checks for our main results from Section \ref{sec:results_main}. First, while our preferred estimation of the extensive margin of trade with a dynamic three-way fixed effects binary choice estimator follows from the stylized facts and our theoretical model, the decision on \textit{which} binary choice estimator to use hinges on the distributional assumption for the error term (and hence for the demand shock in our model). We followed \citet{Eaton2011} and assumed log-normal shocks, leading us to using a probit for our main estimations. We now consider log-logistic shocks and the resulting logit estimator instead. Table \ref{tab:application_logit_ape} displays the results for the same specifications as in Table \ref{tab:application_probit_ape}, but estimated using a logit. Reassuringly, the average partial effects are very similar to the probit case for all variables in all specifications. Introducing the different innovations step-by-step changes the estimated effects in the same way as for the probit. The insights discussed above therefore do not hinge on the specific distributional assumption made in the parametrization of the probability of success.\\

Second, in the estimation of our preferred specification (5) we have discretionary power in one dimension for the exact form of bias correction used. Specifically, the bandwidth parameter $L$ used to estimate the spectral densities has to be chosen. We therefore follow the recommendation by \citet{Fernandez-Val2016a} and investigate the sensitivity of the results with respect to $L$. Table \ref{tab:application_trim_probit_ape} depicts the direct and long-run average partial effects obtained with the bias-corrected dynamic three-way fixed effects probit estimator for $L \in \{1,2,3,4\}$. Again, our results turn out to be very robust. There only appears to be a slight upward trend in the estimated true state dependence for larger $L$.\footnote{In Table \ref{tab:application_trim_logit_ape} in Appendix \ref{app:application}, we show the same sensitivity check for the dynamic three-way logit estimates and find the same robust pattern.}

\begin{table}[!t]
	\centering
	\small
	\caption{\footnotesize Probit Estimation with Different Bandwidths: Bias-Corrected Average Partial Effects}
	\label{tab:application_trim_probit_ape}
	\resizebox{\linewidth}{!}{
		\begin{tabular}{l*{8}{c}}
			\toprule
			& \multicolumn{8}{c}{Dependent variable: $y_{ijt}$} \\
			\cmidrule{2-9}
			& \multicolumn{2}{c}{$L = 1$} 	& \multicolumn{2}{c}{$L = 2$} 	& \multicolumn{2}{c}{$L = 3$} 	& \multicolumn{2}{c}{$L = 4$} \\
			\cmidrule(lr){2-3}\cmidrule(lr){4-5}\cmidrule(lr){6-7}\cmidrule(lr){8-9}
			&\textit{direct}&\textit{long-run}&\textit{direct}&\textit{long-run}&\textit{direct}&\textit{long-run}&\textit{direct}&\textit{long-run}\\
			\midrule
$y_{ij(t-1)}$ & 0.181*** & - & 0.187*** & - & 0.191*** & - & 0.192*** & - \\
& (0.001) & (-) & (0.001) & (-) & (0.001) & (-) & (0.001) & (-)\\
Currency union & 0.022*** & 0.034*** & 0.022*** & 0.034*** & 0.022*** & 0.034*** & 0.022*** & 0.035***\\[0.25em] 
& (0.004) & (0.006) & (0.004) & (0.006) & (0.004) & (0.006) & (0.004) & (0.006)\\
FTA & 0.008*** & 0.012*** & 0.008*** & 0.012*** & 0.008*** & 0.012*** & 0.008*** & 0.012***\\
& (0.002) & (0.002) & (0.002) & (0.002) & (0.002) & (0.002) & (0.002) & (0.002)\\[0.25em] 
WTO/GATT & 0.008*** & 0.013*** & 0.008*** & 0.013*** & 0.009*** & 0.014*** & 0.009*** & 0.014***\\
& (0.001) & (0.002) & (0.001) & (0.002) & (0.001) & (0.002) & (0.001) & (0.002)\\
			\midrule
			\multicolumn{9}{p{1.07\linewidth}}{\footnotesize \textit{Notes:} All columns include Origin $\times$ Year, Destination $\times$ Year and Origin $\times$ Destination fixed effects. Standard errors in parenthesis. $^{***}p<0.01$, $^{**}p<0.05$, $^*p<0.1$} \\
			\bottomrule
		\end{tabular}
	}
\end{table}




\section{Conclusion}
\label{sec:conclusions}

In this paper we reexamine the determinants of the extensive margin of international trade. We set up a model that exhibits a dynamic component and allows for time-invariant unobserved bilateral trade cost factors, generating persistence --- a feature in the data that has so far been given little attention. For the estimation of our model, we propose new bias-corrected fixed effects binary choice estimators to account for the incidental parameters problem and to deal with weakly exogenous regressors.
Finally, we show that our estimates of the determinants of the extensive margin of trade differ significantly from previous ones. This highlights the importance of  true state dependence and unobserved heterogeneity and therefore strongly supports the use of our bias-corrected dynamic fixed effects estimator.\\

The extensive margin of trade obviously extends beyond the aggregate level, warranting further research at lower levels of aggregation, in particular in the context of firms. While our model's prediction and its empirical specification rely on some abstractions, it provides a very tractable and flexible framework that can be estimated with recently established estimation procedures, when combined with the bias correction technique we introduce.\\

A further natural extension for future research would be to define the extensive margin as the number of exporting sectors leading to a fractional instead of a binary outcome \citep[as in][]{SantosSilva2014,Anderson2020}. 
Following the insights for such conditional moment models by \citet{Fernandez-Val2016a}, our split-panel jackknife estimator can be directly applied and the analytical bias correction has to only be slightly adjusted for the fractional outcome setting.




\clearpage

{
\renewcommand{\baselinestretch}{1.15}
\printbibliography
}


\newpage
\setcounter{table}{0}
\renewcommand{\thetable}{A\arabic{table}}
\setcounter{figure}{0}
\renewcommand{\thefigure}{A\arabic{figure}}
\setcounter{equation}{0}
\renewcommand{\theequation}{A\arabic{equation}}
\appendix


\textbf{\LARGE{Online Appendix}}
\section{Stylized facts}
\label{app:stylized_facts}

\begin{figure}[!th]
    \centering
    \caption{Determinants of the Extensive margin of Trade --- Gravity and Persistence (2000–2001).}
    \label{fig:share_non_zeros_2000_2001}
    \includegraphics[width=\linewidth]{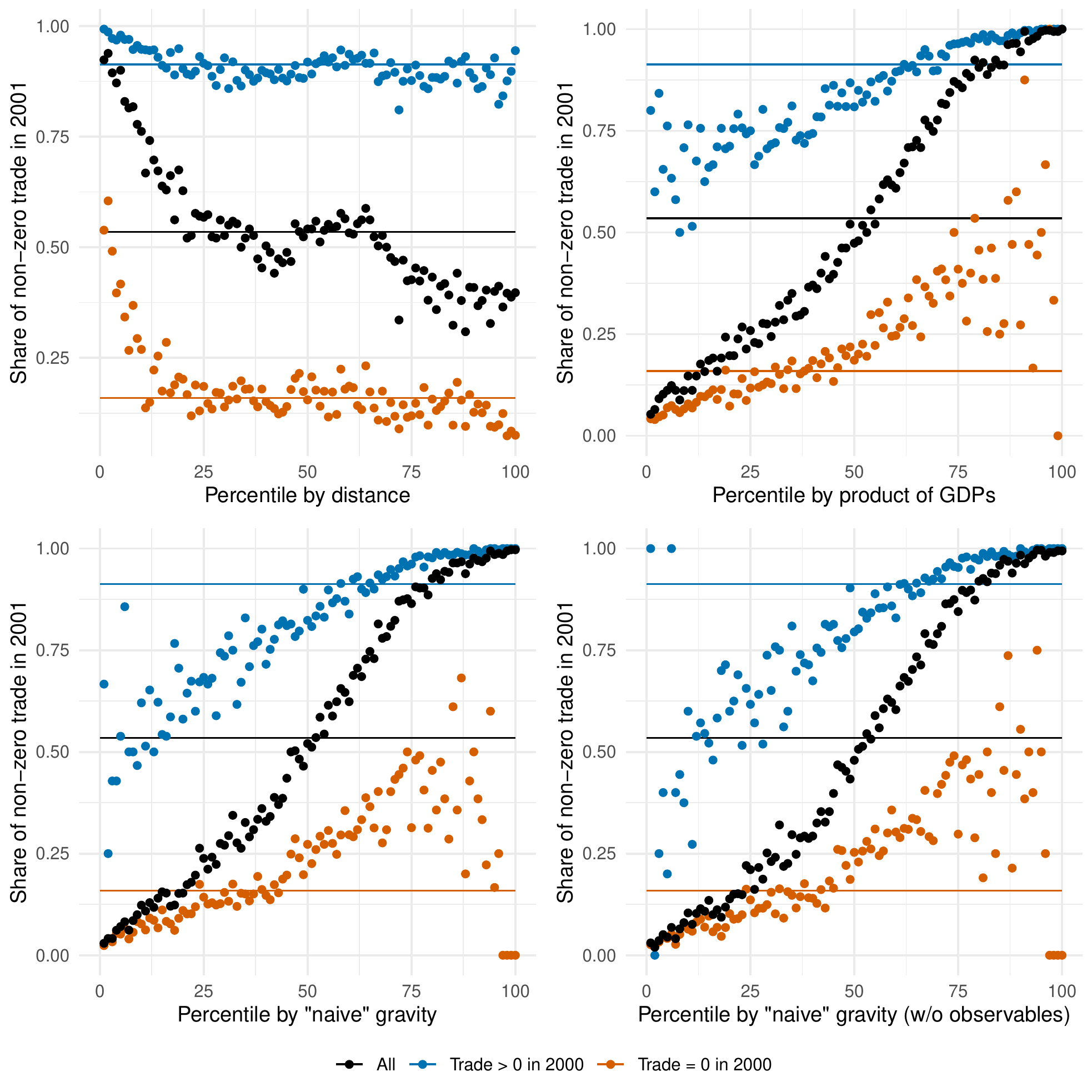}
\end{figure}

\begin{figure}[!t]
    \centering
    \caption{Determinants of the Extensive Margin of Trade --- Gravity and Persistence (2000–2019).}
    \label{fig:share_non_zeros_2000_2019}
    \includegraphics[width=\linewidth]{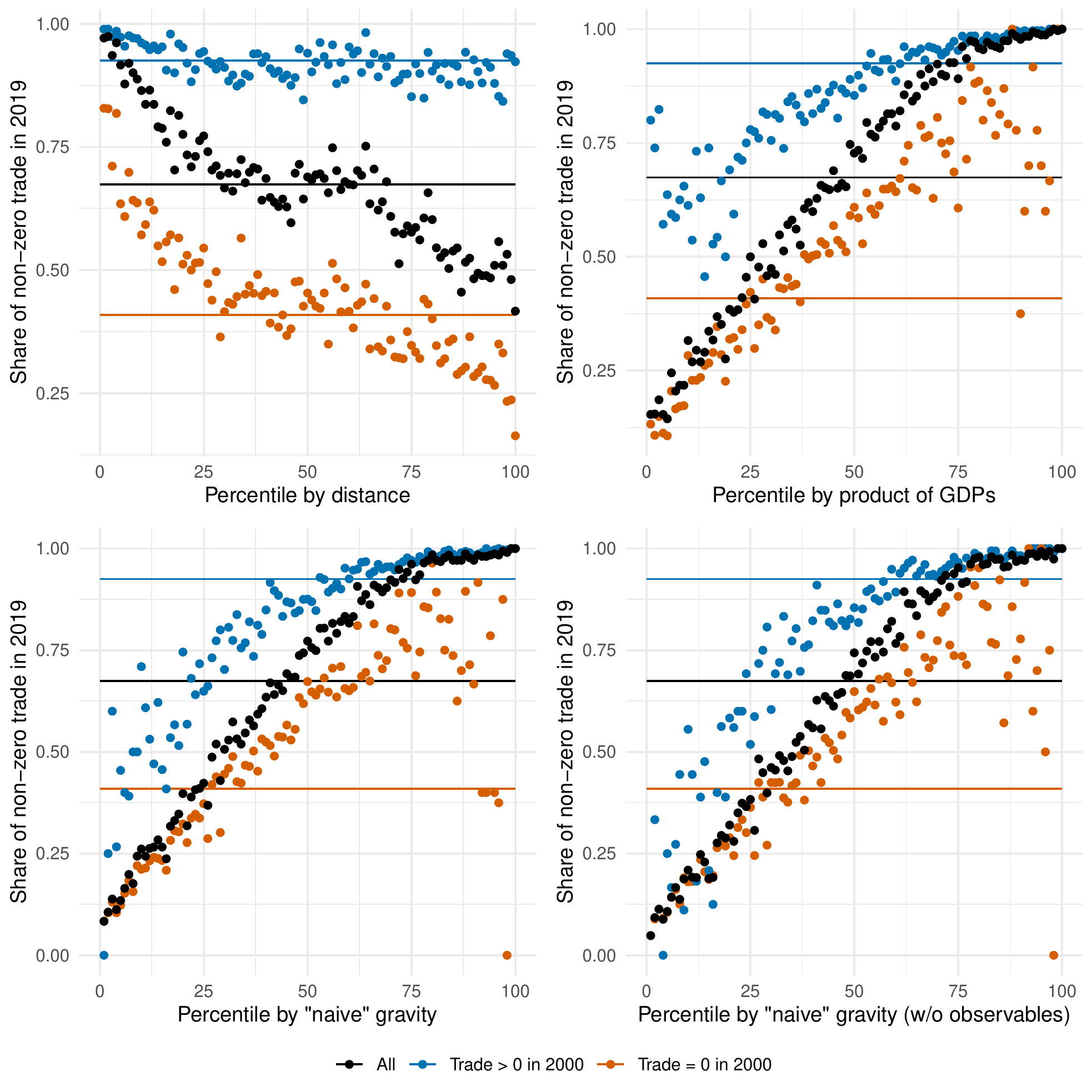}
\end{figure}

\FloatBarrier
\clearpage


\section{Econometric Details}



 \subsection{Computational Details}\label{sec:compdetails}

 In this section we briefly demonstrate how the method of alternating projections (MAP) works in the context of logit and probit models with a two- or three-way error component, and how it can be efficiently embedded into a standard Newton-Raphson optimization routine \citep[see][for further details]{Stammann2018}.\\

 First, note that  $\MX \mathbf{v}$ is essentially a weighted within transformation, where $\mathbf{v}$ is an arbitrary  $n \times 1$ vector, and $\MX =  \mathbf{I}_n - \PX = \mathbf{I}_n - \mathbf{D}(\mathbf{D}^{\prime} \boldsymbol{\Omega} \mathbf{D})^{-1} \mathbf{D}^{\prime} \boldsymbol{\Omega}$.  The computation of $\MX$ is problematic even in moderately large data sets, and since $\MX$ is non-sparse, there is also no general scalar expression to compute $\MX \mathbf{v}$. Thus \citet{Stammann2018} proposes to calculate $\MX \mathbf{v}$ using a simple iterative approach based on the MAP tracing back to \citet{VonNeumann1950} and \citet{Halperin1962}.\footnote{The MAP has been introduced to econometrics by \citet{Guimaraes2010} and \citet{Gaure2013} in the context of  linear models with multi-way fixed effects.} Let  $\mathbf{D}_k$,  denote the dummy variables corresponding to the $k$-th group, $k \in \{1,2,3\}$. Further, let  $\MX_{\mathbf{D}_k} \mathbf{v}$, with  $\MX_{\mathbf{D}_k} =  \mathbf{I}_n - \mathbf{D}_k(\mathbf{D}_k^{\prime} \boldsymbol{\Omega} \mathbf{D}_k)^{-1} \mathbf{D}_k^{\prime} \boldsymbol{\Omega}$. 
 The corresponding scalar expressions of $\MX_{\mathbf{D}_k} \mathbf{v}$ are summarized in Table (\ref{tab:ScalarTransformation}).

 \begin{table}[H]
 	\centering
 	\caption{Scalar Transformations}
 		\label{tab:ScalarTransformation}
 	\renewcommand{\arraystretch}{1.5}
 	\begin{tabular}{ll}
 		\hline
 		group & $\MX_{\mathbf{D}_k} \mathbf{v}$  \\ 
 		\hline
 		importer-time ($k=1$) & $\mathbf{v}_{ijt} - \frac{\sum_{j=1}^{J}   \omega_{ijt}v_{ijt}}{\sum_{j=1}^{J}   \omega_{ijt}}$ \\ 
 		exporter-time ($k=2$) & $\mathbf{v}_{ijt} - \frac{\sum_{i=1}^{I}  \omega_{ijt}v_{ijt}}{\sum_{i=1}^{I}   \omega_{ijt}}$ \\
 		dyadic ($k=3$) & $\mathbf{v}_{ijt} - \frac{  \sum_{t=1}^{T}  \omega_{ijt}v_{ijt}}{\sum_{t=1}^{T}  \omega_{ijt}}$ \\ 
 		\hline
 	\end{tabular}
 \end{table}

 The MAP can be summarized by algorithm 1, where $K=2$ in the case of two-way fixed effects and $K=3$ in the case of three-way fixed effects. Thus,  the MAP only requires to repeatedly apply weighted one-way within transformations \citep[see][]{Stammann2018}. The entire optimization routine is sketched by algorithm 2.

 \begin{algorithm}
	\caption{MAP: Neumann-Halperin}
 	\begin{algorithmic}[1]
 		\State Initialize $\MX \mathbf{v} = \mathbf{v}$. 
		\Repeat 
		\For{$k=1, \dots, K$}
 		\State Compute $\MX_{\mathbf{D}_k} \MX \mathbf{v}$ and update $\MX \mathbf{v}$ such that $\MX \mathbf{v} = \MX_{\mathbf{D}_k}  \MX \mathbf{v}$
		\EndFor	
		\Until{\textbf{convergence}.}
 	\end{algorithmic}
 \end{algorithm}
 \begin{algorithm}[H]
 	\caption{Efficient Newton-Raphson using the MAP}
 	\begin{algorithmic}[1]
 		\State Initialize $\boldsymbol{\beta}^{0}$, $\boldsymbol{\eta}^{0}$, and  $r = 0$. 
 		\Repeat
 		\State Set $r = r + 1$.
 		\State Given $\hat{\boldsymbol{\eta}}^{r - 1}$ compute $\hat{\boldsymbol{\nu}}$ and $\widehat{\boldsymbol{\Omega}}$.
 		\State Given $\hat{\boldsymbol{\nu}}$ and $\widehat{\boldsymbol{\Omega}}$ compute $\widehat{\MX}\hat{\boldsymbol{\nu}}$ and $\widehat{\MX}\mathbf{X}$ using the MAP
		\State Compute $\boldsymbol{\beta}^r - \boldsymbol{\beta}^{r-1} = \left((\widehat{\MX}\mathbf{X})^{\prime} \widehat{\boldsymbol{\Omega}} (\widehat{\MX}\mathbf{X})\right)^{-1} (\widehat{\MX}\mathbf{X})^{\prime} \widehat{\boldsymbol{\Omega}} (\widehat{\MX}\hat{\boldsymbol{\nu}})$
 		\State Compute $\hat{\boldsymbol{\eta}}^{r} = \hat{\boldsymbol{\eta}}^{r-1} +  \hat{\boldsymbol{\nu}} - \widehat{\MX}\hat{{\boldsymbol{\nu}}} + \widehat{\MX}\mathbf{X} (\boldsymbol{\beta}^r - \boldsymbol{\beta}^{r-1})$ 
		\Until{\textbf{convergence}.}
	\end{algorithmic}
 \end{algorithm}


\subsection{Neyman-Scott Variance Example}\label{sec:neyman}

In this section we study two variants of the classical \citet{Neyman1948} variance example to support the form of the bias terms, and to illustrate the functionality of the bias corrections. To the best of our knowledge,  the variance example of \citet{Neyman1948}  has not been investigated for our specific error components. We start with the more general three-way fixed effects case, which nests the two-way error structure. To make the dependence of the estimators on the sample size explicit, we use the subscript $I,J,T$ throughout the Appendix.


\subsubsection{Three-way Fixed Effects}

Let $i=1,\dots, I$, $j=1,\dots, J$ and $t=1,\dots, T$. Consider the following linear three-way fixed effects model

\begin{equation}
\label{eq:linear3way}
    y_{ijt} = \mathbf{x}_{ijt}^{\prime} \boldsymbol{\beta} + \lambda_{it} + \psi_{jt} + \mu_{ij} + u_{ijt} \; .
\end{equation}

According to \citet{Balazsi2018}, the appropriate within transformation corresponding to equation \eqref{eq:linear3way} is given by

\begin{equation*}
z_{ijt} -  \bar{z}_{ij\cdot} -  \bar{z}_{\cdot jt} -  \bar{z}_{i \cdot t} +  \bar{z}_{ \cdot \cdot t} + \bar{z}_{\cdot j \cdot} +  \bar{z}_{i \cdot \cdot} - \bar{z}_{\cdot \cdot \cdot} \; ,
\end{equation*}

where $\bar{z}_{ij\cdot} = \frac{1}{T} \sum_{t=1}^{T} z_{ijt}$, $\bar{z}_{\cdot jt} = \frac{1}{I} \sum_{i=1}^{I} z_{ijt}$, $\bar{z}_{i \cdot t} = \frac{1}{J} \sum_{j=1}^{J} z_{ijt}$, $\bar{z}_{ \cdot \cdot t} = \frac{1}{IJ} \sum_{i=1}^{I} \sum_{j=1}^{J} z_{ijt}$, $\bar{z}_{\cdot j \cdot} = \frac{1}{IT} \sum_{i=1}^{I} \sum_{t=1}^{T} z_{ijt}$, $\bar{z}_{i \cdot \cdot} = \frac{1}{JT} \sum_{j=1}^{J} \sum_{t=1}^{T} z_{ijt}$, and $\bar{z}_{\cdot \cdot \cdot} = \frac{1}{IJT} \sum_{i=1}^{I} \sum_{j=1}^{J} \sum_{t=1}^{T} z_{ijt}$.
\newline

This result is helpful to study the following variant of the \citet{Neyman1948} variance example

\begin{equation*}
    y_{ijt} | \boldsymbol{\lambda}, \boldsymbol{\psi}, \boldsymbol{\mu}  \sim  \mathcal{N}(\lambda_{it} + \psi_{jt} + \mu_{ij}, \beta) \; ,
\end{equation*}

where we can now easily form the uncorrected variance estimator 

\begin{equation}
    \label{eq:ucbeta}
    \hat{\beta}_{I, J, T} = \frac{1}{IJT} \sum_{i=1}^{I} \sum_{j=1}^{J} \sum_{t=1}^{T} (y_{ijt} - \bar{y}_{ij\cdot} -  \bar{y}_{\cdot jt} -  \bar{y}_{i \cdot t} +  \bar{y}_{ \cdot \cdot t} + \bar{y}_{\cdot j \cdot} +  \bar{y}_{i \cdot \cdot} - \bar{y}_{\cdot \cdot \cdot})^2 
\end{equation}

and the (degrees-of-freedom)-corrected counterpart

\begin{equation*}
\hat{\beta}_{I, J, T}^{cor} = \frac{IJT}{(I-1)(J-1)(T-1)} \hat{\beta}_{I, J, T} \; .
\end{equation*}

Taking the expectation  of (\ref{eq:ucbeta}) (conditional on the fixed effects) yields

\begin{align}
    \label{eq:neymanexpansion3way}
    \bar{\beta}_{I,J,T} = \EX_{\alpha}[\hat{\beta}_{I,J,T}] &= \beta^0 \left(\frac{(I-1)(J-1)(T-1)}{IJT}\right)\\ 
        &= \beta^0 \left( 1 - \frac{1}{I} - \frac{1}{J} - \frac{1}{T}  + \frac{1}{IT} + \frac{1}{JT} + \frac{1}{IJ} - \frac{1}{IJT}  \right) \; , \notag
\end{align}

where $\beta^0$ is the true variance parameter. Thus, the three leading  bias terms, which drive the main part of the asymptotic bias, are $\overline{\mathbf{B}}_{1, \infty}^{\beta} = - \beta^0$, $\overline{\mathbf{B}}_{2, \infty}^{\beta} = - \beta^0$, and $\overline{\mathbf{B}}_{3, \infty}^{\beta} = - \beta^0$.


\subsubsubsection{Analytical Bias Correction}

Using equation \eqref{eq:neymanexpansion3way}, we can form the analytically bias-corrected estimator

\begin{align}
\label{eq:ana3way}
    \tilde{\beta}_{I,J,T}^a &= \hat{\beta}_{I,J,T} - \frac{\widehat{\mathbf{B}}^{\beta}_{1,I,J,T}}{I} - \frac{\widehat{\mathbf{B}}^{\beta}_{2,I,J,T}}{J} - \frac{\widehat{\mathbf{B}}^{\beta}_{3,I,J,T}}{T} \; ,
\end{align}

where we set $\widehat{\mathbf{B}}^{\beta}_{1,I,J,T} = - \hat{\beta}_{I,J,T}$, $\widehat{\mathbf{B}}^{\beta}_{2,I,J,T} = - \hat{\beta}_{I,J,T}$, and $\widehat{\mathbf{B}}^{\beta}_{3,I,J,T} = - \hat{\beta}_{I,J,T}$ to reduce the order of the bias in equation \eqref{eq:neymanexpansion3way} at costs of introducing higher order terms (see equation \eqref{eq:expanabeta}). Thus, we can rewrite the analytically bias-corrected estimator \eqref{eq:ana3way}

\begin{equation}
    \label{eq:anabeta}
    \tilde{\beta}_{I,J,T}^a = \hat{\beta}_{I,J,T} \left(1 + \frac{1}{I} + \frac{1}{J} + \frac{1}{T} \right) \; .
\end{equation}

Taking the expectation of (\ref{eq:anabeta}) yields

\begin{align}
    \label{eq:expanabeta}
    \bar{\beta}_{I,J,T}^a =  \EX_{\alpha}[\tilde{\beta}_{I,J,T}^a] &= \beta^0 \left( 1 - \frac{1}{I} - \frac{1}{J} - \frac{1}{T} + \frac{1}{IT}  + \frac{1}{JT} + \frac{1}{IJ} - \frac{1}{IJT}  \right) \left(1 + \frac{1}{I} + \frac{1}{J} + \frac{1}{T} \right)\\
        &= \beta^0 \left( 1 - \frac{1}{IT} - \frac{1}{JT} - \frac{1}{T^2} - \frac{3}{IJ} + \frac{1}{I^3} + \frac{1}{J^3}  + \frac{4}{IJT} + \frac{1}{IT^2}  + \frac{1}{JT^2} \right. \notag\\
        & \qquad \qquad \left. - \frac{1}{I^3T} - \frac{1}{J^3T} - \frac{1}{IJT^2} \right)\notag \; .
\end{align}

\newpage

\subsubsubsection{Split-Panel Jackknife}

%
%
%

As an alternative to equation \eqref{eq:anabeta} we can also form the following SPJ estimator

\begin{equation*}
    \hat{\beta}_{I,J,T}^{spj} = 4 \hat{\beta}_{I,J,T} - \hat{\beta}_{I/2,J,T} - \hat{\beta}_{I,J/2,T} - \hat{\beta}_{I,J,T/2} \; ,
\end{equation*}

where $\hat{\beta}_{I/2,J,T}$ denotes the half panel estimator based on splitting the panel by  exporters.
This estimator also reduces the order of the bias in equation \eqref{eq:neymanexpansion3way} as we see from its  expected value

\begin{align}
    \label{eq:expspj2beta}
    \bar{\beta}_{I,J,T}^{spj} =  E_{\phi}[\hat{\beta}_{I,J,T}^{spj}] &=  4 \bar{\beta}_{I,J,T} - \bar{\beta}_{I/2, J,T} - \bar{\beta}_{I, J/2,T} - \bar{\beta}_{I,J,T/2}\\
        &= \beta^0 \left( 1 - \frac{1}{IT} - \frac{1}{JT} - \frac{1}{IJ} + \frac{2}{IJT}\right) \; . \notag
\end{align}



\subsubsubsection{Numerical Results}

Table \ref{tab:neymanbias3way} shows numerical results for the uncorrected and the bias-corrected estimators in finite samples, where we assume symmetry, i.e.\ $I=J=N$. The results demonstrate that the bias corrections are effective in reducing the bias. 

\begin{table}[ht]	
    \vspace{1em}
	\centering
	\caption{Bias - Three-way Fixed Effects}
	\label{tab:neymanbias3way}
	\begin{tabular}{ccccc}
		\hline
		N & T & $(\bar{\beta}_{I,J,T} - \beta^0)/\beta^0$ & $(\bar{\beta}_{I,J,T}^a - \beta^0)/\beta^0$ &  $(\bar{\beta}_{I,J,T}^{spj} - \beta^0)/\beta^0$\\ 
		\hline
	 10 & 10 & -0.271 & -0.052 & -0.028 \\ 
	 25 & 10 & -0.171 & -0.021  & -0.009 \\ 
	 25 & 25 & -0.115 & -0.009 & -0.005 \\ 
	 50 & 10 & -0.136 & -0.015 & -0.004 \\ 
	 50 & 25 & -0.078 & -0.004  & -0.002 \\ 
	 50 & 50 & -0.059 & -0.002 &  -0.001 \\ 
		\hline
	\end{tabular}
\end{table}


\subsubsection{Two-way Fixed Effects}

In the following we briefly review the example with two-way fixed effects

\begin{equation*}
    y_{ijt} | \boldsymbol{\lambda}, \boldsymbol{\psi} \sim  \mathcal{N}(\lambda_{it} + \psi_{jt}, \beta) \; .
\end{equation*}

Since it  is a subcase of three-way fixed effects example, all previous results simplify by dropping the terms that exhibit $T$.

The uncorrected variance estimator is\footnote{We draw on the appropriate demeaning formula for the two-way fixed effects model	$y_{ijt} = \boldsymbol{x}_{ijt}^{\prime} \boldsymbol{\beta} + \lambda_{it} + \psi_{jt} + u_{ijt}$, which is given by $z_{ijt} - \bar{z}_{\cdot jt} -  \bar{z}_{i \cdot t} +  \bar{z}_{ \cdot \cdot t}$.}

\begin{equation*}
\hat{\beta}_{I, J, T} = \frac{1}{IJT} \sum_{i=1}^{I} \sum_{j=1}^{J} \sum_{t=1}^{T}  (y_{ijt}   -  \bar{y}_{\cdot jt} -  \bar{y}_{i \cdot t} +  \bar{y}_{ \cdot \cdot t})^2
\end{equation*}

and the (degrees-of-freedom)-corrected variance estimator is

\begin{equation*}
    \hat{\beta}_{I, J, T}^{cor} = \frac{IJ}{(I-1)(J-1)} \hat{\beta}_{I, J, T}  \; .
\end{equation*}

Taking the expected value yields

\begin{align}
    \label{eq:neymanexpansion2way}
    \bar{\beta}_{I,J, T} = \EX_{\alpha}[\hat{\beta}_{I,J, T}] &= \beta^0 \left(\frac{(I-1)^2}{IJ}\right)\\
        &= \beta^0 \left( 1  - \frac{1}{I}  - \frac{1}{J}  + \frac{1}{IJ}  \right) \; .\notag
\end{align}


\subsubsubsection{Analytical Bias Correction}

Based on equation \eqref{eq:neymanexpansion2way} we can form the following analytically bias-corrected estimator

\begin{equation*}
    \tilde{\beta}_{I,J, T}^a = \hat{\beta}_{I,J, T} \left(1 + \frac{1}{I} + \frac{1}{J} \right) \; ,
\end{equation*}

which has  the expected value

\begin{equation*}
    \bar{\beta}_{I,J, T}^a =  \EX_{\alpha}[\tilde{\beta}_{I,J, T}^a] = \beta^0 \left( 1-  \frac{3}{IJ} + \frac{1}{I^3} + \frac{1}{J^3} \right) \; .
\end{equation*}
%
%

\clearpage
\subsubsubsection{Split-Panel Jackknife}

A suitable split-panel jackknife estimator is

\begin{equation*}
    \hat{\beta}_{I,J, T}^{spj} = 4 \hat{\beta}_{I,J, T} - \hat{\beta}_{I/2,J, T} - \hat{\beta}_{I,J/2, T} \; ,
\end{equation*}

which has the expected value

\begin{align*}
    \bar{\beta}_{I,J, T}^{spj} = \EX_{\alpha}[\hat{\beta}_{I,J, T}^{spj}] & = 3 \bar{\beta}_{I,J, T} - \bar{\beta}_{I/2, J, T} - \bar{\beta}_{I, J/2, T}\\
        &= \beta^0 \left( 1 -  \frac{1}{IJ}\right) \; .
\end{align*}


\subsubsubsection{Numerical Results}

The numerical results in Table  \ref{tab:neymanbias2way} demonstrate that the bias corrections work.

\begin{table}[ht]	
	\centering
	\caption{Bias - Two-way Fixed Effects}
		\label{tab:neymanbias2way}
	\begin{tabular}{cccc}
		\hline
		N &  $(\bar{\beta}_{I,J, T} - \beta^0)/\beta^0$ & $(\bar{\beta}_{I,J, T}^a - \beta^0)/\beta^0$ & $(\bar{\beta}_{I,J, T}^{spj} - \beta^0)/\beta^0$\\ 
		\hline
		10 & -0.190 & -0.028 & -0.010 \\ 
		25 & -0.078 & -0.005 &  -0.002 \\ 
		50 & -0.040 & -0.001 &  -0.000 \\ 
		100 & -0.020 & -0.000 &  -0.000 \\ 
		\hline
	\end{tabular}
\end{table}


\subsection{Asymptotic Bias Corrections}
\label{sec:asymptotic}

For the following expressions we draw on the results of \citet{Fernandez-Val2016a}, who have already derived the asymptotic distributions of the MLE estimators for  structural parameters and APEs in classical two-way fixed effects models based on $it$-panels. As outlined in \citet{Cruz-Gonzalez2017} the bias corrections of \citet{Fernandez-Val2016a} can easily be adjusted to two-way fixed effects models based on  pseudo-panels  with an $ij$-structure ($i$ corresponds to importer and $j$ to exporter), and importer and exporter fixed effects. We give an intuitive explanation. Since only $J$ observations are informative per exporter fixed effects, we get a bias of order $J$ for including exporter fixed effects, and vice versa a bias of order $I$ for including importer fixed effects. Further, since there are no predetermined regressors in an $ij$-structure, we get two symmetric bias terms

\begin{align}
    \overline{\mathbf{B}}_{1, \infty}&=\text{plim}_{I,J\rightarrow\infty}\left[-\frac{1}{2J}\sum_{j=1}^{J}\frac{\sum_{i=1}^{I} \EX_{\alpha}[H_{ij}\partial_{\eta^2}F_{ij}(\MX \mathbf{X})_{ij}]}{\sum_{i=1}^{I}\EX_{\alpha}[\omega_{ij}]}\right], \label{eq:B1} \\
    \overline{\mathbf{B}}_{2, \infty}&=\text{plim}_{I,J\rightarrow\infty}\left[-\frac{1}{2I}\sum_{i=1}^{I}\frac{\sum_{j=1}^{J} \EX_{\alpha}[H_{ij}\partial_{\eta^2}F_{ij}(\MX \mathbf{X})_{ij}]}{\sum_{j=1}^{J} \EX_{\alpha}[\omega_{ij}]}\right], \label{eq:B2}
\end{align}

where $\omega_{ij}$ is the $ij$-th diagonal entry of $\boldsymbol{\Omega}$, and $\MX =  \mathbf{I}_{IJ} - \mathbf{D}(\mathbf{D}^{\prime} \boldsymbol{\Omega} \mathbf{D})^{-1} \mathbf{D}^{\prime} \boldsymbol{\Omega}$. $\partial_{\iota^2}g(\cdot)$ denotes the second order partial derivative of an arbitrary function $g(\cdot)$ with respect to some parameter $\iota$. The explicit expressions of  $H_{ijt}$ and $\partial_{\eta^{2}} F_{ijt}$ are reported in Table \ref{tab:expression}.
Equations \eqref{eq:B1} and \eqref{eq:B2} are essentially $\overline{\mathbf{D}}_{\infty}$ from \citet{Fernandez-Val2016a} with adjusted  indices. The same adjustment can be transferred to the APEs.\\

In the following we apply the same logic to derive the asymptotic bias terms in our two- and three-way error structure.


\subsubsection{Two-way fixed effects}

We get a bias of order $J$ for including exporter-time fixed effects, since $J$ observations are informative per exporter-time fixed effect. In the same way we get a bias of order $I$ for including importer-time fixed effects. Similar to the case of the $ij$-structure of \citet{Cruz-Gonzalez2017} we  get two symmetric bias terms in the distributions of the structural parameters and the APEs, respectively, because including predetermined regressors  does not violate the strict exogeneity assumption.


\subsubsubsection{Asymptotic distribution of $\hat{\boldsymbol{\beta}}_{I,J,T}$}

\begin{align}
\label{eq:distribution_beta_2way}
    &\sqrt{IJ}(\widehat{\fmbeta}_{I, J, T}-\boldsymbol{\beta}^0)\rightarrow_{d}\overline{\mathbf{W}}_{\infty}^{-1}\mathcal{N}(\kappa \overline{\mathbf{B}}_{1, \infty}+ \kappa^{-1}\overline{\mathbf{B}}_{2, \infty},\overline{\mathbf{W}}_{\infty}) , \quad\text{with} \\ 
    \overline{\mathbf{B}}_{1, \infty}&=\text{plim}_{I,J\rightarrow\infty}\left[-\frac{1}{2J}\sum_{t=1}^{T}\sum_{j=1}^{J}\frac{\sum_{i=1}^{I} \EX_{\alpha}[H_{ijt}\partial_{\eta^2}F_{ijt}(\MX \mathbf{X})_{ijt}]}{\sum_{i=1}^{I}\EX_{\alpha} [\omega_{ijt}]}\right], \notag\\
    \overline{\mathbf{B}}_{2, \infty}&=\text{plim}_{I,J\rightarrow\infty}\left[-\frac{1}{2I}\sum_{t=1}^{T}\sum_{i=1}^{I}\frac{\sum_{j=1}^{J} \EX_{\alpha} [H_{ijt}\partial_{\eta^2}F_{ijt}(\MX \mathbf{X})_{ijt}]}{\sum_{j=1}^{J} \EX_{\alpha} [\omega_{ijt}]}\right], \notag\\
    \overline{\mathbf{W}}_{\infty} & = \text{plim}_{I,J\rightarrow\infty} \left[\frac{1}{IJ}\sum_{i=1}^{I}\sum_{j=1}^{J}\sum_{t=1}^{T} \EX_{\alpha}[\omega_{ijt}(\MX \mathbf{X})_{ijt}(\MX \mathbf{X})_{ijt}^{\prime}]\right], \notag
\end{align}

where $\sqrt{J / I} \rightarrow \kappa$ as $I,J \rightarrow  \infty$.


\subsubsubsection{Asymptotic distribution of $\hat{\boldsymbol{\delta}}_{I,J,T}$}

\begin{align}
\label{eq:distribution_ape_2way}
    &r(\hat{\boldsymbol{\delta}}_{I,J,T} - \boldsymbol{\delta} - I^{-1}\overline{\mathbf{B}}_{1, \infty}^{\delta} - J^{-1}\overline{\mathbf{B}}_{2, \infty}^{\delta})  \rightarrow_{d}  \mathcal{N}(0,\overline{\mathbf{V}}_{\infty}) , \quad\text{with} \\
    \overline{\mathbf{B}}_{1, \infty}^{\delta}&=\text{plim}_{I,J\rightarrow\infty}\left[\frac{1}{2JT}\sum_{t=1}^{T}\sum_{j=1}^{J} \frac{\sum_{i = 1}^{I} - \EX_{\alpha}[ H_{ijt} \partial_{\eta^{2}} F_{ijt}] \EX_{\alpha}[\left(\PX \boldsymbol{\Psi}\right)_{ijt}] + \EX_{\alpha}[\partial_{\eta^{2}} \boldsymbol{\Delta}_{ijt}]}{\sum_{i = 1}^{I}  \EX_{\alpha}[\omega_{ijt}]}\right], \notag\\
    \overline{\mathbf{B}}_{2, \infty}^{\delta}&=\text{plim}_{I,J\rightarrow\infty}\left[\frac{1}{2IT}\sum_{t=1}^{T}\sum_{i=1}^{I} \frac{\sum_{j = 1}^{J} - \EX_{\alpha}[ H_{ijt} \partial_{\eta^{2}} F_{ijt}] \EX_{\alpha}[\left(\PX\boldsymbol{\Psi}\right)_{ijt}] +\EX_{\alpha}[\partial_{\eta^{2}} \boldsymbol{\Delta}_{ijt}]}{\sum_{j = 1}^{J}  \EX_{\alpha}[\omega_{ijt}]}\right], \notag\\
    \overline{\mathbf{V}}_{\infty}^{\delta} &= \text{plim}_{I,J\rightarrow\infty} \frac{r^{2}}{I^{2}J^{2}T^{2}} \EX_{\alpha}\left[ \left(\sum_{i = 1}^{I} \sum_{j =  1}^{J} \sum_{t = 1}^{T}  \bar{\boldsymbol{\Delta}}_{ijt}\right)\left( \sum_{i = 1}^{I} \sum_{j =  1}^{J} \sum_{t = 1}^{T}  \bar{\boldsymbol{\Delta}}_{ijt} \right)^{\prime}  + \sum_{i = 1}^{I} \sum_{j = 1}^{J} \sum_{t = 1}^{T}\boldsymbol{\Gamma}_{ijt} \boldsymbol{\Gamma}_{ijt}^{\prime} \right]  \, ,\notag
\end{align}

where $\bar{\boldsymbol{\Delta}}_{ijt}=\boldsymbol{\Delta}_{ijt}-\boldsymbol{\delta}$, $\boldsymbol{\Delta}_{ijt} = [\Delta_{ijt}^1, \dots, \Delta_{ijt}^m]^{\prime}$, $\boldsymbol{\delta} = [\delta_1, \dots, \delta_m]^{\prime}$, $\delta_k =\frac{1}{IJT} \sum_{i=1}^{I} \sum_{j=1}^{J} \sum_{t=1}^{T} \Delta_{ijt}^k$,   $\boldsymbol{\Psi}_{ijt} = \partial_{\eta} \boldsymbol{\Delta}_{ijt}/\omega_{ijt}$, $r$ is a convergence rate, and

\begin{align*}
\boldsymbol{\Gamma}_{ijt} =& \EX_{\alpha} \left[(IJ)^{-1} \sum_{i = 1}^{I} \sum_{j = 1}^{J} \sum_{t = 1}^{T} \partial_{\beta} \boldsymbol{\Delta}_{ijt} - \left(\PX \mathbf{X}\right)_{ijt} \partial_{\eta} \boldsymbol{\Delta}_{ijt} \right]^{\prime} \overline{\mathbf{W}}_{\infty}^{- 1} \EX_{\alpha}\left[\left(\MX \mathbf{X}\right)_{ijt} \omega_{ijt} \boldsymbol{\nu}_{ijt}\right]\\
&- \EX_{\alpha}\left[\left(\PX \boldsymbol{\Psi}\right)_{ijt} \partial_{\eta} \ell_{ijt}\right] \, .
\end{align*}
$\partial_{\iota}g(\cdot)$ denotes the first order partial derivative of an arbitrary function $g(\cdot)$ with respect to some parameter $\iota$. 
The expression $\overline{\mathbf{V}}_{\infty}^{\delta}$ can be modified by assuming  that $\{\lambda_{it}\}_{IT}$ and $\{\psi_{jt}\}_{JT}$ are independent sequences, and $\lambda_{it}$ and $\psi_{jt}$ are independent for all $it$, $jt$:

\begin{align*}
    \overline{\mathbf{V}}_{\infty}^{\delta} &= \text{plim}_{I,J\rightarrow\infty} \frac{r^{2}}{I^{2}J^{2}T^{2}} \EX_{\alpha} \left( \sum_{i = 1}^{I} \sum_{t = 1}^{T} \sum_{j = 1}^{J} \sum_{r = 1}^{J} \bar{\boldsymbol{\Delta}}_{ijt} \bar{\boldsymbol{\Delta}}_{irt}^{\prime} + \sum_{j = 1}^{J} \sum_{t = 1}^{T} \sum_{i \neq p}^{I} \bar{\boldsymbol{\Delta}}_{ijt} \bar{\boldsymbol{\Delta}}_{pjt}^{\prime} \right. \\
        & \qquad \qquad \left. + \sum_{i = 1}^{I} \sum_{j = 1}^{J} \sum_{t = 1}^{T}\boldsymbol{\Gamma}_{ijt} \boldsymbol{\Gamma}_{ijt}^{\prime} \right)  \, .
\end{align*}


\subsubsubsection{Bias-corrected estimators}


The  form of the bias suggests to separately split the panel by $I$ and $J$, leading to the  following split-panel corrected estimator for the structural parameters:

\begin{align}
    \widehat{\fmbeta}_{I,J,T}^{sp} &= 3\widehat{\fmbeta}_{I,J,T}-\widehat{\fmbeta}_{I/2,J,T}-\widehat{\fmbeta}_{I,J/2,T}, \quad\text{with} \label{eq:spj_beta_2way}\\
    \widehat{\fmbeta}_{I/2,J,T} &= \frac{1}{2}\Big[\widehat{\fmbeta}_{\{i:i\leq \lceil I/2 \rceil \},J,T}+\widehat{\fmbeta}_{\{i:i \geq \lfloor I/2 + 1\rfloor\},J,T}\Big], \notag\\
    \widehat{\fmbeta}_{I,J/2,T} &= \frac{1}{2}\Big[\widehat{\fmbeta}_{I,\{j:j\leq \lceil J/2 \rceil,T\}}+\widehat{\fmbeta}_{I,\{j:j \geq \lfloor J/2 + 1 \rfloor,T\}}\Big] \; , \notag
\end{align}

where $\lfloor \cdot \rfloor$ and $\lceil \cdot \rceil$ denote the floor and ceiling functions.
To clarify the notation, the subscript ${\{i:i\leq \lceil I/2 \rceil \},J,T}$ denotes that the estimator is based on a subsample, which contains all importers and time periods, but only the first half of all exporters.\\

In order to form the appropriate analytical bias correction, we make use of the asymptotic distribution of the MLE, which we have described above. The analytical bias-corrected estimator $\tilde{\boldsymbol{\beta}}^a$ is formed from estimators of the leading bias terms that are subtracted from the MLE of the full sample $\widehat{\fmbeta}_{I,J,T}$. More precisely:

\begin{align*}
    \tilde{\boldsymbol{\beta}}_{I,J,T}^a &= \hat{\boldsymbol{\beta}}_{I, J, T} - \frac{\widehat{\mathbf{B}}_1^{\beta}}{I} - \frac{\widehat{\mathbf{B}}_2^{\beta}}{J} ,
    \quad\text{with} \quad \widehat{\mathbf{B}}_1^{\beta} = \widehat{\mathbf{W}}^{-1}  \widehat{\mathbf{B}}_1, \widehat{\mathbf{B}}_2^{\beta} =\widehat{\mathbf{W}}^{-1}  \widehat{\mathbf{B}}_2, \quad \text{and} \\
    \widehat{\mathbf{B}}_1 &= - \frac{1}{2JT} \sum_{j = 1}^{J} \sum_{t = 1}^{T} \frac{\sum_{i = 1}^{I} \widehat{H}_{ijt} \partial_{\eta^{2}} \widehat{F}_{ijt} \left(\widehat{\MX}\mathbf{X}\right)_{ijt}}{\sum_{i = 1}^{I} \hat{\omega}_{ijt}} \, , \\
    \widehat{\mathbf{B}}_2 &= - \frac{1}{2IT} \sum_{i = 1}^{I} \sum_{t = 1}^{T} \frac{\sum_{j = 1}^{J} \widehat{H}_{ijt} \partial_{\eta^{2}} \widehat{F}_{ijt} \left(\widehat{\MX}\mathbf{X}\right)_{ijt}}{\sum_{j = 1}^{J} \hat{\omega}_{ijt}} \, , \\
    \widehat{\mathbf{W}} &= \frac{1}{IJT} \sum_{i = 1}^{I} \sum_{j = 1}^{J} \sum_{t = 1}^{T} \hat{\omega}_{ijt} \left(\widehat{\MX}\mathbf{X}\right)_{ijt} \left(\widehat{\MX}\mathbf{X}\right)_{ijt}^{\prime} \, ,
\end{align*}

where $\partial_{\iota^2}g(\cdot)$ denotes the second order partial derivative of an arbitrary function $g(\cdot)$ with respect to some parameter $\iota$. The explicit expressions of  $H_{ijt}$ and $\partial_{\eta^{2}} F_{ijt}$ are reported in Table \ref{tab:expression}.\\

The split-panel jackknife estimator works similarly with APEs as with structural parameters. We simply replace in formula \eqref{eq:spj_beta_2way} the estimators for the structural parameters with estimators for the APEs.
The following analytically bias-corrected estimator for the APEs is formed based on the asymptotic distribution  presented in Appendix \ref{sec:asymptotic}:

\begin{align*}
    \tilde{\boldsymbol{\delta}}_{I,J,T}^a & = \hat{\boldsymbol{\delta}}_{I,J,T} - \frac{\widehat{\mathbf{B}}_1^{\delta}}{I} - \frac{\widehat{\mathbf{B}}_2^{\delta}}{J}, \quad \text{with}\\
    \widehat{\mathbf{B}}_1^{\delta} &= \frac{1}{2JT} \sum_{j = 1}^{J} \sum_{t = 1}^{T}  \frac{\sum_{i = 1}^{I} - \widehat{H}_{ijt} \partial_{\eta^{2}} \widehat{F}_{ijt} \left(\widehat{\PX}\widehat{\boldsymbol{\Psi}}\right)_{ijt} + \partial_{\eta^{2}} \widehat{\boldsymbol{\Delta}}_{ijt}}{\sum_{i = 1}^{I}  \hat{\omega}_{ijt}} \, ,\\
    \widehat{\mathbf{B}}_2^{\delta} &= \frac{1}{2IT} \sum_{i = 1}^{I} \sum_{t = 1}^{T}  \frac{\sum_{j = 1}^{J} - \widehat{H}_{ijt} \partial_{\eta^{2}} \widehat{F}_{ijt} \left(\widehat{\PX}\widehat{\boldsymbol{\Psi}}\right)_{ijt} + \partial_{\eta^{2}} \widehat{\boldsymbol{\Delta}}_{ijt}}{\sum_{j = 1}^{J}  \hat{\omega}_{ijt}} \, .
\end{align*}

The covariance can be estimated according to this simplified two-way fixed effects counterpart of equation \eqref{eq:estcov3way} in the main text:

\begin{align}
    \label{eq:estcov2way}
    \widehat{\mathbf{V}}^{\delta} &= 
    \frac{1}{I^{2}J^{2}T^{2}} \left( 
\left(\sum_{i = 1}^{I} \sum_{j =  1}^{J} \sum_{t = 1}^{T}  \widehat{\bar{\boldsymbol{\Delta}}}_{ijt}\right)\left( \sum_{i = 1}^{I} \sum_{j =  1}^{J} \sum_{t = 1}^{T}  \widehat{\bar{\boldsymbol{\Delta}}}_{ijt} \right)^{\prime}
        + \sum_{i = 1}^{I} \sum_{j = 1}^{J} \sum_{t = 1}^{T}\widehat{\boldsymbol{\Gamma}}_{ijt} \widehat{\boldsymbol{\Gamma}}_{ijt}^{\prime} \right)  \, .
\end{align}


\subsubsection{Three-way fixed effects}


With the inclusion of pair fixed effects, we introduce an additional bias of order $T$, since only $T$ observations are informative per pair fixed effect. Another difference that occurs in contrast to the two-way fixed effects case is that predetermined regressors lead to a violation of the strict exogeneity assumption. To deal with this issue we adapt the asymptotic bias terms $\overline{\mathbf{B}}_{\infty}$ and $\overline{\mathbf{B}}_{\infty}^{\delta}$ of \citet{Fernandez-Val2016a} to the new structure.

\clearpage
\subsubsubsection{Conjectured asymptotic distribution of $\hat{\boldsymbol{\beta}}_{I,J,T}$}

\begin{align*}
    &\sqrt{IJT}(\widehat{\fmbeta}_{I, J, T}-\boldsymbol{\beta}^0)\rightarrow_{d}\overline{\mathbf{W}}_{\infty}^{-1}\mathcal{N}(\kappa_1 \overline{\mathbf{B}}_{1, \infty}+ \kappa_2\overline{\mathbf{B}}_{2, \infty} + \kappa_3 \overline{\mathbf{B}}_{3, \infty},\overline{\mathbf{W}}_{\infty}), \quad\text{with}\\
    \overline{\mathbf{B}}_{1, \infty}&=\text{plim}_{I,J,T \rightarrow\infty}\left[-\frac{1}{2JT}\sum_{t=1}^{T}\sum_{j=1}^{J}\frac{\sum_{i=1}^{I} \EX_{\alpha}[H_{ijt}\partial_{\eta^2}F_{ijt}(\MX \mathbf{X})_{ijt}]}{\sum_{i=1}^{I}\EX_{\alpha} [\omega_{ijt}]}\right], \notag\\
    \overline{\mathbf{B}}_{2, \infty}&=\text{plim}_{I,J,T\rightarrow\infty}\left[-\frac{1}{2IT}\sum_{t=1}^{T}\sum_{i=1}^{I}\frac{\sum_{j=1}^{J} \EX_{\alpha} [H_{ijt}\partial_{\eta^2}F_{ijt}(\MX \mathbf{X})_{ijt}]}{\sum_{j=1}^{J} \EX_{\alpha} [\omega_{ijt}]}\right], \notag\\
    \overline{\mathbf{B}}_{3, \infty} &= \text{plim}_{I,J,T\rightarrow\infty} \left[-\frac{1}{2IJ} \sum_{i=1}^{I} \sum_{j=1}^{J} \left( \sum_{t=1}^{T} \EX_{\alpha} [\omega_{ijt}] \right)^{-1} \left(\sum_{t=1}^{T} \EX_{\alpha} [H_{ijt}\partial_{\eta^2}F_{ijt}(\MX \mathbf{X})_{ijt}] \right.\right. \\
        & \qquad \qquad \left.\left. + 2\sum_{\tau=t+1}^{T} \EX_{\alpha} [H_{ijt}(Y_{ijt}-F_{ijt})\omega_{ijt}(\MX \mathbf{X})_{ijt}] \right) \right], \notag\\
    \overline{\mathbf{W}}_{\infty} & = \text{plim}_{I,J,T\rightarrow\infty} \left[\frac{1}{IJT}\sum_{i=1}^{I}\sum_{j=1}^{J}\sum_{t=1}^{T} \EX_{\alpha}[\omega_{ijt}(\MX \mathbf{X})_{ijt}(\MX \mathbf{X})_{ijt}^{\prime}]\right]\; .
\end{align*}

where  $\kappa_1, \kappa_2, \kappa_3$ are constants.
The second term in the numerator of $\overline{\mathbf{B}}_{3, \infty}$ is dropped if all regressors are assumed to be strictly exogenous.


\clearpage
\subsubsubsection{Conjectured asymptotic distribution of $\hat{\boldsymbol{\delta}}_{I,J,T}$}

\begin{align*}
    &r(\hat{\boldsymbol{\delta}}_{I,J,T} - \boldsymbol{\delta} - I^{-1}\overline{\mathbf{B}}_{1, \infty}^{\delta} - J^{-1}\overline{\mathbf{B}}_{2, \infty}^{\delta}  - T^{-1}\overline{\mathbf{B}}_{3, \infty}^{\delta})  \rightarrow_{d}  \mathcal{N}(0,\overline{\mathbf{V}}_{\infty}^{\delta}) , \quad\text{with} \\
    \overline{\mathbf{B}}_{1, \infty}^{\delta}&=\text{plim}_{I,J,T\rightarrow\infty}\left[\frac{1}{2JT}\sum_{t=1}^{T}\sum_{j=1}^{J} \frac{\sum_{i = 1}^{I} - \EX_{\alpha}[ H_{ijt} \partial_{\eta^{2}} F_{ijt}] \EX_{\alpha}[\left(\PX \boldsymbol{\Psi}\right)_{ijt}] + \EX_{\alpha}[\partial_{\eta^{2}} \boldsymbol{\Delta}_{ijt}]}{\sum_{i = 1}^{I}  \EX_{\alpha}[\omega_{ijt}]}\right], \notag\\
    \overline{\mathbf{B}}_{2, \infty}^{\delta}&=\text{plim}_{I,J,T\rightarrow\infty}\left[\frac{1}{2IT}\sum_{t=1}^{T}\sum_{i=1}^{I} \frac{\sum_{j = 1}^{J} - \EX_{\alpha}[ H_{ijt} \partial_{\eta^{2}} F_{ijt}] \EX_{\alpha}[\left(\PX\boldsymbol{\Psi}\right)_{ijt}] +\EX_{\alpha}[\partial_{\eta^{2}} \boldsymbol{\Delta}_{ijt}]}{\sum_{j = 1}^{J}  \EX_{\alpha}[\omega_{ijt}]}\right], \notag\\
    \overline{\mathbf{B}}_{3, \infty}^{\delta} &=  \text{plim}_{I,J,T\rightarrow\infty} \left[ \frac{1}{2IJ} \sum_{i = 1}^{I} \sum_{j = 1}^{J} \left( \sum_{t = 1}^{T} \EX_{\alpha}[\omega_{ijt}] \right)^{-1} \left( \sum_{t = 1}^{T} - \EX_{\alpha}[H_{ijt} \partial_{\eta^{2}} F_{ijt}] \EX_{\alpha}[\left(\PX \boldsymbol{\Psi}\right)_{ijt}] \right.\right. \\
        & \qquad \qquad \left.\left. + \EX_{\alpha}[\partial_{\eta^{2}} \boldsymbol{\Delta}_{ijt}] + 2  \sum_{\tau = t + 1}^{T} \EX_{\alpha}[\partial_{\eta} \ell_{ijt-l} \omega_{ijt} \left(\MX\boldsymbol{\Psi}\right)_{ijt}]\right) \right] \, .\\
    \overline{\mathbf{V}}_{\infty}^{\delta} &= \text{plim}_{I,J,T\rightarrow\infty} \frac{r^{2}}{I^{2}J^{2}T^{2}} \EX_{\alpha} \left[ \left(\sum_{i = 1}^{I}  \sum_{j =  1}^{J} \sum_{t = 1}^{T} \bar{\boldsymbol{\Delta}}_{ijt}\right)\left( \sum_{i = 1}^{I}  \sum_{j =  1}^{J} \sum_{t = 1}^{T} \bar{\boldsymbol{\Delta}}_{ijt} \right)^{\prime} \right. \\
        & \qquad \qquad \left. + \sum_{i = 1}^{I} \sum_{j = 1}^{J} \sum_{t = 1}^{T}\boldsymbol{\Gamma}_{ijt} \boldsymbol{\Gamma}_{ijt}^{\prime} + 2 \sum_{i = 1}^{I} \sum_{j = 1}^{J} \sum_{s > t}^{T} \bar{\boldsymbol{\Delta}}_{ijt} \boldsymbol{\Gamma}_{ijs}^{\prime} \right]  \, ,\\
\boldsymbol{\Gamma}_{ijt} &= \EX_{\alpha} \left[(IJT)^{-1} \sum_{i = 1}^{I} \sum_{j = 1}^{J} \sum_{t = 1}^{T} \partial_{\beta} \boldsymbol{\Delta}_{ijt} - \left(\PX \mathbf{X}\right)_{ijt} \partial_{\eta} \boldsymbol{\Delta}_{ijt} \right]^{\prime} \overline{\mathbf{W}}_{\infty}^{- 1} \EX_{\alpha}\left[\left(\MX \mathbf{X}\right)_{ijt} \omega_{ijt} \boldsymbol{\nu}_{ijt}\right]\\
&- \EX_{\alpha}\left[\left(\PX \boldsymbol{\Psi}\right)_{ijt} \partial_{\eta} \ell_{ijt}\right] \, ,
\end{align*}
and $r$ is a convergence rate.  The last term in the numerator of $\overline{\mathbf{B}}_{3, \infty}$ and  $\overline{\mathbf{V}}_{\infty}^{\delta}$ are dropped if all regressors are assumed to be strictly exogenous. The expression $\overline{\mathbf{V}}_{\infty}^{\delta}$ can be further modified by assuming  that $\{\lambda_{it}\}_{IT}$, $\{\psi_{jt}\}_{JT}$ and $\{\mu_{ij}\}_{IJ}$ are independent sequences, and $\lambda_{it}$, $\psi_{jt}$ and $\mu_{ij}$  are independent for all $it$, $jt$, $ij$:

\begin{align*}
    \widehat{\mathbf{V}}^{\delta} &= \text{plim}_{I,J,T\rightarrow\infty} \frac{r^{2}}{I^{2}J^{2}T^{2}} \EX_{\alpha}  \left( \sum_{i = 1}^{I} \sum_{t = 1}^{T} \sum_{j = 1}^{J} \sum_{r = 1}^{J} \bar{\boldsymbol{\Delta}}_{ijt} \bar{\boldsymbol{\Delta}}_{irt}^{\prime} + \sum_{j = 1}^{J} \sum_{t = 1}^{T} \sum_{i \neq p}^{I} \bar{\boldsymbol{\Delta}}_{ijt} \bar{\boldsymbol{\Delta}}_{pjt}^{\prime} \right. \\
        & \qquad \qquad \left. + \sum_{i = 1}^{I} \sum_{j = 1}^{J} \sum_{s \neq t}^{T} \bar{\boldsymbol{\Delta}}_{ijt} \bar{\boldsymbol{\Delta}}_{ijs}^{\prime} + \sum_{i = 1}^{I} \sum_{j = 1}^{J} \sum_{t = 1}^{T}\boldsymbol{\Gamma}_{ijt} \boldsymbol{\Gamma}_{ijt}^{\prime} + 2  \sum_{i = 1}^{I} \sum_{j = 1}^{J} \sum_{s > t}^{T} \bar{\boldsymbol{\Delta}}_{ijt} \boldsymbol{\Gamma}_{ijs}^{\prime} \right)  \, ,
\end{align*}

\newpage
\section{Monte Carlo Results}

\subsection{Three-way Fixed Effects: Dynamic}\label{app:monte_carlo_dynamic_threeway}

This subsection provides the detailed results corresponding to the graphical representation and verbal discussion in Section \ref{sec:simulation} of the main text.


\begin{table}[!ht]
	\begin{minipage}{.5\linewidth}
		\centering
		\caption{\small Dynamic: Three-way FEs --- $z$, $N = 50$}
		\label{tab:dyn3way_xn50}
		\resizebox{0.95\linewidth}{!}{
			\begin{tabular}{lrrrrrrrr}
				\toprule
&\multicolumn{4}{c}{Coefficients}&\multicolumn{4}{c}{APE}\\
\cmidrule(lr){2-5}\cmidrule(lr){6-9}
&Bias&Bias/SE&SE/SD&CP .95&Bias&Bias/SE&SE/SD&CP .95\\
\midrule
&\multicolumn{8}{c}{N = 50; T = 10}\\
\cmidrule(lr){2-9}
MLE & 29.12 & 13.86 & 0.83 & 0.00 & 4.11 & 2.66 & 0.98 & 0.25 \\ 
ABC (1) & -0.44 & -0.24 & 1.04 & 0.95 & -0.84 & -0.53 & 1.04 & 0.93 \\ 
ABC (2) & -0.60 & -0.33 & 1.02 & 0.94 & -0.97 & -0.61 & 1.02 & 0.92 \\ 
SPJ & -14.13 & -8.11 & 0.61 & 0.00 & 4.50 & 2.79 & 0.78 & 0.26 \\ 
\midrule
&\multicolumn{8}{c}{N = 50; T = 20}\\
\cmidrule(lr){2-9}
MLE & 16.07 & 12.44 & 0.88 & 0.00 & 2.57 & 2.47 & 0.92 & 0.31 \\ 
ABC (1) & -0.12 & -0.10 & 0.99 & 0.95 & -0.15 & -0.14 & 0.95 & 0.93 \\ 
ABC (2) & -0.27 & -0.23 & 0.99 & 0.94 & -0.26 & -0.24 & 0.94 & 0.93 \\ 
SPJ & -4.78 & -4.07 & 0.81 & 0.04 & 0.58 & 0.55 & 0.84 & 0.87 \\ 
\midrule
&\multicolumn{8}{c}{N = 50; T = 30}\\
\cmidrule(lr){2-9}
MLE & 12.24 & 12.11 & 0.89 & 0.00 & 1.84 & 2.14 & 0.98 & 0.43 \\ 
ABC (1) & -0.10 & -0.11 & 0.98 & 0.94 & -0.06 & -0.07 & 0.99 & 0.95 \\ 
ABC (2) & -0.23 & -0.25 & 0.98 & 0.93 & -0.16 & -0.19 & 0.99 & 0.94 \\ 
SPJ & -2.86 & -3.03 & 0.90 & 0.16 & 0.04 & 0.05 & 0.94 & 0.94 \\ 
\midrule
&\multicolumn{8}{c}{N = 50; T = 40}\\
\cmidrule(lr){2-9}
MLE & 10.36 & 12.08 & 0.94 & 0.00 & 1.41 & 1.86 & 1.00 & 0.54 \\ 
ABC (1) & -0.14 & -0.18 & 1.02 & 0.95 & -0.06 & -0.08 & 1.01 & 0.95 \\ 
ABC (2) & -0.25 & -0.31 & 1.02 & 0.94 & -0.16 & -0.21 & 1.01 & 0.94 \\ 
SPJ & -2.14 & -2.65 & 0.91 & 0.26 & -0.13 & -0.17 & 0.94 & 0.93 \\ 
\midrule
&\multicolumn{8}{c}{N = 50; T = 50}\\
\cmidrule(lr){2-9}
MLE & 9.28 & 12.24 & 0.95 & 0.00 & 1.15 & 1.67 & 1.01 & 0.61 \\ 
ABC (1) & -0.16 & -0.22 & 1.02 & 0.95 & -0.06 & -0.09 & 1.01 & 0.95 \\ 
ABC (2) & -0.25 & -0.34 & 1.02 & 0.94 & -0.14 & -0.21 & 1.01 & 0.94 \\ 
SPJ & -1.78 & -2.48 & 0.92 & 0.31 & -0.18 & -0.26 & 0.95 & 0.94 \\ 
				\bottomrule
			\end{tabular}
		}
	\end{minipage}
	\begin{minipage}{.5\linewidth}
		\centering
		\caption{\small Dynamic: Three-way FEs --- $z$, $N = 100$}
		\label{tab:dyn3way_xn100}
		\resizebox{0.95\linewidth}{!}{
			\begin{tabular}{lrrrrrrrr}
				\toprule
&\multicolumn{4}{c}{Coefficients}&\multicolumn{4}{c}{APE}\\
\cmidrule(lr){2-5}\cmidrule(lr){6-9}
&Bias&Bias/SE&SE/SD&CP .95&Bias&Bias/SE&SE/SD&CP .95\\
\midrule
&\multicolumn{8}{c}{N = 100; T = 10}\\
\cmidrule(lr){2-9}
MLE & 24.51 & 24.40 & 0.83 & 0.00 & 3.59 & 4.33 & 0.99 & 0.01 \\ 
ABC (1) & 0.39 & 0.43 & 1.00 & 0.93 & -0.45 & -0.54 & 1.02 & 0.91 \\ 
ABC (2) & 0.21 & 0.24 & 0.99 & 0.95 & -0.58 & -0.70 & 1.01 & 0.89 \\ 
SPJ & -8.49 & -9.71 & 0.67 & 0.00 & 5.57 & 6.74 & 0.83 & 0.00 \\ 
\midrule
&\multicolumn{8}{c}{N = 100; T = 20}\\
\cmidrule(lr){2-9}
MLE & 12.55 & 20.17 & 0.88 & 0.00 & 2.32 & 4.29 & 0.98 & 0.01 \\ 
ABC (1) & 0.21 & 0.35 & 0.97 & 0.93 & 0.03 & 0.06 & 0.99 & 0.95 \\ 
ABC (2) & 0.06 & 0.10 & 0.97 & 0.94 & -0.07 & -0.14 & 0.99 & 0.96 \\ 
SPJ & -2.72 & -4.68 & 0.88 & 0.01 & 1.05 & 1.94 & 0.92 & 0.50 \\ 
\midrule
&\multicolumn{8}{c}{N = 100; T = 30}\\
\cmidrule(lr){2-9}
MLE & 8.96 & 18.38 & 0.92 & 0.00 & 1.62 & 3.64 & 0.95 & 0.06 \\ 
ABC (1) & 0.09 & 0.19 & 0.98 & 0.94 & 0.04 & 0.10 & 0.95 & 0.94 \\ 
ABC (2) & -0.04 & -0.08 & 0.98 & 0.95 & -0.06 & -0.14 & 0.95 & 0.94 \\ 
SPJ & -1.48 & -3.19 & 0.91 & 0.14 & 0.31 & 0.70 & 0.91 & 0.87 \\ 
\midrule
&\multicolumn{8}{c}{N = 100; T = 40}\\
\cmidrule(lr){2-9}
MLE & 7.27 & 17.56 & 0.92 & 0.00 & 1.25 & 3.21 & 0.98 & 0.12 \\ 
ABC (1) & 0.05 & 0.13 & 0.97 & 0.94 & 0.05 & 0.12 & 0.98 & 0.95 \\ 
ABC (2) & -0.05 & -0.13 & 0.97 & 0.93 & -0.05 & -0.12 & 0.98 & 0.94 \\ 
SPJ & -0.97 & -2.44 & 0.92 & 0.32 & 0.12 & 0.32 & 0.95 & 0.92 \\ 
\midrule
&\multicolumn{8}{c}{N = 100; T = 50}\\
\cmidrule(lr){2-9}
MLE & 6.30 & 17.19 & 0.90 & 0.00 & 1.03 & 2.90 & 0.96 & 0.20 \\ 
ABC (1) & 0.05 & 0.13 & 0.95 & 0.92 & 0.05 & 0.14 & 0.96 & 0.94 \\ 
ABC (2) & -0.04 & -0.13 & 0.95 & 0.93 & -0.03 & -0.09 & 0.96 & 0.94 \\ 
SPJ & -0.71 & -2.00 & 0.91 & 0.49 & 0.06 & 0.16 & 0.93 & 0.93 \\ 
\bottomrule
			\end{tabular}
		}
	\end{minipage}
	\begin{minipage}{.25\linewidth}~\end{minipage}
	\begin{minipage}{.5\linewidth}
		\vspace{1em}
		\centering
		\caption{\small Dynamic: Three-way FEs --- $z$, $N = 150$}
		\label{tab:dyn3way_xn150}
		\resizebox{0.95\linewidth}{!}{
			\begin{tabular}{lrrrrrrrr}
				\toprule
&\multicolumn{4}{c}{Coefficients}&\multicolumn{4}{c}{APE}\\
\cmidrule(lr){2-5}\cmidrule(lr){6-9}
&Bias&Bias/SE&SE/SD&CP .95&Bias&Bias/SE&SE/SD&CP .95\\
\midrule
&\multicolumn{8}{c}{N = 150; T = 10}\\
\cmidrule(lr){2-9}
MLE & 23.14 & 35.04 & 0.83 & 0.00 & 3.45 & 5.84 & 0.94 & 0.00 \\ 
ABC (1) & 0.65 & 1.08 & 0.97 & 0.80 & -0.35 & -0.60 & 0.95 & 0.90 \\ 
ABC (2) & 0.47 & 0.79 & 0.96 & 0.87 & -0.47 & -0.82 & 0.94 & 0.87 \\ 
SPJ & -7.12 & -12.24 & 0.67 & 0.00 & 5.81 & 10.11 & 0.78 & 0.00 \\ 
\midrule
&\multicolumn{8}{c}{N = 150; T = 20}\\
\cmidrule(lr){2-9}
MLE & 11.45 & 27.94 & 0.92 & 0.00 & 2.24 & 6.00 & 0.94 & 0.00 \\ 
ABC (1) & 0.29 & 0.75 & 1.00 & 0.89 & 0.07 & 0.19 & 0.94 & 0.94 \\ 
ABC (2) & 0.15 & 0.38 & 1.00 & 0.94 & -0.03 & -0.08 & 0.94 & 0.93 \\ 
SPJ & -2.21 & -5.73 & 0.89 & 0.00 & 1.17 & 3.14 & 0.88 & 0.15 \\ 
\midrule
&\multicolumn{8}{c}{N = 150; T = 30}\\
\cmidrule(lr){2-9}
MLE & 7.96 & 24.76 & 0.92 & 0.00 & 1.58 & 5.14 & 0.91 & 0.00 \\ 
ABC (1) & 0.15 & 0.49 & 0.97 & 0.92 & 0.08 & 0.26 & 0.91 & 0.93 \\ 
ABC (2) & 0.03 & 0.09 & 0.97 & 0.94 & -0.02 & -0.07 & 0.91 & 0.94 \\ 
SPJ & -1.16 & -3.74 & 0.89 & 0.05 & 0.40 & 1.31 & 0.87 & 0.71 \\ 
\midrule
&\multicolumn{8}{c}{N = 150; T = 40}\\
\cmidrule(lr){2-9}
MLE & 6.34 & 23.20 & 0.94 & 0.00 & 1.24 & 4.60 & 0.96 & 0.01 \\ 
ABC (1) & 0.13 & 0.49 & 0.99 & 0.92 & 0.10 & 0.37 & 0.96 & 0.92 \\ 
ABC (2) & 0.02 & 0.09 & 0.99 & 0.95 & 0.01 & 0.03 & 0.96 & 0.94 \\ 
SPJ & -0.70 & -2.66 & 0.94 & 0.25 & 0.20 & 0.76 & 0.93 & 0.86 \\ 
\midrule
&\multicolumn{8}{c}{N = 150; T = 50}\\
\cmidrule(lr){2-9}
MLE & 5.35 & 22.13 & 0.96 & 0.00 & 1.00 & 4.08 & 0.92 & 0.03 \\ 
ABC (1) & 0.08 & 0.35 & 1.00 & 0.93 & 0.07 & 0.31 & 0.92 & 0.92 \\ 
ABC (2) & -0.01 & -0.04 & 1.00 & 0.94 & -0.01 & -0.03 & 0.92 & 0.93 \\ 
SPJ & -0.51 & -2.19 & 0.94 & 0.42 & 0.10 & 0.42 & 0.90 & 0.90 \\ 
\bottomrule
			\end{tabular}
		}
	\end{minipage}
\end{table}

\begin{table}[!ht]
	\begin{minipage}{.5\linewidth}
		\centering
		\caption{\small Dynamic: Three-way FEs --- $y_{t-1}$, $N = 50$}
		\label{tab:dyn3way_yn50}
		\resizebox{.95\linewidth}{!}{
			\begin{tabular}{lrrrrrrrr}
  \toprule
&\multicolumn{4}{c}{Coefficients}&\multicolumn{4}{c}{APE}\\
\cmidrule(lr){2-5}\cmidrule(lr){6-9}
&Bias&Bias/SE&SE/SD&CP .95&Bias&Bias/SE&SE/SD&CP .95\\
\midrule
&\multicolumn{8}{c}{N = 50; T = 10}\\
\cmidrule(lr){2-9}
MLE & -61.77 & -11.92 & 0.96 & 0.00 & -70.30 & -17.25 & 0.94 & 0.00 \\ 
ABC (1) & -5.58 & -1.12 & 1.15 & 0.85 & -6.57 & -1.40 & 1.02 & 0.71 \\ 
ABC (2) & -7.22 & -1.45 & 1.06 & 0.70 & -8.24 & -1.76 & 0.94 & 0.56 \\ 
SPJ & 24.96 & 5.08 & 0.78 & 0.01 & -10.71 & -2.14 & 0.87 & 0.43 \\ 
\midrule
&\multicolumn{8}{c}{N = 50; T = 20}\\
\cmidrule(lr){2-9}
MLE & -26.96 & -7.92 & 0.96 & 0.00 & -36.61 & -11.97 & 0.96 & 0.00 \\ 
ABC (1) & -3.17 & -0.95 & 1.07 & 0.86 & -3.34 & -1.02 & 1.01 & 0.83 \\ 
ABC (2) & -1.03 & -0.31 & 1.03 & 0.95 & -1.07 & -0.32 & 0.96 & 0.93 \\ 
SPJ & 4.66 & 1.41 & 0.90 & 0.71 & -1.85 & -0.55 & 0.88 & 0.88 \\ 
\midrule
&\multicolumn{8}{c}{N = 50; T = 30}\\
\cmidrule(lr){2-9}
MLE & -15.77 & -5.79 & 0.97 & 0.00 & -24.44 & -9.60 & 0.98 & 0.00 \\ 
ABC (1) & -2.11 & -0.79 & 1.06 & 0.89 & -2.14 & -0.80 & 1.02 & 0.88 \\ 
ABC (2) & -0.12 & -0.04 & 1.03 & 0.96 & -0.03 & -0.01 & 0.99 & 0.95 \\ 
SPJ & 1.89 & 0.71 & 0.95 & 0.88 & -0.45 & -0.17 & 0.92 & 0.93 \\ 
\midrule
&\multicolumn{8}{c}{N = 50; T = 40}\\
\cmidrule(lr){2-9}
MLE & -10.46 & -4.48 & 0.97 & 0.00 & -18.41 & -8.27 & 0.98 & 0.00 \\ 
ABC (1) & -1.72 & -0.75 & 1.04 & 0.89 & -1.70 & -0.73 & 1.02 & 0.90 \\ 
ABC (2) & -0.02 & -0.01 & 1.02 & 0.96 & 0.08 & 0.04 & 1.00 & 0.94 \\ 
SPJ & 0.89 & 0.39 & 0.96 & 0.92 & -0.18 & -0.08 & 0.95 & 0.94 \\ 
\midrule
&\multicolumn{8}{c}{N = 50; T = 50}\\
\cmidrule(lr){2-9}
MLE & -7.39 & -3.56 & 0.94 & 0.07 & -14.86 & -7.41 & 0.94 & 0.00 \\ 
ABC (1) & -1.54 & -0.75 & 1.01 & 0.88 & -1.51 & -0.73 & 0.97 & 0.87 \\ 
ABC (2) & -0.10 & -0.05 & 0.99 & 0.94 & 0.00 & 0.00 & 0.95 & 0.93 \\ 
SPJ & 0.27 & 0.13 & 0.94 & 0.93 & -0.23 & -0.11 & 0.91 & 0.92 \\ 
\bottomrule
			\end{tabular}
		}
	\end{minipage}
	\begin{minipage}{.5\linewidth}
		\centering
		\caption{\small Dynamic: Three-way FEs --- $y_{t-1}$, $N = 100$}
		\label{tab:dyn3way_yn100}
		\resizebox{.95\linewidth}{!}{
			\begin{tabular}{lrrrrrrrr}
  \toprule
&\multicolumn{4}{c}{Coefficients}&\multicolumn{4}{c}{APE}\\
\cmidrule(lr){2-5}\cmidrule(lr){6-9}
&Bias&Bias/SE&SE/SD&CP .95&Bias&Bias/SE&SE/SD&CP .95\\
\midrule
&\multicolumn{8}{c}{N = 100; T = 10}\\
\cmidrule(lr){2-9}
MLE & -62.99 & -25.11 & 0.93 & 0.00 & -70.36 & -34.54 & 0.91 & 0.00 \\ 
ABC (1) & -6.44 & -2.65 & 1.08 & 0.22 & -7.87 & -3.39 & 0.97 & 0.09 \\ 
ABC (2) & -7.99 & -3.29 & 0.98 & 0.09 & -9.44 & -4.07 & 0.88 & 0.04 \\ 
SPJ & 21.28 & 8.82 & 0.76 & 0.00 & -11.06 & -4.51 & 0.82 & 0.02 \\ 
\midrule
&\multicolumn{8}{c}{N = 100; T = 20}\\
\cmidrule(lr){2-9}
MLE & -29.01 & -17.53 & 0.98 & 0.00 & -36.61 & -23.90 & 0.98 & 0.00 \\ 
ABC (1) & -3.49 & -2.14 & 1.06 & 0.42 & -3.81 & -2.33 & 1.02 & 0.36 \\ 
ABC (2) & -1.37 & -0.84 & 1.02 & 0.87 & -1.57 & -0.96 & 0.97 & 0.82 \\ 
SPJ & 4.24 & 2.61 & 0.91 & 0.28 & -1.78 & -1.07 & 0.90 & 0.79 \\ 
\midrule
&\multicolumn{8}{c}{N = 100; T = 30}\\
\cmidrule(lr){2-9}
MLE & -18.28 & -13.79 & 0.95 & 0.00 & -24.65 & -19.30 & 0.94 & 0.00 \\ 
ABC (1) & -2.48 & -1.89 & 1.01 & 0.54 & -2.61 & -1.95 & 0.96 & 0.52 \\ 
ABC (2) & -0.51 & -0.39 & 0.98 & 0.92 & -0.54 & -0.41 & 0.94 & 0.91 \\ 
SPJ & 1.83 & 1.40 & 0.93 & 0.70 & -0.49 & -0.36 & 0.90 & 0.90 \\ 
\midrule
&\multicolumn{8}{c}{N = 100; T = 40}\\
\cmidrule(lr){2-9}
MLE & -13.00 & -11.44 & 0.98 & 0.00 & -18.58 & -16.62 & 0.96 & 0.00 \\ 
ABC (1) & -1.94 & -1.73 & 1.03 & 0.57 & -2.02 & -1.74 & 0.98 & 0.58 \\ 
ABC (2) & -0.28 & -0.25 & 1.01 & 0.95 & -0.27 & -0.24 & 0.96 & 0.94 \\ 
SPJ & 1.01 & 0.90 & 0.96 & 0.84 & -0.19 & -0.17 & 0.92 & 0.92 \\ 
\midrule
&\multicolumn{8}{c}{N = 100; T = 50}\\
\cmidrule(lr){2-9}
MLE & -9.85 & -9.73 & 0.99 & 0.00 & -14.89 & -14.80 & 0.98 & 0.00 \\ 
ABC (1) & -1.59 & -1.58 & 1.03 & 0.66 & -1.64 & -1.58 & 1.00 & 0.66 \\ 
ABC (2) & -0.17 & -0.17 & 1.02 & 0.94 & -0.16 & -0.15 & 0.98 & 0.93 \\ 
SPJ & 0.63 & 0.63 & 0.97 & 0.90 & -0.10 & -0.09 & 0.93 & 0.93 \\ 
\bottomrule
			\end{tabular}
		}
	\end{minipage}
	\begin{minipage}{.25\linewidth}~\end{minipage}
	\begin{minipage}{.5\linewidth}
		\vspace{1em}
		\centering
		\caption{\small Dynamic: Three-way FEs --- $y_{t-1}$, $N = 150$}
		\label{tab:dyn3way_yn150}
		\resizebox{.95\linewidth}{!}{
			\begin{tabular}{lrrrrrrrr}
  \toprule
&\multicolumn{4}{c}{Coefficients}&\multicolumn{4}{c}{APE}\\
\cmidrule(lr){2-5}\cmidrule(lr){6-9}
&Bias&Bias/SE&SE/SD&CP .95&Bias&Bias/SE&SE/SD&CP .95\\
\midrule
&\multicolumn{8}{c}{N = 150; T = 10}\\
\cmidrule(lr){2-9}
MLE & -63.56 & -38.38 & 0.95 & 0.00 & -70.53 & -51.93 & 0.93 & 0.00 \\ 
ABC (1) & -6.90 & -4.30 & 1.10 & 0.01 & -8.49 & -5.48 & 0.99 & 0.00 \\ 
ABC (2) & -8.44 & -5.26 & 1.00 & 0.00 & -10.05 & -6.49 & 0.91 & 0.00 \\ 
SPJ & 20.03 & 12.55 & 0.78 & 0.00 & -11.35 & -6.93 & 0.84 & 0.00 \\ 
\midrule
&\multicolumn{8}{c}{N = 150; T = 20}\\
\cmidrule(lr){2-9}
MLE & -29.71 & -27.18 & 0.98 & 0.00 & -36.66 & -35.82 & 0.95 & 0.00 \\ 
ABC (1) & -3.65 & -3.39 & 1.05 & 0.08 & -4.03 & -3.69 & 0.98 & 0.05 \\ 
ABC (2) & -1.55 & -1.43 & 1.01 & 0.71 & -1.80 & -1.65 & 0.94 & 0.61 \\ 
SPJ & 4.07 & 3.78 & 0.93 & 0.06 & -1.75 & -1.58 & 0.89 & 0.62 \\ 
\midrule
&\multicolumn{8}{c}{N = 150; T = 30}\\
\cmidrule(lr){2-9}
MLE & -19.05 & -21.76 & 0.97 & 0.00 & -24.71 & -28.94 & 0.96 & 0.00 \\ 
ABC (1) & -2.60 & -3.00 & 1.03 & 0.15 & -2.77 & -3.11 & 0.98 & 0.14 \\ 
ABC (2) & -0.65 & -0.75 & 1.00 & 0.88 & -0.72 & -0.80 & 0.96 & 0.85 \\ 
SPJ & 1.76 & 2.04 & 0.94 & 0.48 & -0.47 & -0.52 & 0.91 & 0.90 \\ 
\midrule
&\multicolumn{8}{c}{N = 150; T = 40}\\
\cmidrule(lr){2-9}
MLE & -13.81 & -18.37 & 0.96 & 0.00 & -18.63 & -24.92 & 0.95 & 0.00 \\ 
ABC (1) & -2.04 & -2.73 & 1.00 & 0.22 & -2.14 & -2.77 & 0.97 & 0.22 \\ 
ABC (2) & -0.39 & -0.52 & 0.98 & 0.91 & -0.41 & -0.52 & 0.95 & 0.90 \\ 
SPJ & 0.98 & 1.31 & 0.94 & 0.72 & -0.19 & -0.25 & 0.91 & 0.92 \\ 
\midrule
&\multicolumn{8}{c}{N = 150; T = 50}\\
\cmidrule(lr){2-9}
MLE & -10.71 & -16.01 & 0.93 & 0.00 & -14.97 & -22.23 & 0.93 & 0.00 \\ 
ABC (1) & -1.70 & -2.55 & 0.97 & 0.29 & -1.77 & -2.55 & 0.95 & 0.28 \\ 
ABC (2) & -0.29 & -0.44 & 0.95 & 0.91 & -0.29 & -0.42 & 0.93 & 0.90 \\ 
SPJ & 0.57 & 0.85 & 0.91 & 0.84 & -0.15 & -0.21 & 0.89 & 0.92 \\ 
\bottomrule
			\end{tabular}
		}
	\end{minipage}
\end{table}

\begin{table}[!ht]
	\begin{minipage}{.5\linewidth}
		\centering
		\caption{\small Dynamic: Three-way FEs --- $z$, $N = 50$ long-run}
		\label{tab:dyn3waylong_xn50}
		\resizebox{0.6\linewidth}{!}{
			\begin{tabular}{lrrrr}
 \toprule
 &\multicolumn{4}{c}{APE}\\
 \cmidrule(lr){2-5}
 &Bias&Bias/SE&SE/SD&CP .95\\
 \midrule
 &\multicolumn{4}{c}{N = 50; T = 10}\\
 \cmidrule(lr){2-5}
 MLE & -4.15 & -2.88 & 0.97 & 0.19 \\ 
 ABC (1) & -1.31 & -0.85 & 1.02 & 0.87 \\ 
 ABC (2) & -1.66 & -1.08 & 1.00 & 0.80 \\ 
 SPJ & 3.24 & 1.99 & 0.81 & 0.49 \\ 
 \midrule
 &\multicolumn{4}{c}{N = 50; T = 20}\\
 \cmidrule(lr){2-5}
 MLE & -1.90 & -1.91 & 0.93 & 0.53 \\ 
 ABC (1) & -0.47 & -0.45 & 0.94 & 0.92 \\ 
 ABC (2) & -0.34 & -0.33 & 0.94 & 0.92 \\ 
 SPJ & 0.25 & 0.24 & 0.84 & 0.89 \\ 
 \midrule
 &\multicolumn{4}{c}{N = 50; T = 30}\\
 \cmidrule(lr){2-5}
 MLE & -1.26 & -1.51 & 0.99 & 0.67 \\ 
 ABC (1) & -0.27 & -0.32 & 1.00 & 0.94 \\ 
 ABC (2) & -0.15 & -0.18 & 0.99 & 0.95 \\ 
 SPJ & -0.11 & -0.13 & 0.94 & 0.94 \\ 
 \midrule
 &\multicolumn{4}{c}{N = 50; T = 40}\\
 \cmidrule(lr){2-5}
 MLE & -1.01 & -1.39 & 1.02 & 0.72 \\ 
 ABC (1) & -0.24 & -0.32 & 1.01 & 0.95 \\ 
 ABC (2) & -0.15 & -0.20 & 1.01 & 0.95 \\ 
 SPJ & -0.24 & -0.33 & 0.94 & 0.92 \\ 
 \midrule
 &\multicolumn{4}{c}{N = 50; T = 50}\\
 \cmidrule(lr){2-5}
 MLE & -0.88 & -1.32 & 1.04 & 0.74 \\ 
 ABC (1) & -0.23 & -0.34 & 1.04 & 0.95 \\ 
 ABC (2) & -0.15 & -0.23 & 1.04 & 0.95 \\ 
 SPJ & -0.29 & -0.43 & 0.96 & 0.92 \\ 
 \bottomrule
			\end{tabular}
		}
	\end{minipage}
	\begin{minipage}{.5\linewidth}
		\centering
		\caption{\small Dynamic: Three-way FEs --- $z$, $N = 100$ long-run}
		\label{tab:dyn3waylong_xn100}
		\resizebox{0.6\linewidth}{!}{
			\begin{tabular}{lrrrr}
		 \toprule
		&\multicolumn{4}{c}{APE}\\
		\cmidrule(lr){2-5}
		&Bias&Bias/SE&SE/SD&CP .95\\
		\midrule
		&\multicolumn{4}{c}{N = 100; T = 10}\\
		\cmidrule(lr){2-5}
		MLE & -4.62 & -5.98 & 0.99 & 0.00 \\ 
		ABC (1) & -1.06 & -1.30 & 1.01 & 0.76 \\ 
		ABC (2) & -1.40 & -1.72 & 1.00 & 0.59 \\ 
		SPJ & 4.14 & 4.93 & 0.86 & 0.01 \\ 
		\midrule
		&\multicolumn{4}{c}{N = 100; T = 20}\\
		\cmidrule(lr){2-5}
		MLE & -2.06 & -3.96 & 0.97 & 0.02 \\ 
		ABC (1) & -0.32 & -0.61 & 0.98 & 0.90 \\ 
		ABC (2) & -0.19 & -0.36 & 0.98 & 0.93 \\ 
		SPJ & 0.78 & 1.44 & 0.90 & 0.67 \\ 
		\midrule
		&\multicolumn{4}{c}{N = 100; T = 30}\\
		\cmidrule(lr){2-5}
		MLE & -1.38 & -3.19 & 0.96 & 0.11 \\ 
		ABC (1) & -0.21 & -0.48 & 0.96 & 0.91 \\ 
		ABC (2) & -0.10 & -0.22 & 0.96 & 0.93 \\ 
		SPJ & 0.21 & 0.48 & 0.91 & 0.91 \\ 
		\midrule
		&\multicolumn{4}{c}{N = 100; T = 40}\\
		\cmidrule(lr){2-5}
		MLE & -1.06 & -2.77 & 0.98 & 0.22 \\ 
		ABC (1) & -0.16 & -0.40 & 0.98 & 0.92 \\ 
		ABC (2) & -0.06 & -0.17 & 0.98 & 0.95 \\ 
		SPJ & 0.07 & 0.17 & 0.94 & 0.93 \\ 
		\midrule
		&\multicolumn{4}{c}{N = 100; T = 50}\\
		\cmidrule(lr){2-5}
		MLE & -0.86 & -2.48 & 0.99 & 0.30 \\ 
		ABC (1) & -0.12 & -0.34 & 0.99 & 0.94 \\ 
		ABC (2) & -0.04 & -0.12 & 0.99 & 0.96 \\ 
		SPJ & 0.01 & 0.03 & 0.95 & 0.94 \\ 
		\bottomrule
			\end{tabular}
		}
	\end{minipage}
	\begin{minipage}{.25\linewidth}~\end{minipage}
	\begin{minipage}{.5\linewidth}
		\vspace{1em}
		\centering
		\caption{\small Dynamic: Three-way FEs --- $z$, $N = 150$ long-run}
		\label{tab:dyn3waylong_xn150}
		\resizebox{0.6\linewidth}{!}{
			\begin{tabular}{lrrrr}
\toprule
&\multicolumn{4}{c}{APE}\\
\cmidrule(lr){2-5}
&Bias&Bias/SE&SE/SD&CP .95\\
\midrule
	&\multicolumn{4}{c}{N = 150; T = 10}\\
	\cmidrule(lr){2-5}
	MLE & -4.76 & -8.63 & 0.91 & 0.00 \\ 
	ABC (1) & -1.03 & -1.80 & 0.92 & 0.55 \\ 
	ABC (2) & -1.37 & -2.39 & 0.90 & 0.34 \\ 
	SPJ & 4.33 & 7.35 & 0.80 & 0.00 \\ 
	\midrule
	&\multicolumn{4}{c}{N = 150; T = 20}\\
	\cmidrule(lr){2-5}
	MLE & -2.10 & -5.79 & 0.93 & 0.00 \\ 
	ABC (1) & -0.31 & -0.83 & 0.94 & 0.86 \\ 
	ABC (2) & -0.18 & -0.48 & 0.93 & 0.90 \\ 
	SPJ & 0.91 & 2.43 & 0.89 & 0.33 \\ 
	\midrule
	&\multicolumn{4}{c}{N = 150; T = 30}\\
	\cmidrule(lr){2-5}
	MLE & -1.39 & -4.60 & 0.91 & 0.01 \\ 
	ABC (1) & -0.19 & -0.63 & 0.92 & 0.88 \\ 
	ABC (2) & -0.08 & -0.26 & 0.92 & 0.92 \\ 
	SPJ & 0.31 & 1.02 & 0.87 & 0.78 \\ 
	\midrule
	&\multicolumn{4}{c}{N = 150; T = 40}\\
	\cmidrule(lr){2-5}
	MLE & -1.03 & -3.86 & 0.97 & 0.04 \\ 
	ABC (1) & -0.12 & -0.43 & 0.97 & 0.92 \\ 
	ABC (2) & -0.03 & -0.10 & 0.97 & 0.94 \\ 
	SPJ & 0.16 & 0.60 & 0.93 & 0.89 \\ 
	\midrule
	&\multicolumn{4}{c}{N = 150; T = 50}\\
	\cmidrule(lr){2-5}
	MLE & -0.85 & -3.52 & 0.93 & 0.07 \\ 
	ABC (1) & -0.11 & -0.43 & 0.93 & 0.91 \\ 
	ABC (2) & -0.03 & -0.13 & 0.93 & 0.93 \\ 
	SPJ & 0.07 & 0.27 & 0.90 & 0.91 \\ 
	\bottomrule
			\end{tabular}
		}
	\end{minipage}
\end{table}

\FloatBarrier



\subsection{Two-way fixed effects}\label{app:monte_carlo_dynamic_twoway}

The simulations in this section correspond to a theory-consistent estimation of the extensive margin outlined in Section \ref{sec:theory} of the main text, taking into account unobserved time-varying exporter- and importer-specific terms as well as dynamics, but not allowing for bilateral unobserved heterogeneity. Specifically, we generate data according to

\begin{align*}
    y_{ijt} &= \ind\left[\beta_{y} y_{ijt-1} + \beta_{z} z_{ijt} +  \psi_{jt} + \lambda_{it} +  \geq \epsilon_{ijt} \right] \, , \\
    y_{ij0} &= \ind\left[\beta_{z} z_{ij0}  + \psi_{j0}  + \lambda_{i0} +  \geq \epsilon_{ij0} \right] \, ,
\end{align*}

where  $i = 1, \dots, N$, $j = 1, \dots, N$, $t=1, \dots, T $,  $\psi_{jt} \sim \iid \N(0, 1 / 16)$, $\lambda_{it} \sim \iid \N(0, 1 / 16)$, and $\epsilon_{ijt} \sim \iid \N(0, 1)$.\footnote{Since  $\{\lambda_{it}\}_{IT}$ and $\{\psi_{jt}\}_{JT}$ are  independent sequences, and $\lambda_{it}$ and $\psi_{jt}$ are independent for all $it$,  $jt$, we follow \citet{Fernandez-Val2016a} and incorporate this information in the covariance estimator for the APEs. The explicit expression is provided in the Appendix \ref{sec:asymptotic}.} Further, $\beta_{y} = 0.5$, $\beta_{z} = 1$, and $z_{ijt} = 0.5z_{ijt-1} + \psi_{jt} + \lambda_{it} +  \nu_{ijt}$, where $\nu_{ijt} \sim \iid \N(0, 0.5)$, $z_{ij0} \sim \iid \N(0, 1)$. To get an impression of how the different statistics evolve with changing panel dimensions, we consider  all possible combinations of $N \in \{50, 100, 150\}$ and $T \in \{10, 20, 30, 40 , 50\}$. For each of these combinations we generate $1,000$ samples.\\


\begin{figure}[p]
	\centering
	\caption{Dynamic: Two-way Fixed Effects --- Predetermined Regressor}
	\label{fig:dyn2way_y}
	\includegraphics[height = 0.45\textheight]{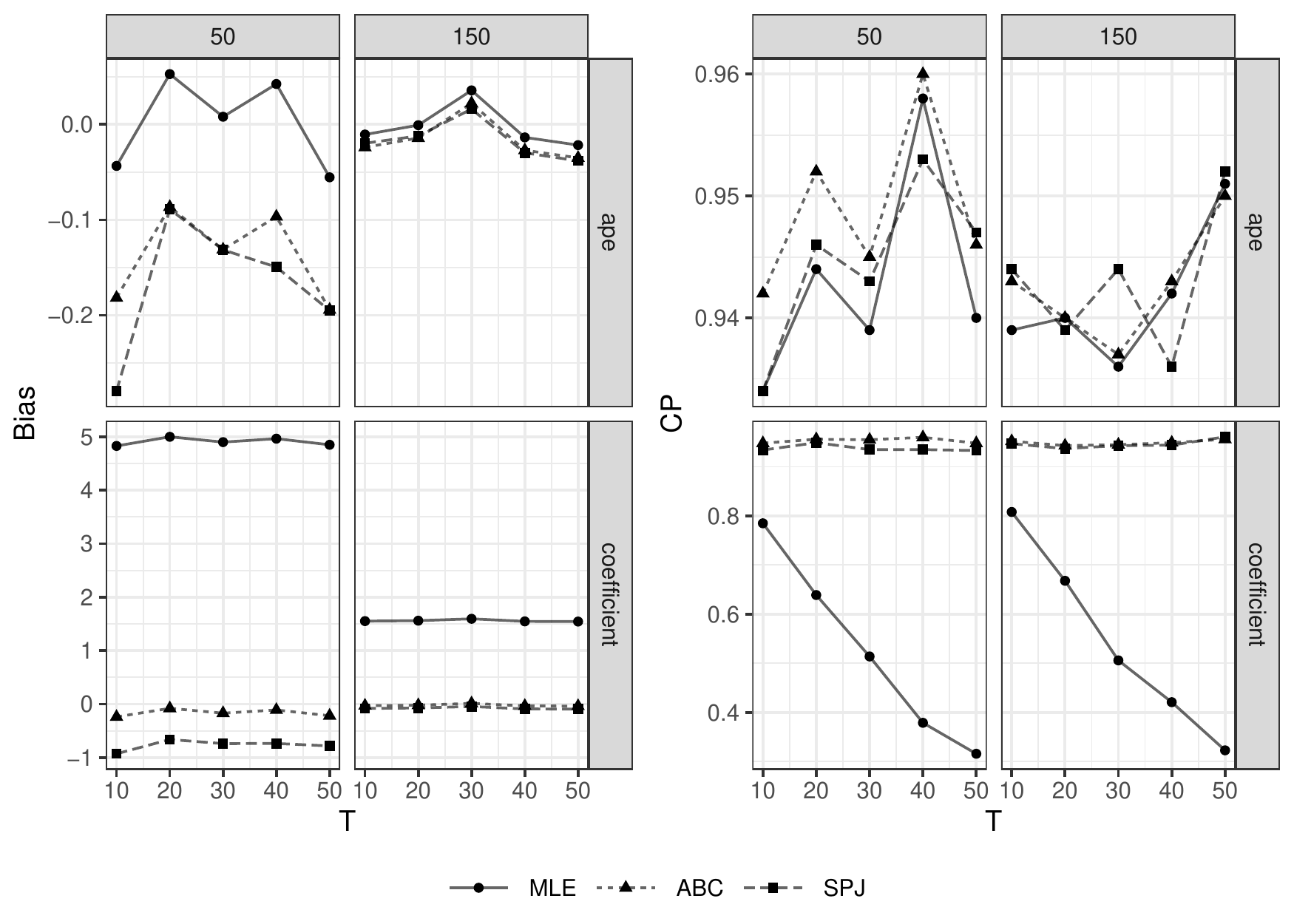}
	\vspace{1em}
    \caption{Dynamic: Two-way Fixed Effects --- Exogenous Regressor}
	\label{fig:dyn2way_x}
	\includegraphics[height = 0.45\textheight]{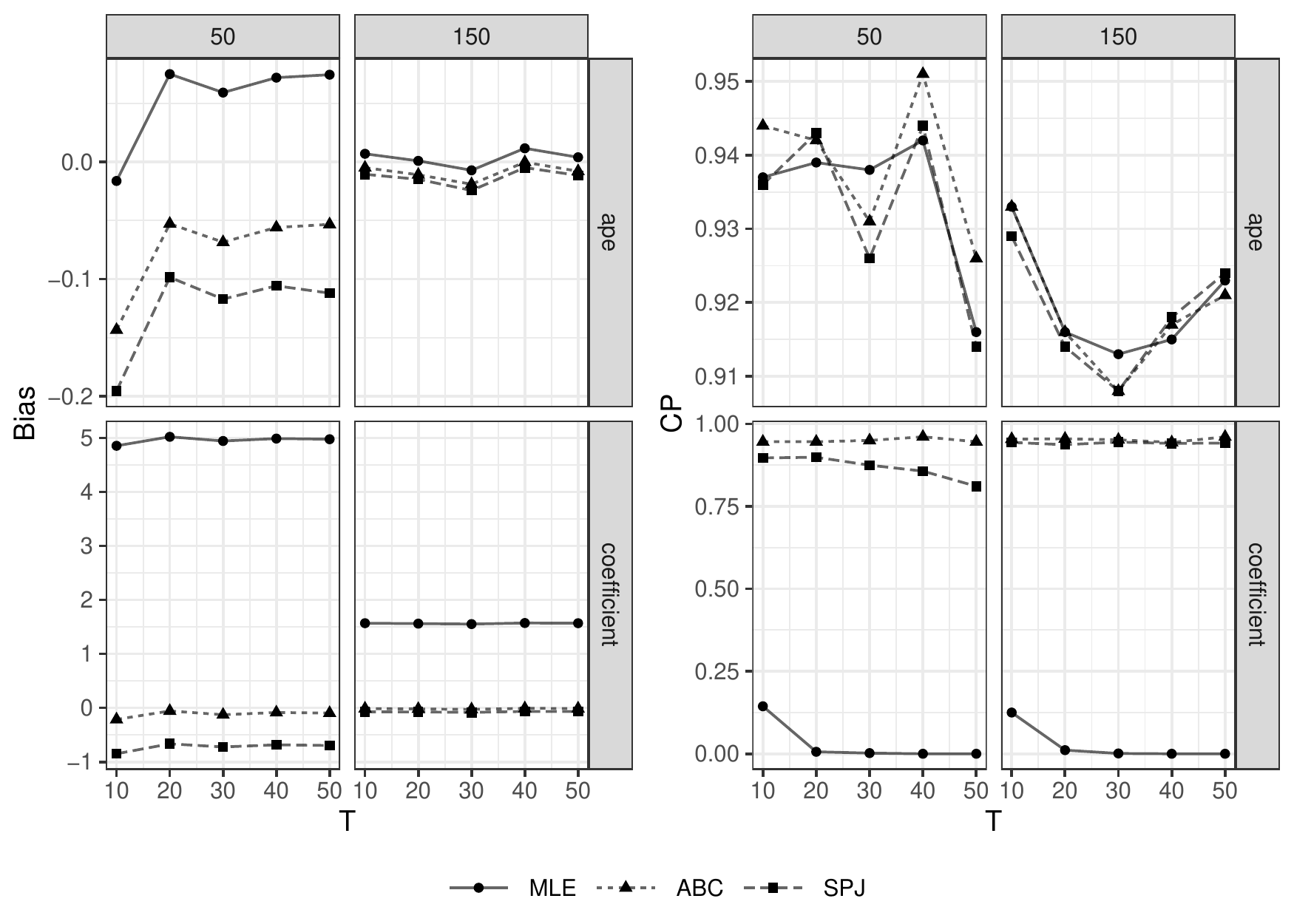}
\end{figure}

\newpage

\begin{figure}[ht]
	\centering
	\caption{Dynamic: Two-way Fixed Effects ---  Exogenous Regressor (Long-Run)}
	\label{fig:dyn2way_xlong}
	\includegraphics[height = 0.45\textheight]{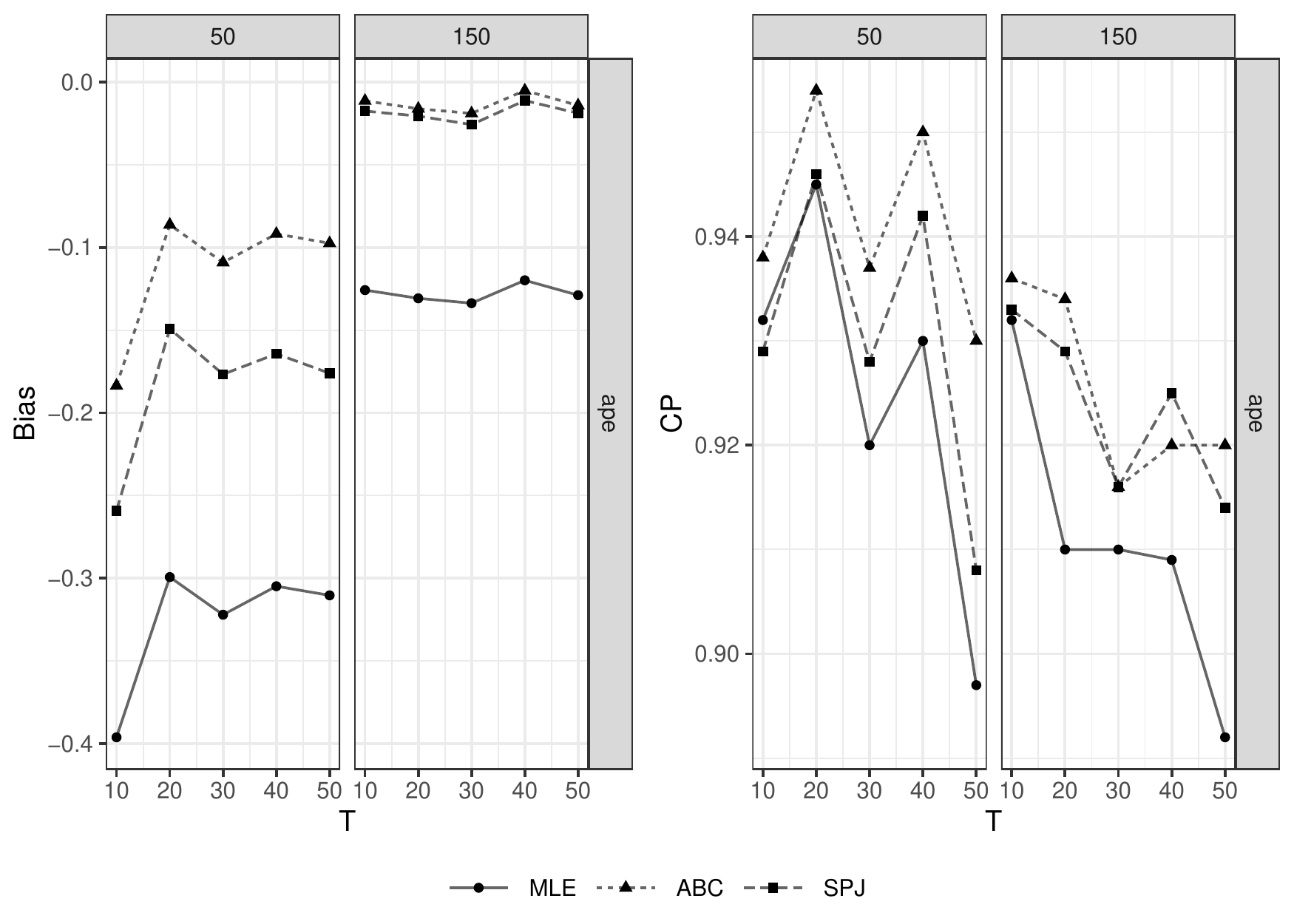}
\end{figure}

\begin{table}[!ht]
	\begin{minipage}{.5\linewidth}
		\centering
		\caption{\small Dynamic: Two-way FEs --- $z$, $N = 50$}
		\label{tab:dyn2way_xn50}
		\resizebox{\linewidth}{!}{
			\begin{tabular}{lrrrrrrrr}
			\toprule
&\multicolumn{4}{c}{Coefficients}&\multicolumn{4}{c}{APE}\\
\cmidrule(lr){2-5}\cmidrule(lr){6-9}
&Bias&Bias/SE&SE/SD&CP .95&Bias&Bias/SE&SE/SD&CP .95\\
\midrule
&\multicolumn{8}{c}{N = 50; T = 10}\\
\cmidrule(lr){2-9}
MLE & 4.86 & 3.10 & 0.95 & 0.14 & -0.02 & -0.01 & 0.96 & 0.94 \\ 
ABC & -0.21 & -0.14 & 0.99 & 0.95 & -0.14 & -0.11 & 0.97 & 0.94 \\ 
SPJ & -0.85 & -0.56 & 0.96 & 0.90 & -0.20 & -0.14 & 0.96 & 0.94 \\ 
\midrule
&\multicolumn{8}{c}{N = 50; T = 20}\\
\cmidrule(lr){2-9}
MLE & 5.02 & 4.53 & 0.97 & 0.01 & 0.08 & 0.08 & 0.96 & 0.94 \\ 
ABC & -0.06 & -0.05 & 1.01 & 0.95 & -0.05 & -0.05 & 0.97 & 0.94 \\ 
SPJ & -0.67 & -0.62 & 0.97 & 0.90 & -0.10 & -0.10 & 0.95 & 0.94 \\ 
\midrule
&\multicolumn{8}{c}{N = 50; T = 30}\\
\cmidrule(lr){2-9}
MLE & 4.94 & 5.46 & 0.96 & 0.00 & 0.06 & 0.08 & 0.93 & 0.94 \\ 
ABC & -0.13 & -0.14 & 1.00 & 0.95 & -0.07 & -0.09 & 0.94 & 0.93 \\ 
SPJ & -0.72 & -0.82 & 0.96 & 0.88 & -0.12 & -0.15 & 0.93 & 0.93 \\ 
\midrule
&\multicolumn{8}{c}{N = 50; T = 40}\\
\cmidrule(lr){2-9}
MLE & 4.99 & 6.36 & 1.00 & 0.00 & 0.07 & 0.11 & 0.95 & 0.94 \\ 
ABC & -0.09 & -0.11 & 1.04 & 0.96 & -0.06 & -0.08 & 0.96 & 0.95 \\ 
SPJ & -0.68 & -0.90 & 1.03 & 0.86 & -0.11 & -0.15 & 0.94 & 0.94 \\ 
\midrule
&\multicolumn{8}{c}{N = 50; T = 50}\\
\cmidrule(lr){2-9}
MLE & 4.98 & 7.09 & 0.97 & 0.00 & 0.07 & 0.12 & 0.91 & 0.92 \\ 
ABC & -0.10 & -0.14 & 1.01 & 0.95 & -0.05 & -0.09 & 0.92 & 0.93 \\ 
SPJ & -0.70 & -1.02 & 0.96 & 0.81 & -0.11 & -0.18 & 0.89 & 0.91 \\ 
				\bottomrule
			\end{tabular}
		}
	\end{minipage}
	\begin{minipage}{.5\linewidth}
		\centering
		\caption{\small Dynamic: Two-way FEs --- $z$, $N = 100$}
		\label{tab:dyn2way_xn100}
		\resizebox{\linewidth}{!}{
			\begin{tabular}{lrrrrrrrr}
			\toprule
&\multicolumn{4}{c}{Coefficients}&\multicolumn{4}{c}{APE}\\
\cmidrule(lr){2-5}\cmidrule(lr){6-9}
&Bias&Bias/SE&SE/SD&CP .95&Bias&Bias/SE&SE/SD&CP .95\\
\midrule
&\multicolumn{8}{c}{N = 100; T = 10}\\
\cmidrule(lr){2-9}
MLE & 2.34 & 3.08 & 0.97 & 0.15 & -0.02 & -0.02 & 0.91 & 0.92 \\ 
ABC & -0.07 & -0.09 & 0.99 & 0.95 & -0.05 & -0.06 & 0.91 & 0.92 \\ 
SPJ & -0.22 & -0.29 & 0.98 & 0.94 & -0.06 & -0.08 & 0.91 & 0.92 \\ 
\midrule
&\multicolumn{8}{c}{N = 100; T = 20}\\
\cmidrule(lr){2-9}
MLE & 2.38 & 4.42 & 1.01 & 0.00 & 0.01 & 0.02 & 0.94 & 0.94 \\ 
ABC & -0.03 & -0.06 & 1.03 & 0.95 & -0.02 & -0.04 & 0.94 & 0.93 \\ 
SPJ & -0.17 & -0.32 & 1.02 & 0.94 & -0.03 & -0.06 & 0.93 & 0.93 \\ 
\midrule
&\multicolumn{8}{c}{N = 100; T = 30}\\
\cmidrule(lr){2-9}
MLE & 2.36 & 5.38 & 0.97 & 0.00 & 0.00 & 0.00 & 0.91 & 0.92 \\ 
ABC & -0.05 & -0.10 & 0.99 & 0.95 & -0.03 & -0.06 & 0.91 & 0.93 \\ 
SPJ & -0.18 & -0.41 & 0.96 & 0.92 & -0.04 & -0.09 & 0.90 & 0.92 \\ 
\midrule
&\multicolumn{8}{c}{N = 100; T = 40}\\
\cmidrule(lr){2-9}
MLE & 2.39 & 6.29 & 0.98 & 0.00 & 0.02 & 0.05 & 0.90 & 0.92 \\ 
ABC & -0.02 & -0.05 & 0.99 & 0.95 & -0.01 & -0.03 & 0.91 & 0.92 \\ 
SPJ & -0.15 & -0.40 & 0.97 & 0.93 & -0.02 & -0.06 & 0.89 & 0.92 \\ 
\midrule
&\multicolumn{8}{c}{N = 100; T = 50}\\
\cmidrule(lr){2-9}
MLE & 2.40 & 7.04 & 0.99 & 0.00 & 0.02 & 0.05 & 0.91 & 0.92 \\ 
ABC & -0.01 & -0.04 & 1.01 & 0.95 & -0.01 & -0.03 & 0.91 & 0.92 \\ 
SPJ & -0.14 & -0.43 & 0.99 & 0.92 & -0.02 & -0.06 & 0.90 & 0.92 \\ 
\bottomrule
			\end{tabular}
		}
	\end{minipage}
	\begin{minipage}{.25\linewidth}~\end{minipage}
	\begin{minipage}{.5\linewidth}
		\vspace{1em}
		\centering
		\caption{\small Dynamic: Two-way FEs --- $z$, $N = 150$}
		\label{tab:dyn2way_xn150}
		\resizebox{\linewidth}{!}{
			\begin{tabular}{lrrrrrrrr}
			\toprule
&\multicolumn{4}{c}{Coefficients}&\multicolumn{4}{c}{APE}\\
\cmidrule(lr){2-5}\cmidrule(lr){6-9}
&Bias&Bias/SE&SE/SD&CP .95&Bias&Bias/SE&SE/SD&CP .95\\
\midrule
&\multicolumn{8}{c}{N = 150; T = 10}\\
\cmidrule(lr){2-9}
MLE & 1.57 & 3.13 & 0.99 & 0.12 & 0.01 & 0.01 & 0.89 & 0.93 \\ 
ABC & -0.01 & -0.03 & 1.00 & 0.95 & -0.00 & -0.01 & 0.89 & 0.93 \\ 
SPJ & -0.07 & -0.15 & 0.99 & 0.94 & -0.01 & -0.02 & 0.89 & 0.93 \\ 
\midrule
&\multicolumn{8}{c}{N = 150; T = 20}\\
\cmidrule(lr){2-9}
MLE & 1.56 & 4.39 & 0.98 & 0.01 & 0.00 & 0.00 & 0.87 & 0.92 \\ 
ABC & -0.02 & -0.06 & 0.99 & 0.95 & -0.01 & -0.03 & 0.87 & 0.92 \\ 
SPJ & -0.08 & -0.21 & 0.97 & 0.94 & -0.01 & -0.04 & 0.87 & 0.91 \\ 
\midrule
&\multicolumn{8}{c}{N = 150; T = 30}\\
\cmidrule(lr){2-9}
MLE & 1.55 & 5.35 & 0.98 & 0.00 & -0.01 & -0.02 & 0.86 & 0.91 \\ 
ABC & -0.03 & -0.10 & 0.99 & 0.95 & -0.02 & -0.06 & 0.86 & 0.91 \\ 
SPJ & -0.08 & -0.29 & 0.98 & 0.94 & -0.02 & -0.08 & 0.85 & 0.91 \\ 
\midrule
&\multicolumn{8}{c}{N = 150; T = 40}\\
\cmidrule(lr){2-9}
MLE & 1.57 & 6.26 & 0.98 & 0.00 & 0.01 & 0.04 & 0.89 & 0.92 \\ 
ABC & -0.01 & -0.03 & 0.99 & 0.94 & -0.00 & -0.00 & 0.89 & 0.92 \\ 
SPJ & -0.07 & -0.26 & 0.98 & 0.94 & -0.00 & -0.02 & 0.89 & 0.92 \\ 
\midrule
&\multicolumn{8}{c}{N = 150; T = 50}\\
\cmidrule(lr){2-9}
MLE & 1.57 & 6.98 & 1.01 & 0.00 & 0.00 & 0.02 & 0.91 & 0.92 \\ 
ABC & -0.01 & -0.05 & 1.03 & 0.96 & -0.01 & -0.03 & 0.91 & 0.92 \\ 
SPJ & -0.07 & -0.30 & 1.01 & 0.94 & -0.01 & -0.05 & 0.90 & 0.92 \\ 
\bottomrule
			\end{tabular}
		}
	\end{minipage}
\end{table}

\begin{table}[!ht]
	\begin{minipage}{.5\linewidth}
		\centering
		\caption{\small Dynamic: Two-way FEs --- $y_{t-1}$, $N = 50$}
		\label{tab:dyn2way_yn50}
		\resizebox{\linewidth}{!}{
			\begin{tabular}{lrrrrrrrr}
	\toprule
&\multicolumn{4}{c}{Coefficients}&\multicolumn{4}{c}{APE}\\
\cmidrule(lr){2-5}\cmidrule(lr){6-9}
&Bias&Bias/SE&SE/SD&CP .95&Bias&Bias/SE&SE/SD&CP .95\\
\midrule
&\multicolumn{8}{c}{N = 50; T = 10}\\
\cmidrule(lr){2-9}
MLE & 4.83 & 1.10 & 0.95 & 0.78 & -0.04 & -0.01 & 0.95 & 0.93 \\ 
ABC & -0.24 & -0.05 & 0.99 & 0.95 & -0.18 & -0.04 & 0.97 & 0.94 \\ 
SPJ & -0.93 & -0.21 & 0.96 & 0.93 & -0.28 & -0.06 & 0.94 & 0.93 \\ 
\midrule
&\multicolumn{8}{c}{N = 50; T = 20}\\
\cmidrule(lr){2-9}
MLE & 5.00 & 1.60 & 0.98 & 0.64 & 0.05 & 0.02 & 0.96 & 0.94 \\ 
ABC & -0.08 & -0.03 & 1.02 & 0.96 & -0.09 & -0.03 & 0.98 & 0.95 \\ 
SPJ & -0.66 & -0.21 & 1.00 & 0.95 & -0.09 & -0.03 & 0.96 & 0.95 \\ 
\midrule
&\multicolumn{8}{c}{N = 50; T = 30}\\
\cmidrule(lr){2-9}
MLE & 4.90 & 1.92 & 0.94 & 0.51 & 0.01 & 0.00 & 0.94 & 0.94 \\ 
ABC & -0.17 & -0.07 & 0.98 & 0.95 & -0.13 & -0.05 & 0.96 & 0.94 \\ 
SPJ & -0.74 & -0.29 & 0.95 & 0.94 & -0.13 & -0.05 & 0.94 & 0.94 \\ 
\midrule
&\multicolumn{8}{c}{N = 50; T = 40}\\
\cmidrule(lr){2-9}
MLE & 4.97 & 2.24 & 0.99 & 0.38 & 0.04 & 0.02 & 0.99 & 0.96 \\ 
ABC & -0.11 & -0.05 & 1.03 & 0.96 & -0.10 & -0.04 & 1.01 & 0.96 \\ 
SPJ & -0.74 & -0.34 & 1.00 & 0.94 & -0.15 & -0.07 & 0.99 & 0.95 \\ 
\midrule
&\multicolumn{8}{c}{N = 50; T = 50}\\
\cmidrule(lr){2-9}
MLE & 4.85 & 2.45 & 0.98 & 0.32 & -0.06 & -0.03 & 0.97 & 0.94 \\ 
ABC & -0.22 & -0.11 & 1.02 & 0.95 & -0.19 & -0.10 & 0.98 & 0.95 \\ 
SPJ & -0.78 & -0.40 & 1.01 & 0.93 & -0.19 & -0.10 & 0.97 & 0.95 \\ 
				\bottomrule
			\end{tabular}
		}
	\end{minipage}
	\begin{minipage}{.5\linewidth}
		\centering
		\caption{\small Dynamic: Two-way FEs --- $y_{t-1}$, $N = 100$}
		\label{tab:dyn2way_yn100}
		\resizebox{\linewidth}{!}{
			\begin{tabular}{lrrrrrrrr}
	\toprule
&\multicolumn{4}{c}{Coefficients}&\multicolumn{4}{c}{APE}\\
\cmidrule(lr){2-5}\cmidrule(lr){6-9}
&Bias&Bias/SE&SE/SD&CP .95&Bias&Bias/SE&SE/SD&CP .95\\
\midrule
&\multicolumn{8}{c}{N = 100; T = 10}\\
\cmidrule(lr){2-9}
MLE & 2.44 & 1.14 & 1.03 & 0.80 & 0.08 & 0.04 & 1.04 & 0.95 \\ 
ABC & 0.03 & 0.01 & 1.05 & 0.95 & 0.05 & 0.02 & 1.04 & 0.96 \\ 
SPJ & -0.10 & -0.05 & 1.04 & 0.95 & 0.06 & 0.03 & 1.04 & 0.95 \\ 
\midrule
&\multicolumn{8}{c}{N = 100; T = 20}\\
\cmidrule(lr){2-9}
MLE & 2.35 & 1.54 & 0.98 & 0.66 & -0.03 & -0.02 & 0.99 & 0.94 \\ 
ABC & -0.06 & -0.04 & 1.00 & 0.94 & -0.06 & -0.04 & 1.00 & 0.95 \\ 
SPJ & -0.19 & -0.13 & 0.99 & 0.94 & -0.06 & -0.04 & 0.99 & 0.94 \\ 
\midrule
&\multicolumn{8}{c}{N = 100; T = 30}\\
\cmidrule(lr){2-9}
MLE & 2.39 & 1.92 & 0.97 & 0.52 & 0.03 & 0.02 & 0.99 & 0.95 \\ 
ABC & -0.02 & -0.02 & 0.99 & 0.95 & -0.00 & -0.00 & 0.99 & 0.95 \\ 
SPJ & -0.14 & -0.11 & 0.96 & 0.94 & -0.00 & -0.00 & 0.97 & 0.95 \\ 
\midrule
&\multicolumn{8}{c}{N = 100; T = 40}\\
\cmidrule(lr){2-9}
MLE & 2.38 & 2.20 & 0.97 & 0.41 & 0.00 & 0.00 & 0.97 & 0.94 \\ 
ABC & -0.03 & -0.03 & 0.99 & 0.95 & -0.03 & -0.03 & 0.98 & 0.94 \\ 
SPJ & -0.16 & -0.15 & 0.98 & 0.94 & -0.03 & -0.03 & 0.97 & 0.94 \\ 
\midrule
&\multicolumn{8}{c}{N = 100; T = 50}\\
\cmidrule(lr){2-9}
MLE & 2.40 & 2.48 & 0.98 & 0.29 & 0.02 & 0.02 & 0.97 & 0.94 \\ 
ABC & -0.01 & -0.01 & 1.00 & 0.95 & -0.01 & -0.01 & 0.98 & 0.95 \\ 
SPJ & -0.14 & -0.15 & 0.97 & 0.93 & -0.02 & -0.02 & 0.96 & 0.93 \\ 
\bottomrule
			\end{tabular}
		}
	\end{minipage}
	\begin{minipage}{.25\linewidth}~\end{minipage}
	\begin{minipage}{.5\linewidth}
		\vspace{1em}
		\centering
		\caption{\small Dynamic: Two-way FEs --- $y_{t-1}$, $N = 150$}
		\label{tab:dyn2way_yn150}
		\resizebox{\linewidth}{!}{
			\begin{tabular}{lrrrrrrrr}
	\toprule
&\multicolumn{4}{c}{Coefficients}&\multicolumn{4}{c}{APE}\\
\cmidrule(lr){2-5}\cmidrule(lr){6-9}
&Bias&Bias/SE&SE/SD&CP .95&Bias&Bias/SE&SE/SD&CP .95\\
\midrule
&\multicolumn{8}{c}{N = 150; T = 10}\\
\cmidrule(lr){2-9}
MLE & 1.55 & 1.09 & 0.99 & 0.81 & -0.01 & -0.01 & 0.99 & 0.94 \\ 
ABC & -0.03 & -0.02 & 1.00 & 0.95 & -0.02 & -0.02 & 0.99 & 0.94 \\ 
SPJ & -0.08 & -0.06 & 0.99 & 0.95 & -0.02 & -0.01 & 0.98 & 0.94 \\ 
\midrule
&\multicolumn{8}{c}{N = 150; T = 20}\\
\cmidrule(lr){2-9}
MLE & 1.56 & 1.55 & 0.97 & 0.67 & -0.00 & -0.00 & 0.96 & 0.94 \\ 
ABC & -0.02 & -0.02 & 0.98 & 0.94 & -0.01 & -0.01 & 0.96 & 0.94 \\ 
SPJ & -0.07 & -0.07 & 0.97 & 0.94 & -0.01 & -0.01 & 0.96 & 0.94 \\ 
\midrule
&\multicolumn{8}{c}{N = 150; T = 30}\\
\cmidrule(lr){2-9}
MLE & 1.60 & 1.94 & 1.00 & 0.51 & 0.04 & 0.04 & 0.99 & 0.94 \\ 
ABC & 0.01 & 0.02 & 1.01 & 0.94 & 0.02 & 0.03 & 1.00 & 0.94 \\ 
SPJ & -0.04 & -0.05 & 1.01 & 0.94 & 0.02 & 0.02 & 0.99 & 0.94 \\ 
\midrule
&\multicolumn{8}{c}{N = 150; T = 40}\\
\cmidrule(lr){2-9}
MLE & 1.55 & 2.17 & 0.99 & 0.42 & -0.01 & -0.02 & 0.97 & 0.94 \\ 
ABC & -0.03 & -0.05 & 1.01 & 0.95 & -0.03 & -0.04 & 0.97 & 0.94 \\ 
SPJ & -0.09 & -0.13 & 0.99 & 0.94 & -0.03 & -0.04 & 0.96 & 0.94 \\ 
\midrule
&\multicolumn{8}{c}{N = 150; T = 50}\\
\cmidrule(lr){2-9}
MLE & 1.54 & 2.42 & 1.00 & 0.32 & -0.02 & -0.03 & 0.98 & 0.95 \\ 
ABC & -0.04 & -0.06 & 1.01 & 0.96 & -0.04 & -0.05 & 0.98 & 0.95 \\ 
SPJ & -0.09 & -0.15 & 1.00 & 0.96 & -0.04 & -0.06 & 0.97 & 0.95 \\ 
\bottomrule
			\end{tabular}
		}
	\end{minipage}
\end{table}

\begin{table}[!ht]
	\begin{minipage}{.5\linewidth}
		\centering
		\caption{\small Dynamic: Two-way FEs --- $z$, $N = 50$ long-run}
		\label{tab:dyn2waylong_xn50}
		\resizebox{0.6\linewidth}{!}{
			\begin{tabular}{lrrrr}
\toprule
&\multicolumn{4}{c}{APE}\\
\cmidrule(lr){2-5}
&Bias&Bias/SE&SE/SD&CP .95\\
\midrule
&\multicolumn{4}{c}{N = 50; T = 10}\\
\cmidrule(lr){2-5}
MLE & -0.40 & -0.30 & 0.97 & 0.93 \\ 
ABC & -0.18 & -0.14 & 0.96 & 0.94 \\ 
SPJ & -0.26 & -0.19 & 0.95 & 0.93 \\ 
\midrule
&\multicolumn{4}{c}{N = 50; T = 20}\\
\cmidrule(lr){2-5}
MLE & -0.30 & -0.32 & 1.00 & 0.94 \\ 
ABC & -0.09 & -0.09 & 0.99 & 0.95 \\ 
SPJ & -0.15 & -0.16 & 0.97 & 0.95 \\ 
\midrule
&\multicolumn{4}{c}{N = 50; T = 30}\\
\cmidrule(lr){2-5}
MLE & -0.32 & -0.41 & 0.97 & 0.92 \\ 
ABC & -0.11 & -0.14 & 0.96 & 0.94 \\ 
SPJ & -0.18 & -0.23 & 0.93 & 0.93 \\ 
\midrule
&\multicolumn{4}{c}{N = 50; T = 40}\\
\cmidrule(lr){2-5}
MLE & -0.30 & -0.45 & 1.00 & 0.93 \\ 
ABC & -0.09 & -0.14 & 0.99 & 0.95 \\ 
SPJ & -0.16 & -0.24 & 0.97 & 0.94 \\ 
\midrule
&\multicolumn{4}{c}{N = 50; T = 50}\\
\cmidrule(lr){2-5}
MLE & -0.31 & -0.52 & 0.92 & 0.90 \\ 
ABC & -0.10 & -0.16 & 0.92 & 0.93 \\ 
SPJ & -0.18 & -0.29 & 0.88 & 0.91 \\ 
\bottomrule
			\end{tabular}
		}
	\end{minipage}
	\begin{minipage}{.5\linewidth}
		\centering
		\caption{\small Dynamic: Two-way FEs --- $z$, $N = 100$ long-run}
		\label{tab:dyn2waylong_xn100}
		\resizebox{0.6\linewidth}{!}{
			\begin{tabular}{lrrrr}
	\toprule
	&\multicolumn{4}{c}{APE}\\
	\cmidrule(lr){2-5}
	&Bias&Bias/SE&SE/SD&CP .95\\
	\midrule
	&\multicolumn{4}{c}{N = 100; T = 10}\\
	\cmidrule(lr){2-5}
	MLE & -0.20 & -0.27 & 0.95 & 0.93 \\ 
	ABC & -0.05 & -0.06 & 0.94 & 0.94 \\ 
	SPJ & -0.06 & -0.08 & 0.94 & 0.94 \\ 
	\midrule
	&\multicolumn{4}{c}{N = 100; T = 20}\\
	\cmidrule(lr){2-5}
	MLE & -0.19 & -0.35 & 0.96 & 0.93 \\ 
	ABC & -0.03 & -0.06 & 0.95 & 0.94 \\ 
	SPJ & -0.05 & -0.09 & 0.94 & 0.94 \\ 
	\midrule
	&\multicolumn{4}{c}{N = 100; T = 30}\\
	\cmidrule(lr){2-5}
	MLE & -0.19 & -0.43 & 0.95 & 0.92 \\ 
	ABC & -0.03 & -0.07 & 0.94 & 0.94 \\ 
	SPJ & -0.05 & -0.11 & 0.93 & 0.93 \\ 
	\midrule
	&\multicolumn{4}{c}{N = 100; T = 40}\\
	\cmidrule(lr){2-5}
	MLE & -0.18 & -0.47 & 0.94 & 0.91 \\ 
	ABC & -0.02 & -0.06 & 0.93 & 0.94 \\ 
	SPJ & -0.03 & -0.09 & 0.92 & 0.94 \\ 
	\midrule
	&\multicolumn{4}{c}{N = 100; T = 50}\\
	\cmidrule(lr){2-5}
	MLE & -0.17 & -0.51 & 0.95 & 0.91 \\ 
	ABC & -0.02 & -0.05 & 0.94 & 0.93 \\ 
	SPJ & -0.03 & -0.09 & 0.93 & 0.93 \\ 
				\bottomrule
			\end{tabular}
		}
	\end{minipage}
	\begin{minipage}{.25\linewidth}~\end{minipage}
	\begin{minipage}{.5\linewidth}
		\vspace{1em}
		\centering
		\caption{\small Dynamic: Two-way FEs --- $z$, $N = 150$ long-run}
		\label{tab:dyn2waylong_xn150}
		\resizebox{0.6\linewidth}{!}{
			\begin{tabular}{lrrrr}
\toprule
&\multicolumn{4}{c}{APE}\\
\cmidrule(lr){2-5}
&Bias&Bias/SE&SE/SD&CP .95\\
\midrule
&\multicolumn{4}{c}{N = 150; T = 10}\\
\cmidrule(lr){2-5}
MLE & -0.13 & -0.22 & 0.93 & 0.93 \\ 
ABC & -0.01 & -0.02 & 0.93 & 0.94 \\ 
SPJ & -0.02 & -0.03 & 0.92 & 0.93 \\ 
\midrule
&\multicolumn{4}{c}{N = 150; T = 20}\\
\cmidrule(lr){2-5}
MLE & -0.13 & -0.33 & 0.91 & 0.91 \\ 
ABC & -0.02 & -0.04 & 0.91 & 0.93 \\ 
SPJ & -0.02 & -0.05 & 0.90 & 0.93 \\ 
\midrule
&\multicolumn{4}{c}{N = 150; T = 30}\\
\cmidrule(lr){2-5}
MLE & -0.13 & -0.41 & 0.89 & 0.91 \\ 
ABC & -0.02 & -0.06 & 0.89 & 0.92 \\ 
SPJ & -0.03 & -0.08 & 0.88 & 0.92 \\ 
\midrule
&\multicolumn{4}{c}{N = 150; T = 40}\\
\cmidrule(lr){2-5}
MLE & -0.12 & -0.42 & 0.92 & 0.91 \\ 
ABC & -0.01 & -0.02 & 0.91 & 0.92 \\ 
SPJ & -0.01 & -0.04 & 0.91 & 0.92 \\ 
\midrule
&\multicolumn{4}{c}{N = 150; T = 50}\\
\cmidrule(lr){2-5}
MLE & -0.13 & -0.51 & 0.92 & 0.89 \\ 
ABC & -0.01 & -0.06 & 0.92 & 0.92 \\ 
SPJ & -0.02 & -0.07 & 0.91 & 0.91 \\ 
				\bottomrule
			\end{tabular}
		}
	\end{minipage}
\end{table}

\clearpage
\subsection{Static: Three-way Fixed Effects}\label{sec:montecarlo_extended}








\begin{equation*}
    y_{ijt} = \ind\left[\beta_z z_{ijt} + \psi_{jt}+ \lambda_{it} +   \mu_{ij} \geq \epsilon_{ijt} \right] \, , 
\end{equation*}

where   $\psi_{jt} \sim \iid \N(0, 1 / 24)$, $\lambda_{it} \sim \iid \N(0, 1 / 24)$,  $\mu_{ij} \sim \iid \N(0, 1 / 24)$,  and $\epsilon_{ijt} \sim \iid \N(0, 1)$.  Further, $\beta_{z} = 1$ and $z_{ijt} = 0.5z_{ijt-1} +  \psi_{jt} + \lambda_{it} + \mu_{ij} +   \nu_{ijt}$, where $\nu_{ijt} \sim \iid \N(0, 0.5)$, $z_{ij0} \sim \iid \N(0, 1)$.\\



\begin{table}[!ht]
	\begin{minipage}{.5\linewidth}
		\centering
		\caption{\small Static: Three-way FEs --- $z$, $N = 50$}
		\label{tab:stat3way_xn50}
		\resizebox{\linewidth}{!}{
			\begin{tabular}{lrrrrrrrr}
				\toprule
&\multicolumn{4}{c}{Coefficients}&\multicolumn{4}{c}{APE}\\
\cmidrule(lr){2-5}\cmidrule(lr){6-9}
&Bias&Bias/SE&SE/SD&CP .95&Bias&Bias/SE&SE/SD&CP .95\\
\midrule
&\multicolumn{8}{c}{N = 50; T = 10}\\
\cmidrule(lr){2-9}
MLE & 21.50 & 11.16 & 0.85 & 0.00 & 0.99 & 0.72 & 0.93 & 0.86 \\ 
ABC & -1.46 & -0.84 & 1.02 & 0.87 & -1.33 & -0.92 & 1.00 & 0.84 \\ 
SPJ & -11.69 & -7.10 & 0.69 & 0.00 & -0.35 & -0.24 & 0.80 & 0.88 \\ 
\midrule
&\multicolumn{8}{c}{N = 50; T = 20}\\
\cmidrule(lr){2-9}
MLE & 12.15 & 10.11 & 0.89 & 0.00 & 0.39 & 0.42 & 1.00 & 0.93 \\ 
ABC & -0.59 & -0.53 & 0.99 & 0.91 & -0.45 & -0.47 & 1.04 & 0.92 \\ 
SPJ & -3.99 & -3.60 & 0.85 & 0.10 & -0.65 & -0.68 & 0.91 & 0.87 \\ 
\midrule
&\multicolumn{8}{c}{N = 50; T = 30}\\
\cmidrule(lr){2-9}
MLE & 9.51 & 10.07 & 0.90 & 0.00 & 0.27 & 0.36 & 0.98 & 0.94 \\ 
ABC & -0.40 & -0.44 & 0.98 & 0.93 & -0.25 & -0.32 & 1.00 & 0.94 \\ 
SPJ & -2.34 & -2.63 & 0.91 & 0.28 & -0.45 & -0.58 & 0.93 & 0.90 \\ 
\midrule
&\multicolumn{8}{c}{N = 50; T = 40}\\
\cmidrule(lr){2-9}
MLE & 8.36 & 10.40 & 0.95 & 0.00 & 0.24 & 0.37 & 1.03 & 0.93 \\ 
ABC & -0.24 & -0.32 & 1.02 & 0.94 & -0.15 & -0.22 & 1.04 & 0.95 \\ 
SPJ & -1.65 & -2.17 & 0.94 & 0.42 & -0.31 & -0.47 & 0.96 & 0.92 \\ 
\midrule
&\multicolumn{8}{c}{N = 50; T = 50}\\
\cmidrule(lr){2-9}
MLE & 7.60 & 10.70 & 0.93 & 0.00 & 0.19 & 0.31 & 1.05 & 0.95 \\ 
ABC & -0.23 & -0.34 & 0.99 & 0.94 & -0.14 & -0.23 & 1.06 & 0.95 \\ 
SPJ & -1.39 & -2.05 & 0.93 & 0.44 & -0.28 & -0.46 & 0.97 & 0.92 \\ 
				\bottomrule
			\end{tabular}
		}
	\end{minipage}
	\begin{minipage}{.5\linewidth}
		\centering
		\caption{\small Static: Three-way FEs --- $z$, $N = 100$}
		\label{tab:stat3way_xn100}
		\resizebox{\linewidth}{!}{
			\begin{tabular}{lrrrrrrrr}
				\toprule
&\multicolumn{4}{c}{Coefficients}&\multicolumn{4}{c}{APE}\\
\cmidrule(lr){2-5}\cmidrule(lr){6-9}
&Bias&Bias/SE&SE/SD&CP .95&Bias&Bias/SE&SE/SD&CP .95\\
\midrule
&\multicolumn{8}{c}{N = 100; T = 10}\\
\cmidrule(lr){2-9}
MLE & 17.62 & 19.04 & 0.86 & 0.00 & 0.56 & 0.79 & 1.01 & 0.88 \\ 
ABC & -0.86 & -1.02 & 1.01 & 0.83 & -1.05 & -1.44 & 1.05 & 0.71 \\ 
SPJ & -8.01 & -9.78 & 0.73 & 0.00 & 0.46 & 0.62 & 0.87 & 0.87 \\ 
\midrule
&\multicolumn{8}{c}{N = 100; T = 20}\\
\cmidrule(lr){2-9}
MLE & 9.12 & 15.69 & 0.90 & 0.00 & 0.22 & 0.47 & 1.00 & 0.92 \\ 
ABC & -0.29 & -0.53 & 0.97 & 0.91 & -0.27 & -0.57 & 1.02 & 0.92 \\ 
SPJ & -2.42 & -4.41 & 0.88 & 0.01 & -0.24 & -0.49 & 0.93 & 0.90 \\ 
\midrule
&\multicolumn{8}{c}{N = 100; T = 30}\\
\cmidrule(lr){2-9}
MLE & 6.68 & 14.60 & 0.96 & 0.00 & 0.13 & 0.33 & 1.00 & 0.94 \\ 
ABC & -0.18 & -0.40 & 1.01 & 0.94 & -0.14 & -0.36 & 1.01 & 0.94 \\ 
SPJ & -1.22 & -2.77 & 0.94 & 0.22 & -0.18 & -0.46 & 0.95 & 0.91 \\ 
\midrule
&\multicolumn{8}{c}{N = 100; T = 40}\\
\cmidrule(lr){2-9}
MLE & 5.51 & 14.18 & 0.95 & 0.00 & 0.08 & 0.24 & 0.99 & 0.95 \\ 
ABC & -0.14 & -0.37 & 1.00 & 0.94 & -0.10 & -0.30 & 0.99 & 0.93 \\ 
SPJ & -0.81 & -2.14 & 0.95 & 0.44 & -0.15 & -0.44 & 0.95 & 0.91 \\ 
\midrule
&\multicolumn{8}{c}{N = 100; T = 50}\\
\cmidrule(lr){2-9}
MLE & 4.85 & 14.10 & 0.91 & 0.00 & 0.06 & 0.21 & 0.99 & 0.94 \\ 
ABC & -0.11 & -0.32 & 0.95 & 0.92 & -0.07 & -0.24 & 1.00 & 0.94 \\ 
SPJ & -0.60 & -1.78 & 0.90 & 0.56 & -0.12 & -0.38 & 0.95 & 0.93 \\ 
\midrule
			\end{tabular}
		}
	\end{minipage}
	\begin{minipage}{.25\linewidth}~\end{minipage}
	\begin{minipage}{.5\linewidth}
		\vspace{1em}
		\centering
		\caption{\small Static: Three-way FEs --- $z$, $N = 150$}
		\label{tab:stat3way_xn150}
		\resizebox{\linewidth}{!}{
			\begin{tabular}{lrrrrrrrr}
				\toprule
&\multicolumn{4}{c}{Coefficients}&\multicolumn{4}{c}{APE}\\
\cmidrule(lr){2-5}\cmidrule(lr){6-9}
&Bias&Bias/SE&SE/SD&CP .95&Bias&Bias/SE&SE/SD&CP .95\\
\midrule
&\multicolumn{8}{c}{N = 150; T = 10}\\
\cmidrule(lr){2-9}
MLE & 16.39 & 26.88 & 0.88 & 0.00 & 0.41 & 0.84 & 0.99 & 0.86 \\ 
ABC & -0.73 & -1.29 & 1.02 & 0.76 & -1.01 & -2.01 & 1.02 & 0.49 \\ 
SPJ & -7.12 & -13.07 & 0.73 & 0.00 & 0.63 & 1.24 & 0.85 & 0.73 \\ 
\midrule
&\multicolumn{8}{c}{N = 150; T = 20}\\
\cmidrule(lr){2-9}
MLE & 8.16 & 21.28 & 0.97 & 0.00 & 0.18 & 0.54 & 1.02 & 0.92 \\ 
ABC & -0.22 & -0.60 & 1.04 & 0.92 & -0.24 & -0.72 & 1.03 & 0.89 \\ 
SPJ & -2.06 & -5.66 & 0.91 & 0.00 & -0.15 & -0.45 & 0.92 & 0.90 \\ 
\midrule
&\multicolumn{8}{c}{N = 150; T = 30}\\
\cmidrule(lr){2-9}
MLE & 5.78 & 19.17 & 0.97 & 0.00 & 0.10 & 0.37 & 1.01 & 0.95 \\ 
ABC & -0.12 & -0.41 & 1.02 & 0.94 & -0.11 & -0.41 & 1.02 & 0.94 \\ 
SPJ & -0.98 & -3.36 & 0.95 & 0.09 & -0.13 & -0.47 & 0.97 & 0.92 \\ 
\midrule
&\multicolumn{8}{c}{N = 150; T = 40}\\
\cmidrule(lr){2-9}
MLE & 4.64 & 18.08 & 0.97 & 0.00 & 0.05 & 0.22 & 1.03 & 0.94 \\ 
ABC & -0.10 & -0.40 & 1.01 & 0.93 & -0.08 & -0.36 & 1.03 & 0.94 \\ 
SPJ & -0.62 & -2.46 & 0.96 & 0.32 & -0.11 & -0.46 & 0.99 & 0.92 \\ 
\midrule
&\multicolumn{8}{c}{N = 150; T = 50}\\
\cmidrule(lr){2-9}
MLE & 4.01 & 17.65 & 1.01 & 0.00 & 0.05 & 0.24 & 1.01 & 0.95 \\ 
ABC & -0.06 & -0.27 & 1.05 & 0.95 & -0.05 & -0.22 & 1.01 & 0.95 \\ 
SPJ & -0.42 & -1.89 & 0.98 & 0.53 & -0.07 & -0.36 & 0.97 & 0.92 \\ 
\bottomrule
			\end{tabular}
		}
	\end{minipage}
\end{table}


\clearpage

\section{Application}
\label{app:application}

\begin{figure}[!h]
    \centering
    \caption{Distribution of frequency of switches between zero and non-zero trade.}
    \label{fig:frequency_switches}
    \includegraphics[width=0.9\linewidth]{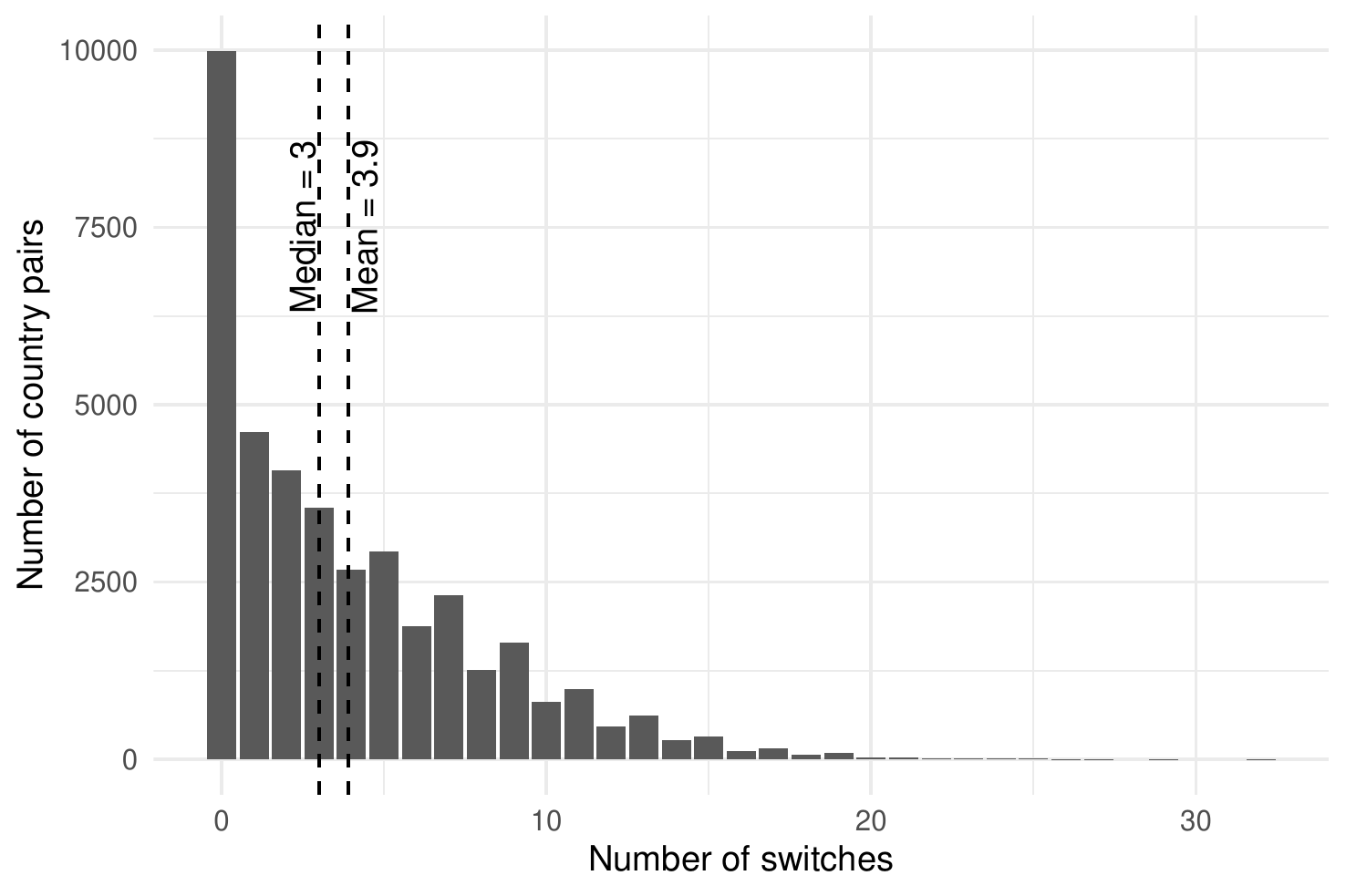}
\end{figure}

\begin{table}[!h]
	\vspace{1em}
	\centering
	\caption{Homogeneity Test: Split-Panel Jackknife}
	\label{tab:homogeneity}
	\begin{tabular}{rrrr}
		\toprule
	 	& $I$ & $J$ & $T$\\
		\midrule
		(2) & 175.055  & 70.700 & - \\ 
		 & (0.000) & (0.000) & (-)\\[0.25em]
	    (3) & 56.681 & 72.546  & -\\ 
		& (0.000) & (0.000) & (-) \\[0.25em]
	    (4) & 14.388 & 23.684 & 73.773 \\ 
		& (0.002)  &  (0.000) & (0.000) \\[0.25em]
	    (5) & 26.085 & 20.669 & 72.056 \\ 
		& (0.000) & (0.000) & (0.000) \\[0.25em]
	\midrule
		\multicolumn{4}{p{0.4\textwidth}}{\footnotesize \textit{Notes:}  Test statistics of Wald tests for equality of structural parameters across sub panels splitted by $I$, $J$ and $T$; p-values in parantheses.} \\
	\bottomrule
	\end{tabular}
\end{table}

\begin{table}[p]
	\centering
	\small
	\caption{Probit Estimation: Coefficients}
		\label{tab:application_probit_coefficient}
\begin{tabular}{l*{6}{c}}
\toprule
& \multicolumn{5}{c}{Dependent variable: $y_{ijt}$} \\
\cmidrule{2-6}
& (1) & (2) & (3) & (4) & (5) \\
\midrule
$y_{ij(t-1)}$ & - &  -& \textbf{1.752}*** &  - & \textbf{1.169}***\\
& (-) & (-) & (\textbf{0.004}) & (-) & (\textbf{0.005})\\		
& - & - & 1.801*** & - & 1.090***\\
& (-) & (-)  & (0.004) & (-)  & (0.005)\\[0.25em] 
log(Distance) & - & \textbf{-0.818}*** & \textbf{-0.530}*** & - & - \\
& (-) & (\textbf{0.003}) & (\textbf{0.003}) & (-) & (-) \\
& -0.711*** & -0.836*** & -0.546*** & - & - \\
& (0.002) & (0.003) & (0.003) & (-)  & (-)\\[0.25em] 
Land border & - & \textbf{0.140}*** & \textbf{0.078}*** & - & - \\
& (-) & (\textbf{0.014}) & (\textbf{0.016}) & (-)  & (-) \\
& 0.169*** & 0.149*** & 0.089*** & - & - \\
& (0.013) & (0.014) & (0.016) & (-)  & (-) \\[0.25em] 
Legal & - & \textbf{0.032}*** & \textbf{0.022}*** & - & - \\
& (-) & (\textbf{0.004}) & (\textbf{0.004}) & (-)  & (-) \\
& 0.024*** & 0.034*** & 0.023*** & - & - \\
& (0.004) & (0.004) & (0.004) & (-) & (-)\\[0.25em] 
Language & - & \textbf{0.410}*** & \textbf{0.271}*** & - & - \\
& (-) & (\textbf{0.005}) & (\textbf{0.006}) & (-)  & (-) \\
& 0.368*** & 0.420*** & 0.280*** & - & - \\
& (0.005) & (0.005) & (0.006) & (-) & (-) \\[0.25em] 
Colonial ties & - & \textbf{0.297}*** & \textbf{0.182}*** & - & - \\
& (-) & (\textbf{0.006}) & (\textbf{0.007}) & (-) & (-) \\
& 0.268*** & 0.302*** & 0.186*** &  - & - \\
& (0.006) & (0.006) & (0.007) & (-)  & (-) \\[0.25em] 
Currency union & - & \textbf{0.555}*** & \textbf{0.355}*** &  \textbf{0.276}*** & \textbf{0.195}***\\
& (-) & (\textbf{0.014}) & (\textbf{0.016}) & (\textbf{0.035}) & (\textbf{0.038})\\
& 0.356*** & 0.571*** & 0.370*** & 0.308*** & 0.228***\\
& (0.012) & (0.014) & (0.017) & (0.036) & (0.038)\\[0.25em] 
FTA & - & \textbf{0.310}*** & \textbf{0.199}*** & \textbf{0.102}*** & \textbf{0.069}***\\
& (-) & (\textbf{0.007}) & (\textbf{0.008}) & (\textbf{0.015}) & (\textbf{0.016})\\
& 0.265*** & 0.313*** & 0.200*** & 0.101*** & 0.075***\\
& (0.006) & (0.007) & (0.008) & (0.015) & (0.015)\\[0.25em] 
WTO / GATT & - & \textbf{0.196}*** & \textbf{0.127}*** & \textbf{0.105}*** & \textbf{0.076}***\\
& (-) & (\textbf{0.007}) & (\textbf{0.008}) & (\textbf{0.014}) & (\textbf{0.015})\\
& 0.258*** & 0.200*** & 0.131*** & 0.113*** & 0.089***\\
& (0.004) & (0.007) & (0.008) & (0.014) & (0.015)\\
\midrule
Fixed effects  & $i,j,t$ & $it,jt$ & $it,jt$ & $it,jt,ij$ & $it,jt,ij$ \\
Sample size & 1652296 & 1652296 & 1613354 & 1652296 & 1613354   \\
Deviance    & 11.65$\times$10\textsuperscript{4}  & 9.90$\times$10\textsuperscript{4}  & 7.17$\times$10\textsuperscript{4}    & 6.50$\times$10\textsuperscript{4}     & 5.76$\times$10\textsuperscript{4}    \\
\midrule
\multicolumn{6}{p{0.88\textwidth}}{\footnotesize \textit{Notes:}  Column (1) uncorrected coefficients, columns (2) - (5) bias-corrected coefficients (bold font) and uncorrected coefficients (standard font). Columns (4) and (5) bias-corrected with $L = 2$. Standard errors in parenthesis. $^{***}p<0.01$, $^{**}p<0.05$, $^*p<0.1$} \\
\bottomrule
\end{tabular}
	
\end{table}

\begin{table}
	\centering
	\small
	\caption{\footnotesize Logit Estimation with Different Bandwidths: Bias-Corrected Average Partial Effects}
	\label{tab:application_trim_logit_ape}
	\resizebox{\linewidth}{!}{
		\begin{tabular}{l*{8}{c}}
			\toprule
			& \multicolumn{8}{c}{Dependent variable: $y_{ijt}$} \\
			\cmidrule{2-9}
			& \multicolumn{2}{c}{$L = 1$} 	& \multicolumn{2}{c}{$L = 2$} 	& \multicolumn{2}{c}{$L = 3$} 	& \multicolumn{2}{c}{$L = 4$} \\
			\cmidrule(lr){2-3}\cmidrule(lr){4-5}\cmidrule(lr){6-7}\cmidrule(lr){8-9}
			&\textit{direct}&\textit{long-run}&\textit{direct}&\textit{long-run}&\textit{direct}&\textit{long-run}&\textit{direct}&\textit{long-run}\\
			\midrule
$y_{ij(t-1)}$ & 0.163*** & - & 0.169*** & - & 0.172*** & - & 0.173*** & \\
& (0.001) & (-) & (0.001) & (-) & (0.001) & (-) & (0.001) & (-)\\[0.25em] 
Currency union  & 0.023*** & 0.035*** & 0.022*** & 0.035*** & 0.022*** & 0.035*** & 0.022*** & 0.035***\\
& (0.004) & (0.006) & (0.004) & (0.006) & (0.004) & (0.006) & (0.004) & (0.006)\\[0.25em] 
FTA & 0.008*** & 0.013*** & 0.008*** & 0.013*** & 0.008*** & 0.013*** & 0.008*** & 0.013***\\
& (0.002) & (0.002) & (0.002) & (0.002) & (0.002) & (0.002) & (0.002) & (0.002)\\[0.25em] 
WTO/GATT & 0.009*** & 0.014*** & 0.009*** & 0.014*** & 0.009*** & 0.014*** & 0.009*** & 0.014***\\[0.25em] 
& (0.001) & (0.002) & (0.001) & (0.002) & (0.001) & (0.002) & (0.001) & (0.002)\\		
			\midrule
			\multicolumn{9}{p{1.07\linewidth}}{\footnotesize \textit{Notes:} All columns include Origin $\times$ Year, Destination $\times$ Year and Origin $\times$ Destination fixed effects. Standard errors in parenthesis. $^{***}p<0.01$, $^{**}p<0.05$, $^*p<0.1$} \\
			\bottomrule
		\end{tabular}
	}
\end{table}

\end{document}